\documentclass{aa}  
\usepackage{graphicx}
\usepackage{natbib}
\usepackage{aalongtable,lscape}
\usepackage[figuresright]{rotating}
\usepackage[]{html}
\bibpunct{(}{)}{;}{a}{}{,}
\usepackage{txfonts}
\begin{document}
   \title{Automated supervised classification of variable stars.}
   \subtitle{II. Application to the OGLE database.\thanks{Variability Catalogue available from the $\aap$ anonymous ftp site.}\fnmsep\thanks{Figures \ref{bulge-rrlyr-f1-r21} to \ref{DSCUT-BULGE-4} , and tables \ref{BCEP-SMC-param} to \ref{DSCUT-BULGE-param-2} are only available in
   the electronic form of the paper.}}

   \author{L. M. Sarro\inst{1}
         \and
          J. Debosscher\inst{2}
          \and
          M. L\'opez\inst{3}
          \and
          C. Aerts\inst{2,4}
          }

   \institute{Dpt. de Inteligencia Artificial , UNED, Juan del Rosal, 16, 28040 Madrid, Spain
         \and
           Instituut voor Sterrenkunde, Catholic University of Leuven, Celestijnenlaan 200D, 3001 Leuven, Belgium 
          \and 
           Laboratorio de Astrof{\'{i}}sica Espacial y F{\'{i}}sica Fundamental, INTA, Apartado de Correos 50727, 28080 Madrid, Spain
          \and
           Department of Astrophysics, Radboud University Nijmegen, POBox 9010, 6500 GL Nijmegen, the Netherlands}
            
 \date{}


  \abstract
   { Scientific exploitation of large variability databases can only be
fully optimized if these archives contain, besides the actual
observations, annotations about the variability class of the objects
they contain. Supervised classification of observations produces these
tags, and makes it possible to generate refined candidate lists and
catalogues suitable for further investigation.}
      {We aim to extend and test the classifiers
presented in a previous work against an independent
dataset. We complement the assessment of the validity of
the classifiers by applying them to the set of OGLE light curves
treated as variable objects of unknown class. The results are compared
to published classification results based on the so-called extractor
methods.}
   {Two complementary analyses are carried out in parallel. In both
cases, the original time series of OGLE observations of the Galactic
bulge and Magellanic Clouds are processed in order to identify and
characterize the frequency components. In the first approach, the
classifiers are applied to the data and the results analyzed in terms
of systematic errors and differences between the definition samples in
the training set and in the extractor rules. In the second approach,
the original classifiers are extended with colour information and,
again, applied to OGLE light curves.}
   {We have constructed a classification system that can process huge
amounts of time series in negligible time and provide reliable samples
of the main variability classes. We have evaluated its strengths and
weaknesses and provide potential users of the classifier with a
detailed description of its characteristics to aid in the
interpretation of classification results. Finally, we apply the
classifiers to obtain object samples of classes not previously
studied in the OGLE database and analyse the results. We pay specific 
attention to the B-stars in the samples, as their pulsations are 
strongly dependent on metallicity.}
{}
   \maketitle
%
%
\section{Introduction}

In the last decade, astronomy witnessed several major advances. The
advent of large detection arrays, the operation of robotic telescopes
and the consolidation of high duty cycle space missions have provided
astronomers with a wealth of observations with unprecedented
sensitivity in virtually the whole electromagnetic spectrum during
long uninterrupted periods of time. At the same time, the ever-growing
storage capacity of digital devices has made it possible to archive
and make these enormous datasets available. The consolidation of the
Virtual Observatory (VO) technology and the interoperability provided
by its services make it possible for the astronomer to work
consistently on large portions of the electromagnetic spectrum,
combining different data models (magnitudes, colours, spectra, radial
velocities, etc).

The traditional procedures for data reduction and
analysis do not scale with the sizes of the available data
warehouses. Some of its components have been automated and can now
be carried out in a systematic way, but it is becoming evident that
optimal scientific exploitation of these databases requires the
addition of information inferred from the observed data to enable the
extraction of homogeneous (in some sense) samples of observations for
further specific studies that could not be applied to the
entire database. The process by which this added value is extracted is
widely known as Knowledge Discovery and relies mostly on recent
advances in the artificial intelligence fields of pattern recognition,
statistical learning or multi-agent systems. 

The use of these new techniques has the particular advantage that,
once accepted that every search for a given type of object is biased
{\sl ab initio} by the adopted definition of that class, automatic
classifiers produce consistent object lists according to the same
objective and stable criteria openly declared in the so-called
training set. We thus eliminate subjective and unquantifiable
considerations inherent to, for example, visual inspection and produce
object samples comparable across different surveys. 

Altogether, the integration of Computer Science techniques (Grid
computing, Artificial Intelligence and VO technology) and domain
knowledge (physics in this case), and the new possibilities that
this synergy offers are known as e-Science. Science proceeds in much
the same way as before; the e- prefix only provides the basis to
approach more ambitious scientific challenges, feasible on the grounds
of more and better quality data.

In \citet[ hereafter paper I]{PaperI} we introduced the problem of the
scientific analysis of variable objects and proposed several methods
to classify new objects on the basis of their photometric time
series. The OGLE database (see section \ref{OGLE} for a summary of its
objectives and characteristics) exemplifies some of the difficulties
described in previous paragraphs. Although not its principal target,
the OGLE survey has produced as a by-product hundreds of thousands of
light curves of objects in the Galactic bulge and in the Large and
Small Magellanic Clouds. These light curves have been analysed using
the so-called extractor methods. Extractor methods can be assimilated
to the classical rule-based systems where the target objects are
identified by defining characteristic attribute ranges (where attribute
is to be interpreted as any of the parameters used to describe the
object light curves such as the significant frequencies, harmonic
amplitudes or phase differences) where these objects must lie. In a
subsequent stage, individual light curves are visually inspected and
the object samples refined on a per object basis.

In this work we also present an extension of the classifiers defined in Paper I, to handle
photometric colours. In section \ref{OGLE} we summarize the objectives
and characteristics of the OGLE survey; section \ref{colours}
describes the sources and criteria used for the assignment of colours
to the training set and section \ref{results} compares the results of
the application of the classifiers (both with and without colours)
to the OGLE database (bulge and Magellanic Clouds) with object lists
available in the literature (obtained by means of extractor methods
and human intervention) for a reduced set of classes. Finally, we
analyse the object lists obtained with our classifiers for special
classes in the realm of multiperiodic variables, not previously
studied in an extensive way (to the best of our knowledge) in the context of the OGLE
database.

\section{\label{OGLE}The OGLE database and its published Catalogues of variables.}

 The Optical Gravitational Lensing Experiment (OGLE) is a long term
joint microlensing survey aimed at detecting the Galaxy dark matter
halo by its bending effect on the light coming from background
stars. As a by-product, the project has been generating light curves
of millions of stars of varying signal-to-noise ratiosx. The project
has undergone several major upgrades. The data treated here belong to
the OGLE-II phase of the project.

The OGLE database at the time of writing contains time series of
several hundred thousand variable objects, all of which have been
analysed by us, using the codes and techniques presented in Paper I.
The bulge, LMC and SMC OGLE catalogues have been searched for
particular variability types in the past (see Table
\ref{OGLE-extractor}) using extractor methods. In the
following sections we briefly describe, where possible, the extraction
rules used in the construction of each of the catalogues in order to
provide a proper framework for the analysis of the classification
results and to facilitate the explanation of possible discrepancies.

\begin{table*}
\caption{Published catalogues used in comparison with the
outcome of our classifiers.}
\label{OGLE-extractor}
\centering
\begin{tabular}{cccc}
  \hline
  \hline
  Variability class  & Source object & Reference  & Number of objects \\
  \hline
RR~Lyrae & LMC & \citet{Soszynski:2003} & 5455 (RRab) ; 1655 (RRc) ; 272 (RRe) ; 230 (RRd) \\
RR~Lyrae & SMC & \citet{Soszynski:2002} & 458 (RRab) ; 56 (RRc) ; 57 (RRd) \\
Cepheids & LMC & \citet{Udalski:1999b}& 1335 \\
Cepheids & SMC & \citet{Udalski:1999c}& 2049 \\
Double mode Cepheids & LMC & \citet{Soszynski:2000}& 81 \\
Double mode Cepheids & SMC & \citet{Udalski:1999d}&  95\\
Pop. II Cepheids & LMC & \citet{Kubiak:2003} & 14\\
Pop. II Cepheids & bulge & \citet{Kubiak:2003} & 54\\
Eclipsing binaries & LMC & \citet{Wyrzykowski:2003}& 2580\\
Eclipsing binaries & SMC & \citet{Wyrzykowski:2004}& 1350\\
Eclipsing binaries & LMC & \citet{Groenewegen:2005} & 178\\
Eclipsing binaries & SMC & \citet{Groenewegen:2005} & 16\\
Eclipsing binaries & Bulge & \citet{Groenewegen:2005} & 2053\\
Long period Variables & LMC & \citet{Soszynski:2005}& 3221\\
Mira & Bulge & \citet{2005MNRAS.364..117M} & 1968\\
Mira & Bulge & \citet{2005AandA...443..143G} &  2691\\
Various & Bulge & \citet{Mizerski:2002} & 4597  \\
$\delta$-Scuti & Bulge & \citet{2006MmSAI..77..223P}&193\\
\hline
\end{tabular}
\end{table*}

In Table \ref{OGLE-extractor} we include information on the number of
objects in each of the published catalogues. These numbers include
double detections in overlapping zones across different fields. We
include these double detections because they are represented by
independent light curves, and we are mainly interested in the
true/false positive/negative detection rates, not so much in the
objects lists themselves (except in the analysis of multiperiodic
variables).

The classifiers presented in Paper I and the colour extensions
presented here and discussed below were applied to the OGLE LMC/SMC
\citep{Zebrun:2001} and Galactic Bulge \citep{Wozniak:2002} catalogues
as downloaded from
\textit{http://bulge.astro.princeton.edu/$\sim$ogle/ogle2/dia/} and
\textit{ftp://bulge.princeton.edu/$\sim$ogle/ogle2/bulge\_dia\_variables}
respectively. Again, these catalogues contain duplicate entries that
we kept for the same reasons as above. According to \cite{Eyer:qso}
and \cite{Eyer:pm}, the catalogues include spurious detections of
variable objects. In \cite{Eyer:qso}, these spurious detections are
discussed and several systematic effects identified (chip
perturbations, mirror realuminization and proximity to bright
objects). In \cite{Eyer:pm}, the authors discover a type of artifact
introduced by the difference image analyses (DIA) consisting of the
occurrence of pairs of monotonic anti-correlated light curves as a
result of the presence of high proper motion stars in dense
fields. The impact of these artifacts is restricted to the Bulge
fields and, since i) they do not result in periodic signals and ii)
systematic trends are removed from the fits to the data (see Paper I),
we do not expect them to affect our results significantly.

The detailed study of the first type of artifacts is out of the scope
of this work. Nevertheless, it would be extremely interesting to
investigate how these artifacts are classified by our algorithms and,
most importantly, the possibility of detecting them as a separate
group by using clustering techniques. This is presently being studied
as part of the Gaia effort to ensure a robust data processing
pipeline.\\ In the five following subsections we discuss the work in
the literature already done from the OGLE light curves.  We regard
these ``human'' classification results as correct and compare our
automated results with them to evaluate the latter.

\subsection{RR~Lyrae variables}

The selection of the RR Lyrae variables in the OGLE catalogues was made
in several stages (see \citeauthor{Soszynski:2002},
\citeyear{Soszynski:2002} for the SMC and \citeauthor{Soszynski:2003},
\citeyear{Soszynski:2003} for the LMC). In the first stage, variable
stars were identified on the basis of the standard deviation of all
individual OGLE PSF measurements. Their light curves were analysed
using the Analysis of Variance (AoV) algorithm and all objects showing statistically
significant periodic signals were then visually inspected and manually
classified into one of several classes. In the second stage, DIA
photometry was used to select candidates with I magnitudes between 15
and 20 for the LMC (18.4 and 19.4 for the SMC), and with standard
deviations at least 0.01 mag above the median value of the
standard deviations of stars of equal brightness for the LMC (0.02 for
the SMC in the I band and 0.05 for the V band). Again, periodic
signals were searched for, and stars with periods longer than 1 day
and/or signal-to-noise ratios below 3.5 were rejected. Then, Fourier
analysis was performed and unspecified rules were applied to extract
each of the RR~Lyrae subtypes. Single mode pulsators were selected
according to their position in the $\log
P$-$R_{21}$($=\frac{A_{12}}{A_{11}}$) and $\log P$-$amplitude$
diagrams, where $R_{21}=\frac{A_{12}}{A_{11}}$ is the amplitude ratio
of the first two harmonics of the first significant frequency. The
separation of first overtone pulsators is based on a threshold of
$P>0.26$ d. Second overtone pulsators were selected amongst stars with
periods below 0.3 days as those with low amplitude sinusoidal light
curves, which involves again visual inspection of the light curves one
by one. Finally, double mode pulsators were sought by selecting those
stars with statistically significant second frequencies at a ratio
close to $0.745$ of the first one. Again, all light curves and power
spectra were carefully inspected before they were included in the
double mode RR~Lyrae stars catalogue.

Bulge RR~Lyrae variables in \citet{2004MNRAS.349..193S} were selected
by fitting an ellipse to the locus of stars in a diagram representing
the ratio of the second to first harmonic amplitude ($R_{21}$) and the
phase difference between these harmonics ($\phi_{21}$ or $PH_{12}$ in
Paper I; see e.g.  Fig. \ref{collinge}), and using a hard threshold
decision boundary, according to the method first proposed by
\citet{Alard:1996}. The ellipse is centered on (4.5 rad, 0.43) with
semi-major axis $a = 0.8$ and semi-minor axis $b = 0.17$, and the
angle between the horizontal and the major axis is $-10 \deg$. These
candidates have been further refined and analysed in a recent study by
\citet{Collinge:2006}.

\subsection{Cepheids}

The catalogues of Cepheid variables in the OGLE database have been
presented in \citet{Kubiak:2003}, \citet{Udalski:1999b} and
\citet{Udalski:1999c}. In the identification of Cepheids, objects with
I magnitudes brighter that 19.5 (LMC) and 20 (SMC) were selected for
further analysis based on the visual inspection of the light curves
and their position in the colour-magnitude diagram (CMD). The region
occupied by Cepheid pulsators has been defined by the authors to be upper
bounded by $I < 18.5$ and delimited in colour by $0.25 < V-I <
1.3$. Objects with no available colours or colours to the right of the
red boundary were recovered if their light curves were conspicuously
of the Cepheid type. Again, visual inspection of all light curves was
a main ingredient of the classification process.

Double mode Cepheids were identified amongst Cepheids by fixing the
range of allowed frequency ratios to $0.735\pm0.02$ (first overtone to
fundamental mode) or $0.805\pm0.02$ (second to first overtone) in the
case of the prewhitened search for second periods from Fourier
Analysis, and the same ratios $\pm0.015$ for the application of the
CLEAN algorithm \citep{Roberts:CLEAN}. See \citet{Udalski:1999d} and
\citet{Soszynski:2000} for the SMC and LMC catalogues respectively.

The catalogue of Population II Cepheids in the bulge has been presented in
\citet{Kubiak:2003}. It is defined in the period range between 0.6 and
a few days and, again, the selection was based on the visual inspection
of the light curve shapes and their similarities to those described by
\citet{1983A&A...124..108D}.

\subsection{Eclipsing binaries}

Eclipsing binaries in the Large and Small Magellanic Clouds have been
extracted using different methods. While the SMC eclipsing binaries
were identified on the basis of visual inspection of the folded light
curves of all variable objects \citep{Wyrzykowski:2004}, LMC eclipsing
binaries were preselected by a neural network
\citep{Wyrzykowski:2003}. An artificial neural network was trained on
two dimensional images of folded light curves of the first field
(LMC\_SC1), selected to separate unseen light curves into three main
types: eclipsing, sinusoidal and saw-shape. The training proceeded
until the mean training error\footnote{This is the resampling error
estimate mentioned in Paper I. Assessing the error rates of a
classifier by judging its performance on the same examples used in its
training produces overly optimistic estimates of the error. These
unrealistic estimates cannot be reproduced when the classifier is
applied to previously unseen objects.}  was below $10^{-8}$. Then, the
refinement and subclassification of the eclipsing candidates was
carried out by visual inspection of the folded light curves.

\citet{Groenewegen:2005} has constructed a catalogue of candidate eclipsing
binary systems in the Galactic bulge suitable for distance estimation
(mainly detached systems), based on the statistical properties of
the phased light curve and subsequent visual inspection. Furthermore,
\citet{Mizerski:2002} have provided a list of candidate W~UMa systems
based on typical values of the Fourier coefficients of their light
curve decompositions calculated by \citet{rucinski:1993}.

\subsection{Long period variables}

Catalogues of Mira and semiregular Variables in the LMC have been
presented in Soszynski et al. (2004, 2005). The frequency analysis was
similar to the one described for all previous variability types and
the selection criteria were based on the I band magnitude ($I < 17$)
and on the position in the period--NIR Wessenheit index diagram. In
this diagram, sequences C and C' (see \citeauthor{Wood:1999},
\citeyear{Wood:1999}) were identified as Miras and semiregular
Variables. Furthermore, stars in the B sequence can also be assigned
to the Mira-semiregular category if the secondary period falls in any
sequence except sequence A. No quantitative criterion was given to
separate sequences in the plot, so the assignment of a star to any of
the sequences is subjective.

Recently, \citet{2005AandA...443..143G} and \citet{2005MNRAS.364..117M}
have published catalogues of Mira variables in the Galactic bulge. The
selection criterion in the first case was simply based on $I$-band
light curve amplitudes (in the sense of peak-to-peak range) above 0.9 mag followed by visual inspection, and resulted in a sample of 2691
objects. In the second catalogue, the selection criteria were periods
above 100 days, amplitudes larger than 1.0 in the $V$ magnitude and
$\theta$ values (the phase dispersion minimization regularity
indicator) below 0.6, followed by visual inspection. This resulted in a
sample of 1968 Mira variables in the OGLE bulge fields.

\subsection{Delta Scuti stars in the Galactic bulge.}

\citet{Mizerski:2002} and \citet{2006MmSAI..77..223P} have published
lists of high amplitude $\delta$~Scuti (HADS) stars in the bulge fields. In
the first case (only the first bulge field), no criterion was given for the selection of the 11 HADS
candidates but reference is made to the use of luminosities in the
identification process. In the second work, a HADS star was defined as a
star with a period less than 0.25 days for which at least one harmonic
of the main mode was detected and which was not an RR~Lyrae star or
W~UMa system (distinguished by means of the Fourier coefficients and
visual inspection of the light curve).\\

\section{The extended classifier: using colour information}
\label{colours}

All objects in the bulge, LMC and SMC OGLE catalogues were subject to
the frequency analysis described in Paper I. The final numbers of
objects analysed with this method are 50708 in the LMC, 14473 in the
SMC and 214786 in the Galactic bulge.

In an effort to improve the performance of the classifiers presented
in Paper I, we have constructed alternative ones with colour
information added to the basic time series parameters described
therein. This is not a mere upgrade
making the previous release obsolete since many archives provide no
colour information for classification. This is the case, for example,
for the Optical Monitoring Camera onboard INTEGRAL, that has returned
thousands of light curves, only a small fraction of which have
diachronic colours available. 

The process to incorporate photometric colours in the classifiers
followed the same scheme described in Paper I for the time series
classifiers. For the training set presented there, a search was
conducted in the Hipparcos catalogue \citep{HIPPARCOS} and SIMBAD in
order to retrieve magnitudes in the Johnson photometric
system. Johnson's colours for training set objects from the OGLE
database (double mode pulsators and eclipsing binaries) were
preferentially retrieved from the catalogues by
\citet{Wyrzykowski:2003}, \citet{Soszynski:2000}, and
\citet{Soszynski:2002}. Additionally, the 2MASS catalogue
of \citet{2mass} was searched for counterparts in order to add the $J-H$
and $H-K$ colour attributes to the original training set. The search
was conducted imposing a 3 arcsec search radius and quality flags A
and/or B in the three bands. 

Synchronicity between the observations in the different passbands
cannot be assured when only SIMBAD colours were available. This is
especially relevant for the case of large amplitude variables where
observations in opposite phases of the light curve can lead to totally
erroneous colour indices. Fortunately, the vast majority of training
examples of large amplitude classes are taken either from the
HIPPARCOS/Tycho catalogue or from the OGLE database itself, thus
minimizing the impact of diachronic observations in our training set.

The inclusion of colour information was done separately for several
colour sets. In order to assess the relevance of the infrared colours
for the classification task, two versions of the training set (with
and without 2MASS colours) were constructed. Additionally, two
versions of each training set (with and without the $B-V$ colour) were
created. The reason for this is the fact that we were not able to
obtain $B-V$ colours for a large fraction of the OGLE bulge
variables. Therefore, the assessment of the classifier results
conducted on bulge variables (see below) only incorporates the $V-I$
and 2MASS colours.

As a result, $B-V$, $V-I$, $J-H$ and $H-K$ colours were obtained for
at least 77\% of the stars in the training set (1344 of 1754 instances). The
exact sizes of each training set are as follows:
\begin{enumerate}
\item $V-I$ : 1602 instances
\item $B-V$ and $V-I$: 1592 instances
\item $V-I$, $J-H$ and $H-K$: 1348 instances
\item $B-V$, $V-I$, $J-H$ and $H-K$: 1344 instances
\end{enumerate}

\begin{figure*}
   \centering
   \includegraphics[angle=-90,scale=.70]{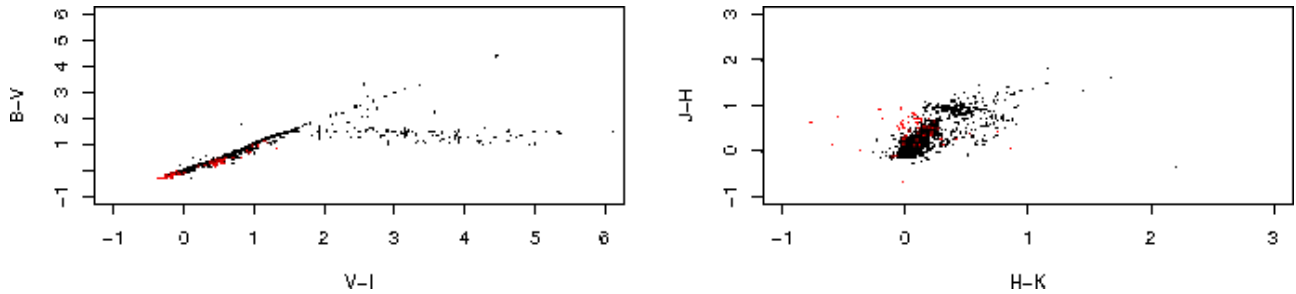}
   \caption{Training set colours. Black dots indicate training set
   objects in the Galaxy whereas red circles correspond to the classes
   defined with OGLE members, i.e. eclipsing binaries and double mode
   Cepheids and RR~Lyrae stars.}
   \label{ts-colours}
\end{figure*}
Figure \ref{ts-colours} shows two colour-colour diagrams for Johnson
and 2MASS photometry of the training set.

Str\"omgren colours were also searched in the catalogue by
\citet{Hauck:1998}. Unfortunately, they were only found to be
available for a much smaller fraction (less than 50\%) of the training
set and covering only certain variability classes, leaving the less
frequent ones almost unrepresented. A complete classifier with
ability to predict classes using Str\"omgren colours has been
developed only for multiperiodic variables, where the impact of such
information was found to be optimal, but will not be the subject of
analysis in the following.

Colours for the OGLE Galactic bulge, LMC and SMC objects used for
testing were obtained from the 2MASS and OGLE databases. 2MASS objects
within a search radius of 3 arcseconds and quality flags A or B were
assumed to be counterparts of the OGLE objects. With these parameters,
we retrieve 43351 instances (objects) with Johnson colours ($B-V$ and
$V-I$) amongst the 50708 LMC objects (see section \ref{OGLE}), and
26720 with combined Johnson and 2MASS colours; 12425 SMC objects with
Johnson colours and 6937 with Johnson and 2MASS colours; and 146034
bulge objects with $V-I$ (all of which have 2MASS photometry too). The
fraction of bulge objects with $B-V$ colours available was so small
that we preferred to work with $V-I$ and 2MASS photometry alone.

\begin{figure*}
   \centering
   \includegraphics[angle=-90,scale=.70]{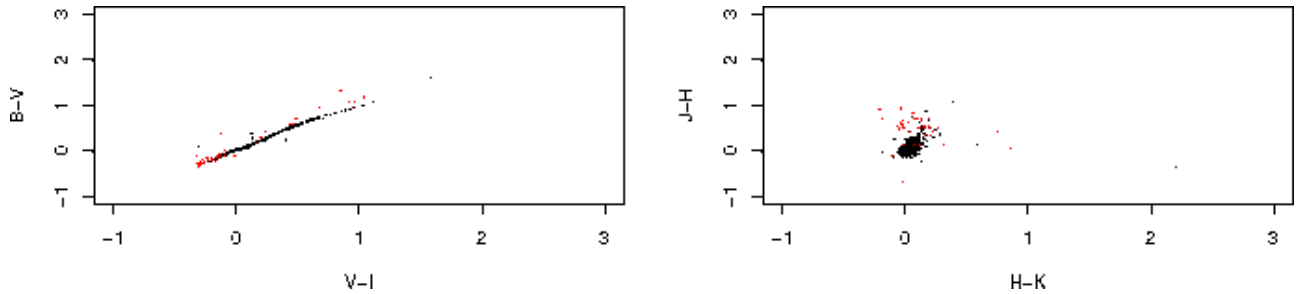}
   \caption{Colour-colour diagrams of the eclipsing binaries in the
   Hipparcos (black dots) and OGLE (red circles) catalogues. }
   \label{eclb}
\end{figure*}

We have found a systematic difference in the
$J-H$ colours of eclipsing binaries in the Hipparcos sample and in the
OGLE LMC catalogue. Figure \ref{eclb} shows two colour-colour diagrams
of Hipparcos and OGLE LMC eclipsing binaries in Johnson and 2MASS
photometric bands respectively.

Visual inspection of the plots reveals what seems a selection effect
in the choice of eclipsing binaries for the training set. The reason
for choosing OGLE eclipsing systems (all from the LMC) is their very
good sampling quality. It seems that favouring high signal-to-noise
ratios has biased the sample towards blue objects with an unexplained
excess in the $J-H$ colour. We have not found a plausible explanation
for the concurrence of both effects but we expect to improve the
eclipsing binaries prototypes in the training set with new examples
from the CoRoT database.

All objects from the OGLE database (either in the training set or in
the test set) have been dereddened using OGLE extinction maps:
\citet{Udalski:1999b} for the LMC and \citet{Udalski:1999c} for the
SMC. Objects in the Galactic bulge were dereddened using the
extinction maps by \citet{2004MNRAS.349..193S}. The extiction values
of OGLE field number 44 (missing in the original work due to the lack
of red clump giants well above the V band detection limit) are
approximated by the corresponding values in the closest OGLE field
(number 5). All extinction maps were combined with the classical CCM
extinction curve by \citet{1989ApJ...345..245C}. For bulge variables
this extinction curve produces corrections indistinguishable from
those of \citet{2003ARA&A..41..241D} used by
\citet{2005AandA...443..143G} in their analysis of Mira
variables. \citet{2003ApJ...594..279G} have studied the validity of
the classical CCM relationship for the Magellanic Clouds. Figures 2--6
in their work seem to suggest that the CCM curve is a safe
approximation (to within the measurement errors) of the Magellanic
Clouds extinction curves in the infrared bands considered here.

Unfortunately, the reddening correction applied to the colour
indices and described above will only produce strictly valid results
for stars at the mean distance of the red clump giants used in the
derivation of the extinction maps.Our correction may be less accurate
for other stars, but we do not have a better one available at present.

\section{\label{results}Classification results}

In the following, we will refer to the sets of objects classified in
one of the categories described in section \ref{OGLE} as class samples
(e.g., the RR~Lyrae sample or the Cepheids sample). In this section we
will compare the results obtained by the automatic classifiers with
those found in the literature. We have applied the battery of
classifiers presented in Paper I and their extensions to treat colour
information, to the OGLE LMC/SMC \citep{Zebrun:2001} and bulge
\citep{Wozniak:2002} variability archives. Full results of the
comparison of the statistical performance of the different classifiers
will be published in a specialized journal. Here we only report on the
overall best performing algorithm, the multi-stage classifier based on
Bayesian Networks (MSBN) as well as on the Gaussian Mixtures
classifier (GM), which was described in Paper I. The latter is simpler
in its design and interpretation and works better than the former for
the low-amplitude multiperiodic pulsator classes SPB and
$\gamma$-Doradus. It is thus best suited to retrieve these types of
asteroseismological targets (e.g. \cite{Cunha:2007} for a review). The
MSBN classifier on the other hand works better for the
larger-amplitude monoperiodic variables (including eclipsing
binaries), and for the other types of multiperiodic variables such as
BCEP or DSCUT stars.

The multi-stage classifier based on Bayesian Networks (MSBN) takes
advantage of several feature selection steps adapted to each
classification problem. Trying to select a global feature set for the
classification of the entire set of 35 classes results in a suboptimal
trade-off because attributes crucial for the separation of two classes
close to each other in the parameter space can be irrelevant
in identifying the remaining 33. On the contrary, dividing the
classification problem in several stages where smaller problems are
tackled allows for the particularized selection of feature sets
that are optimal in each step.

Several alternative groupings and orderings were attempted and
different algorithms tried in each step and the resulting performances
were either equal to or poorer using standard hypothesis testing
procedures. Although the search could never have been exhaustive, the
most reasonable combinations of groups of classes, orderings and
attribute selection techniques have been explored, the one presented
here resulting in the best overall performance. The classification
algorithms tried include neural networks, Bayesian networks, support
vector machines and Bayesian ensembles of neural networks; feature
selection techniques include the wrapper approach for those algorithms
where computation time made it feasible, and attribute set scores
based on correlation, mutual information and symmetrical uncertainty
between attributes and the class.

The MSBN has four stages of dichotomic classifiers, one for each of
the main categories of classical variables: stage 1 to separate
eclipsing from non-eclipsing variables; stage 2 to separate Cepheids
and non-Cepheids; stage 3 to separate the long period variables from
the rest, and stage 4 to separate RR~Lyrae variables from the rest
(stage 6 is also dichotomic, but corresponds to a more specialized
level that separates long period variables into the Mira and
Semiregular types). It starts with a first dichotomic classifier that
attempts to separate eclipsing binaries from all other variability
types. The attribute set used in the first and subsequent stages is
listed in Table \ref{tab-attributes}. The second dichotomic stage
separates the group of classes CLCEP, PTCEP, RVTAU and DMCEP (see
Table \ref{tabel1}, an abridged version of Table 2 in Paper I, for
class abbreviations) from all other classes. Then, a third classifier
attempts to identify the group of long period variables (MIRA and SR)
and a fourth classifier separates RR~Lyrae stars (RRAB, RRC and RRD)
from the rest of the classes. Complementary to these, there are
specialized classifiers that separate classes within groups. There is
a classifier for Cepheids that classifies CLCEP, PTCEP, RVTAU and
DMCEP, and equivalent classifiers for long period variables and
RR~Lyrae stars. The subclassification of eclipsing binaries is made
according to the methodology described in \citet{Sarro:2006}. Finally,
there is a classifier that separates all other classes not included in
the groupings described above, i.e. irregular and most multiperiodic
variables. The complete class probability vector for an object is
computed combining the output from all classifiers. For example, the
probability of belonging to class RRC is the probability of not being
an eclipsing binary (stage 1) times the probability of not being a
Cepheid (stage 2) times the probability of not being a long period
variable (stage 3) times the probability of being an RR~Lyrae pulsator
(stage 4) times the probability of being an RRC pulsating star (stage
7).

\begin{table}
\centering
\begin{tabular}{|c|c|}
  \hline
  Class& Abbreviation \\
  \hline
  Periodically variable supergiants  & PVSG \\
  Pulsating Be-stars & BE\\
  $\beta$-Cephei stars & BCEP \\
  Classical Cepheids & CLCEP \\
  Beat (double-mode)-Cepheids & DMCEP \\
  Population II Cepheids & PTCEP \\
  Chemically peculiar stars & CP \\
  $\delta$-Scuti stars & DSCUT \\
  $\lambda$-Bootis stars & LBOO \\
  SX-Phe stars & SXPHE \\
  $\gamma$-Doradus stars & GDOR \\
  Luminous Blue Variables & LBV \\
  Mira stars & MIRA \\
  Semi-Regular stars & SR \\
  RR-Lyrae, type RRab & RRAB \\
  RR-Lyrae, type RRc & RRC \\
  RR-Lyrae, type RRd & RRD \\
  RV-Tauri stars & RVTAU \\
  Slowly-pulsating B stars & SPB \\
  Solar-like oscillations in red giants & SLR \\
  Pulsating subdwarf B stars & SDBV \\
  Pulsating DA white dwarfs & DAV \\
  Pulsating DB white dwarfs & DBV \\
  GW-Virginis stars & GWVIR \\
  Rapidly oscillating Ap stars & ROAP \\
  T-Tauri stars & TTAU \\
  Herbig-Ae/Be stars & HAEBE \\
  FU-Ori stars & FUORI \\
  Wolf-Rayet stars & WR \\
  X-Ray binaries & XB \\
  Cataclysmic variables & CV \\
  Eclipsing binary, type EA & EA \\
  Eclipsing binary, type EB & EB \\
  Eclipsing binary, type EW & EW \\
  Ellipsoidal binaries & ELL \\
  \hline
\end{tabular}
\caption{Stellar variability classes and the code abbreviation used in
Paper I.}
 \label{tabel1}
\end{table}  

\begin{table*}
\caption{Attributes used in each classification stage by the
sequential classifier. Abbreviations used are as follows: log-fi
represents the logarithm of the i-th frequency; log-fi-fj is the
logarithm of the ratio fi/fj; afi represents the sum of squares of the
harmonic amplitudes in frequency i; log-afihj-t is the logarithm of
the total amplitude of the j-th harmonic of the i-th frequency;
log-crfij is the logarithm of the jth ratio of harmonic amplitudes of
the i-th frequency (j=0 corresponds to the ratio of the amplitude of
the second harmonic over that of the first, j=1, to the ratio of the
amplitude of the third harmonic over that of the first and so on);
log-crfihj-fi'hj' represents the logarithm of the amplitude ratio
between harmonics j and j' of frequencies i and i' respectively; pdfij
is the j-th phase difference between the various harmonics of the i-th
frequency (j=0 corresponds to the first and second harmonics, j=1, to
the third and first harmonics, and so on); varrat represents
the variance ratio defined in Paper I.}
\label{tab-attributes}
\centering
\begin{tabular}{cc}
  \hline
  \hline
  Stage  & Classes \& Attributes \\
  \hline
1 & Eclipsing/non eclipsing \\
 & log-f3, log-f2-f1, log-af1h1-t, log-crf10, log-crf15, log-crf20, log-crf25, log-crf32, log-crf33, pdf12, pdf13, pdf14, pdf23 \\
2 & Cepheids/non Cepheids  \\
 &  log-f1, log-f2-f1, af1, af2, log-af1h1-t, log-af1h2-t, log-crf3h1-f1h1, log-crf10, log-crf11, log-crf14, pdf12, varrat \\
3 & Red giants/non red giants  \\
 &  log-f1, log-f2, log-f3, log-f2-f1, af1, af2, log-af1h1-t, log-af2h3-t, log-crf21, log-crf24, log-crf30 \\
4 & RR~Lyrae/Non RR~Lyrae \\ 
 & log-f1, af1, log-af1h1-t, log-af3h1-t, log-crf3h1-f1h1, log-crf11, log-crf14, pdf12, pdf13 \\ 
5 & CLCEP/DMCEP/PTCEP/RVTAU \\
 & log-f1, log-af1h1-t, log-af2h3-t, log-af2h4-t, log-af3h4-t, log-crf12, log-crf32, pdf12, varrat\\
6 & MIRA/SR \\
 &   af1, log-af1h1-t, log-af1h3-t, log-af2h4-t, log-af3h3-t, varrat\\
7 & RRAB/RRC/RRD\\ 
 & log-f1, log-f2-f1, af1, log-af1h2-t, log-crf2h1-f1h1, log-crf10, pdf12, varrat\\ 
8 & PVSG BE BCEP CP DSCUT ELL GDOR HAEBE HMXB LBOO LBV PTCEP ROAP SPB SXPHE TTAU WR FUORI PSDB \\
 &   log-f1, log-af2h1-t, log-crf2h1-f1h1, log-crf10, log-crf13 \\ 
\hline
\end{tabular}
\end{table*}

In the next sections, the classifiers are applied to the entire list
of objects flagged by the OGLE team as variable. Also,
they are applied to the object samples referenced in section
\ref{OGLE}. Again, it has to be born in mind that not all objects in
the samples have been identified by the algorithms described in Paper
I as having at least a significant frequency and therefore, the column
named 'Total number of objects' in the following tables always refers
to this set of objects fulfilling the two criteria: being identified
in the literature as belonging to a variability class and with a
positive frequency identification. 

In general, the three populations observed by OGLE (the Galactic bulge
and the Large and Small Magellanic Clouds) are very different from a
statistical point of view. In this work we have found it clearer to
illustrate the performance of the classifiers with plots of the LMC
samples since they represent a compromise in the number of stars in
each sample, both sufficient for statistical purposes and, at the same
time, not so large that the plots become uninterpretable. Equivalent
plots for the Galactic Bulge populations are included as online
material (corresponding to the results presented by
\cite{Mizerski:2002} for the first bulge field) while SMC figures can
be obtained upon request from the authors.

\subsection{RR~Lyrae stars}

Table \ref{rrlyrae} summarizes results obtained with each of the
classifiers (GM and MSBN) on OGLE data without colours added (NC),
with $B-V$ and $V-I$ colours (+BVI) and with all colours (+JHK). The
experiments in the bulge did not include $B-V$ for the reasons
explained in section \ref{colours}. The Gaussian Mixtures classifier
only makes use of the $B-V$ colour index except in the bulge where
only the $V-I$ colour index was used.

In the LMC, \citet{Soszynski:2003} found 7612 RR~Lyrae stars.  A search
was performed in the OGLE variability database using the coordinates
provided by the authors in the electronic version of the
catalogue. This search only produced photometric time series for
2734 (plus 56 double mode pulsators published in a separate
catalogue). The situation is analogous to the SMC where
\citet{Soszynski:2002} list a total of 571 RR~Lyrae stars but we are
only able to identify corresponding entries in the variability
database for 89 (plus 4 double mode pulsators that we will not include
in the study since these systems are part of the training set). We
have found no explanation for this large discrepancy and thus, in the
following we compare our detection rate with these total numbers (2790
for the LMC and 89 for the SMC).

In the LMC, the multistage classifier based on Bayesian Networks
correctly identifies as RR~Lyrae 2597 of the 2790 stars (93\%)
classified as such by the OGLE team. The percentage increases to a
96\% when BVI colours are used as attributes for classification. In
the SMC, the percentage increases up to a 95.5\% without colours and
98\% with BVI colours. As could be expected, the low signal to noise
ratios of the 2MASS detections worsens the percentages down to 85\% in
the LMC while the SMC detection rate is too low to draw significant
conclusions. In the bulge, the same classifier has a performance of
87\% working on time series attributes alone (NC) and much poorer
performances when colours are added. We interpret this as the result
of a poor dereddening using field average values of the
extinction. Since the $V-I$ value was obtained from the OGLE project
itself, we believe there is no room for the interpretation of this
performance degradation as being produced by counterpart
misidentifications. 

The largest errors of the sequential classifier in these category of
variable stars are RR~Lyrae systems misclassified as double mode
Cepheids or eclipsing binaries. This is interpreted as the effect of
overfitting to the training set, that is, as a consequence of the fact
that DMCEP (see Table \ref{tabel1} for abbreviations) and eclipsing
binaries are the only classes, together with double mode RR~Lyrae
stars, whose training examples are taken from the OGLE database. In
this sense, the classifier is recognizing similarities likely due to
the observational setup of the OGLE survey and common to the three
classes whose prototypes are taken from its database. The GM
classifier is clearly more robust against overfitting as shown in the
table and in the section devoted to the analysis of Cepheid stars.

\begin{table*}
\caption{Number of RR~Lyrae stars according to the OGLE catalogues and
correctly identified by the Gaussian Mixtures (GM) and multistage
Bayesian networks (MSBN) classifiers presented here. The table lists
the number of stars in the OGLE catalogues with a clear counterpart in
the OGLE variability database and, subsequently, the fraction of these
with available visible and visible+2MASS colours. }

\label{rrlyrae}
\centering
\begin{tabular}{cccccccccc}
  \hline
  Catalogue  & Source & \multicolumn{3}{c}{Potential detections} & \multicolumn{2}{c}{GM} & \multicolumn{3}{c}{MSBN} \\
  \hline 
& & NC & +(B)VI & +JHK& NC & +(B)VI & NC & +(B)VI & +JHK\\ 
  \hline
  \hline
OGLE RR~Lyrae & LMC   & 2790 & 2558 & 137 & 2014 & 1819  & 2597 & 2457 & 117 \\
OGLE RR~Lyrae & SMC   &   93 &   87 &   2 &   63 &   61  &   89 &   85 &   2 \\
OGLE RR~Lyrae & bulge &   70 &   22 &  17 &   61 &    7  &   61 &   12 &   6  \\  
  \hline
\end{tabular}
\end{table*}

The RR~Lyrae sample compiled by the OGLE team also provides subtype
information. Therefore, we can further compare the subclassification
of RR~Lyrae stars into one of its subclasses: RRab, RRc and RRd. Table
\ref{RR-BN-cm} summarizes the confusion matrix obtained with the
sequential classifier based on Bayesian networks and with the GM
classifier when applied to the LMC sample without colours.

\begin{table}
\caption{Confusion matrix for the RR~Lyrae subtypes. Each column lists
the number of objects of a given subtype (shown as column header)
classified as all possible subtypes.}
\label{RR-BN-cm}
\centering
\begin{tabular}{ccccccc}
  \hline
& \multicolumn{3}{c}{GM} & \multicolumn{3}{c}{MSBN} \\
  \hline
    & RRAB & RRC & RRD & RRAB & RRC & RRD\\
  \hline
  \hline
RRAB & 1913 &  1  &  0 & 2420 &  4 &  2 \\
RRC  &    0 & 22  &  0 &    3 & 76 &  0 \\
RRD  &    1 & 21  & 54 &   21 & 14 & 53 \\
  \hline
\end{tabular}
\end{table}

\begin{figure*}
   \centering
   \includegraphics[angle=-90,scale=0.6]{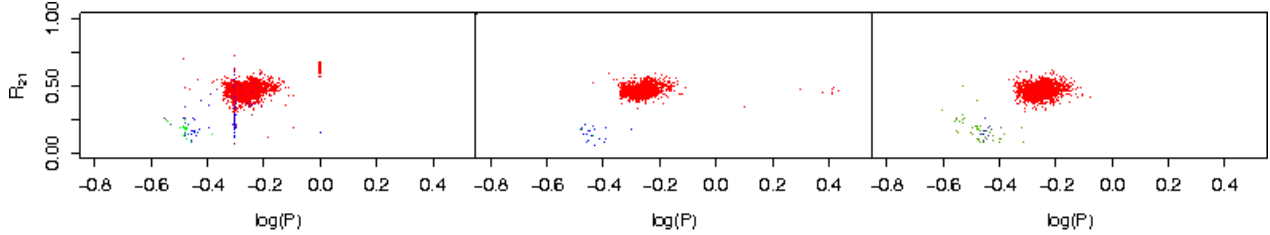}
   \caption{The $R_{21}-\log(P)$ plane of RRAB (red), RRC (green) and
   RRD stars (blue) in the LMC, according to the multistage Bayesian
   networks (left) and Gaussian Mixtures classifiers (middle) and the
   OGLE catalogue (right).}
   \label{rrlyr-f1-r21}
\end{figure*}

\begin{figure*}
   \centering
   \includegraphics[angle=-90,scale=0.6]{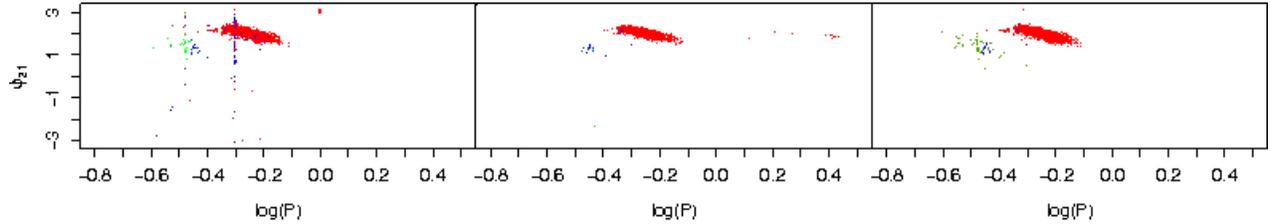}
   \caption{The $\phi_{21}-\log(P)$ plane of RRAB (red), RRC (green) and
   RRD stars (blue)  in the LMC, according to the multistage Bayesian networks
   (left) and Gaussian Mixtures classifiers (middle) and the OGLE
   catalogue (right).}
   \label{rrlyr-f1-phi21}
\end{figure*}

Obviously, the True Positive Rate (TPR) is not the only way to measure
the success of a classifier. The false positive rate (FPR, the number
of non members of the class mistakenly classified as such) for a given
class is also a good measure that quantifies the contamination degree
of the resulting samples. Unfortunately, we can only measure the FPR
coming from the OGLE sample classes other than RR~Lyrae, described in
section \ref{OGLE}. However, we can find useful hints of the true FPR
for example by looking at the definition plots of the RR~Lyrae
class. When applied to the whole of the LMC (SMC) database with 50708
(14473) instances, the sequential classifier finds 3019 (273) RRab
candidates, 131 (18) RRc candidates and 335 (88) RRd candidates. We
again attribute the large numbers of double mode pulsators to the use
of OGLE examples of this class in the training set. Figures
\ref{rrlyr-f1-r21} and \ref{rrlyr-f1-phi21} show the position of the
LMC candidates produced by the Bayesian and Gaussian Mixtures
classifiers in the $\log(P)-R_{21}$ and $\log(P)-\phi_{21}$ diagrams.

The plots were constructed with all
instances that fulfilled the condition that the class probability
given the data ($p({\cal C}_k|{\cal D})$) was higher for RR~Lyrae
subtypes than for any other class. The plots can be adapted to a given
decision threshold: setting $p({\cal C}_k=RR~Lyrae|{\cal D})>0.9$ in
the sequential classifier, for example, removes most of the conspicuous
ghost frequencies around $\log(P)=0, -0.3, -0.5$ (P in days) and most other stars
not in the dense loci of the RR~Lyrae subtypes. Similar thresholds can
be defined for the GM classifier in terms of the Mahalanobis distance
to the center of the cluster. 


\begin{figure}
   \centering
   \includegraphics[angle=-90,scale=0.3]{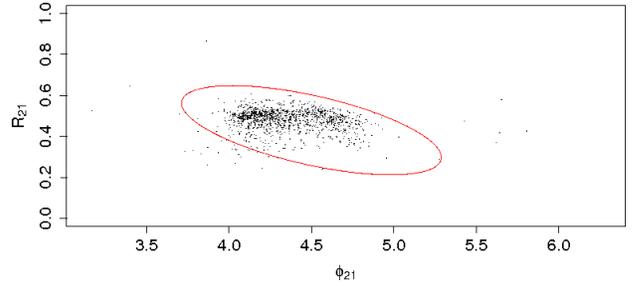}
   \caption{The $\phi_{21}-R_{21}$ plane of RRAB stars (90\% decision
   threshold) in the bulge. The ellipse shows the decision boundary
   adopted by \citet{2004MNRAS.349..193S} and \cite{Collinge:2006}.}
   \label{collinge}
\end{figure}

A comparison with the results by \citet{Collinge:2006} is shown in
Fig. \ref{collinge}. As summarized in section \ref{OGLE}, they
identify 1888 fundamental mode RR~Lyrae candidates in the bulge plus
25 repetitions in overlapping regions between fields. The MSBN
classifier finds 1862 (97\%) candidates inside the ellipse that
defines the RRab locus according to
\citet{2004MNRAS.349..193S}. Besides these, the MSBN classifier
provides 756 new candidates, not all inside the ellipse .

\begin{figure*}
   \centering
   \includegraphics[angle=-90,scale=0.6]{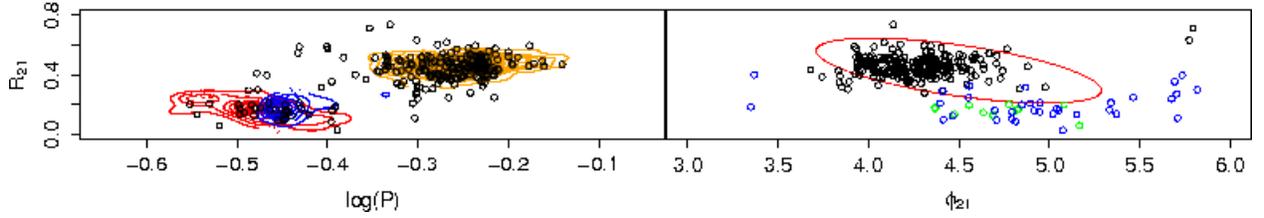}
   \caption{Two projections of three parameters (period, amplitude
   ratio and phase difference) of stars in the LMC classified as
   RR~Lyrae and not in the OGLE RR~Lyrae sample. In the left plot,
   contour lines represent the probability density as obtained from
   the OGLE sample by using kernel methods. Orange corresponds to the
   RRAB sample, red to RRC and blue to RRD. In the right plot, the
   ellipse is that used in \cite{Collinge:2006}.}
   \label{new-rrs-r21}
\end{figure*}

\begin{figure*}
   \centering
   \includegraphics[angle=-90,scale=0.4]{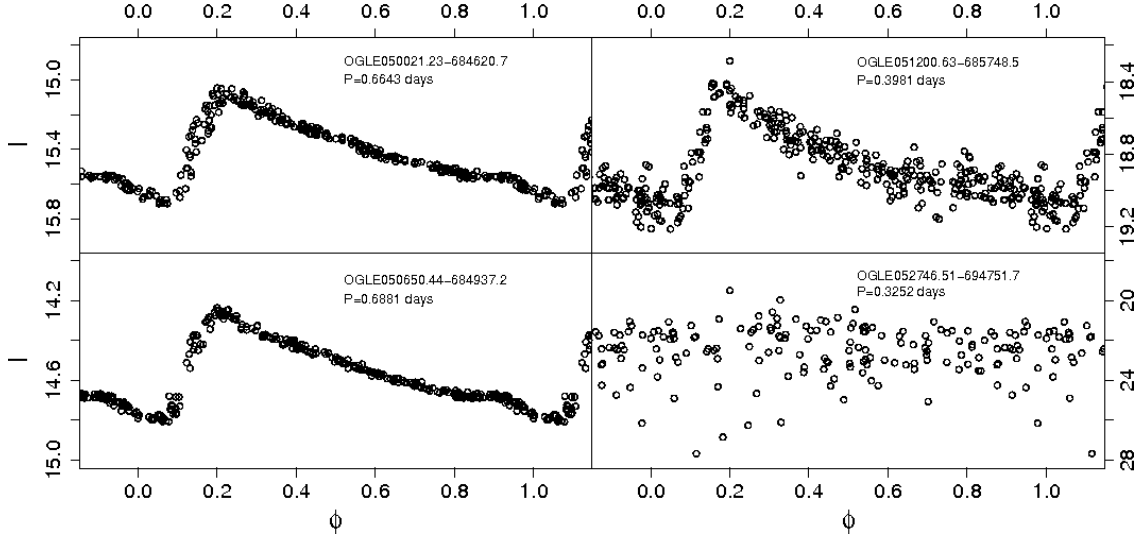}
   \caption{$I$-band light curves of three candidates of the RR~Lyrae
   category not identified as such in the OGLE catalogue (longer
   period candidates in the upper and lower left plots and shorter
   period and lower signal-to-noise ratio in the upper right
   panel). The lower right plot is an example of a low
   probability candidate with no conspicuous modulation of the light
   curve.}
   \label{new-rr-lcs}
\end{figure*}

One may wonder where the new RR~Lyrae candidates are located in the
parameter space. Since this space has a large number of dimensions, it
will prove useful to project it onto planes as with previous
plots. Figure \ref{new-rrs-r21} shows two such projections onto the
$\log(P)$-$R_{21}$ and $\phi_{21}$-$R_{21}$ planes for stars in the
LMC classified by the MSBN classifier as RR~Lyrae, the latter plane
being the one used by \cite{Collinge:2006} to define the bulge sample of RR~Lyrae
stars. The first plot shows superimposed the contours of the
probability density functions constructed using standard kernel
methods applied to the RR~Lyrae samples provided by the OGLE
team. Both plots clearly show how the new candidates (with
probabilities above 90\%) fall mostly in the RR~Lyrae locus. Although a
detailed analysis of all new candidates in all the following
categories is beyond the scope of this article, we have randomly
checked some folded light curves of the new candidates such as those
shown in Figure \ref{new-rr-lcs}. Most of the new candidates have
folded light curves similar to those in the left and upper right
panels of the figure with varying signal-to-noise ratios. We show,
completeness, the folded light curve of a star with a class assignment
of RR~Lyrae (with a low probability, though) and characterized by a
low statistical significance of the frequency detection. It helps us
exemplify why and how, imposing more stringent significance thresholds
on the frequency detection, we can remove poor quality candidates from
the lists.

\subsection{Cepheids}

Table \ref{cep-table} lists the results obtained for the LMC with the
same classifiers tested in the previous section.  In this case, the
best performances (achieved by the MSBN classifier) in the LMC are of
94\% without colours, 99\% with BVI photometry and 98\% with BVI plus
JHK photometry. These performances are around 85\% in the SMC although
the use of 2MASS photometry increases the true positive rate back to
95\%. In the bulge, the results confirm the problem with inadequate
dereddening.

\begin{figure*}
   \centering
   \includegraphics[angle=-90,scale=0.6]{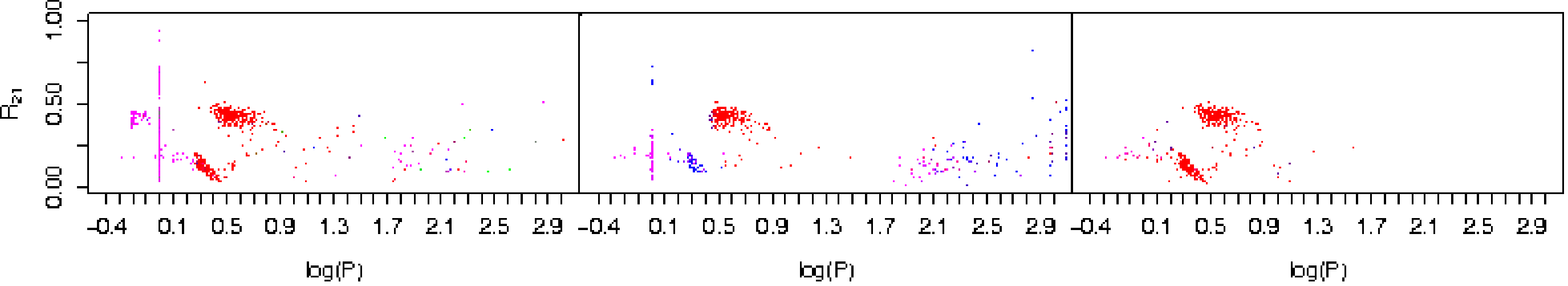}
   \caption{The $R_{21}-\log(P)$ plane of classical Cepheids (red),
   RVTAU (green), PTCEP (blue) and DMCEP (magenta) in the LMC
   according to the multistage (left) and GM (middle) classifiers and the OGLE team
   sample (right, only fundamental and first overtone classical Cepheids in
   red, and DMCEP in magenta).}
   \label{lmccep-f1-r21}
\end{figure*}

\begin{figure*}
   \centering
   \includegraphics[angle=-90,scale=0.6]{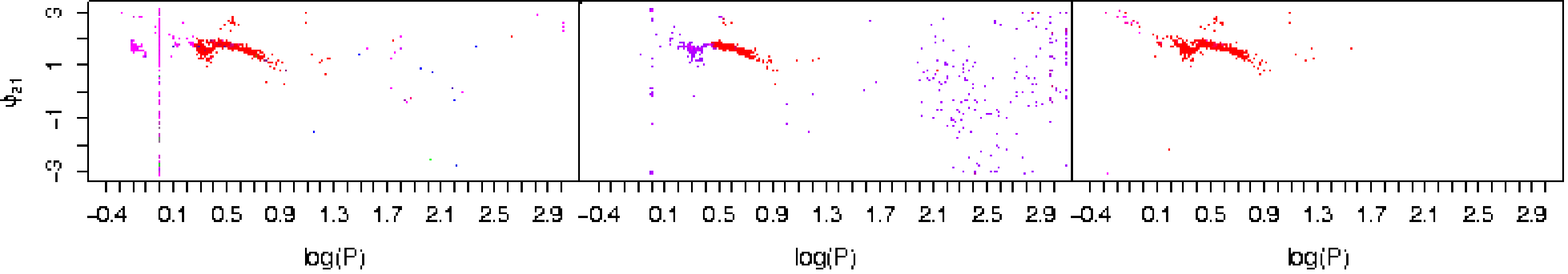}
   \caption{The $\phi_{21}-\log(P)$ plane of classical Cepheids (red),
   RVTAU (green), PTCEP (blue) and DMCEP (magenta) in the LMC
   according to the multistage and GM classifiers and the OGLE team
   sample (only fundamental and first overtone classical Cepheids in
   red, and DMCEP in magenta).}
   \label{lmccep-f1-phi21}
\end{figure*}

While the OGLE Cepheids sample only contains and distinguishes
fundamental and first overtone pulsators, our classifier identifies
RVTAU and PTCEP systems. These are included in the plots describing
the automatic classifiers but not in the OGLE sample plot (see Figures
\ref{lmccep-f1-r21}, \ref{lmccep-f1-phi21}, \ref{bulcep-f1-r21}, and
\ref{bulcep-f1-phi21}). It is evident from these plots that, as was
indeed the case with the RR~Lyrae systems, the MSBN classifier is
overfitted to the training set and tends to overestimate the
probability of the classes represented in the training set with
examples taken from the OGLE database (double mode Cepheids in this
case, double mode RR~Lyrae pulsators in the previous one). This
overfitting can also be detected in the analysis of the new DMCEP
candidates according to the MSBN classifier, which are mostly first
overtone classical Cepheids close in the hyperparameter space to the
DMCEP locus, but lacking the characteristic frequency ratio. Apart
from this effect (that can only be corrected when more examples of
double mode pulsators from other surveys are available) we see that
the MSBN classifier incorrectly assigns the DMCEP class to a cluster
of RR~Lyrae stars at $\log(P) \approx -0.2$ (P in days).  This effect
can be traced back to the density of DMCEP and RRAB training examples
in that region, but it is evident that this classifier is not robust
enough and requires a better sampling of the density of examples
there. We have tried several modifications of the design presented in
section \ref{colours} in order to redraw the boundary between double
mode Cepheids and RR~Lyrae stars. This seemingly simple task (both
classes are linearly separable in several attributes according to the
training set) turned out to result in undesired performance
degradation (of the order of 15\%) in the classical Cepheid detection
(or true positive) rate. Solutions to this problem included new
hierarchy designs (separating Cepheids and RR~Lyrae systems at the
same time), reordering of the partial classifiers and several
different attribute selection techniques. In our opinion, the MSBN
classifier described in section \ref{colours} represents a better
global solution to the problem of automatic classification of variable
objects that needs further refinement at the forementioned
boundary. The GM classifier on the contrary, has
no RR~Lyrae contamination in the DMCEP candidate list despite being
constructed upon the same training set.

Table \ref{cep-BN-GM-cm} shows the confusion matrices for the subtypes
of Cepheids common to the classifiers and OGLE catalogues. We see how
the MSBN higher detection rate has, as an undesired side effect,
a large number of misclassifications of classical Cepheids as double
mode. Also, it is unable to correctly identify Population II
Cepheids. Although there is also a sizable contamination of CLCEP
stars in the DMCEP group produced by the GM classifier, the
overfitting is less serious than in the MSBN case. Unfortunately this
improvement is also accompanied in the GM classifier by a large FPR
(False Positive Rate) in the PTCEP class.

\begin{figure}
   \centering
   \includegraphics[angle=-90,scale=0.22]{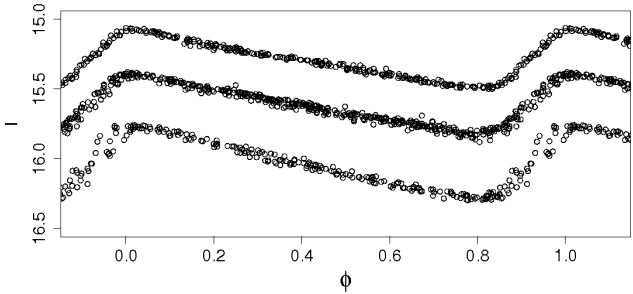}
   \caption{$I$-band light curves of OGLE050131.82-692319.0,
   OGLE051759.78-691602.5 and OGLE053643.23-701030.7 folded with the
   periods $P=3.3622$, $P=3.5656$ and $P=3.7675$ respectively
   and displaced vertically for clarity.}
   \label{new-clceps}
\end{figure}

Even though no new DMCEP star has been found (most high probability
candidates turn out to be first overtone Cepheids), at least some of
the MSBN classifier candidates for the CLCEP category seem
promising. Again, a full detailed study of the new candidates is
beyond the scope of this work, but Figure \ref{new-clceps}, showing
the folded light curves of three systems lying at the core of the
CLCEP locus, seems to suggest that there can be classical Cepheids
missed by the OGLE team. The number of CLCEPs missed by the
traditional method cannot be too large because there are only 20 new
candidates with a probability above 90\%. Of course, lowering the
probability threshold can provide more extended (but less safe)
candidate lists.

\begin{table*}
\caption{Number of Cepheids according to the OGLE catalogues and
correctly identified by the Gaussian Mixtures (GM) and multistage
Bayesian networks (MSBN) classifiers presented here. The table lists
the number of stars in the OGLE catalogues with a clear counterpart in
the OGLE variability database and, subsequently, the fraction of these
with available visible and visible+2MASS colours.}
\label{cep-table}
\centering
\begin{tabular}{cccccccccc}
  \hline
  Catalogue  & Source & \multicolumn{3}{c}{Potential detections} & \multicolumn{2}{c}{GM} & \multicolumn{3}{c}{MSBN} \\
  \hline 
& & NC & +(B)VI & +JHK& NC & +(B)VI & NC & +(B)VI & +JHK\\ 
  \hline 
  \hline
OGLE Cepheids & LMC   &   1443 & 1313 & 1022  & 1065 & 891 & 1363 & 1298 & 1001 \\
OGLE Cepheids & SMC   &   1914 & 1838 &  598  & 1034 & 829 & 1617 & 1559 &  567 \\
OGLE Cepheids & bulge &     54 &   39 &   23  &   44 &  14 &   50 &   19 &   15 \\
\end{tabular}
\end{table*}

\begin{table*}
\caption{Confusion matrix for the various Cepheids subtypes and the
classifiers applied to the LMC without using photometric colours. Each
column lists the number of objects of a given subtype according to the
OGLE catalogue (shown as column header) classified as all possible
subtypes.}
\label{cep-BN-GM-cm}
\centering
\begin{tabular}{ccccccc}
  \hline
 & \multicolumn{3}{c}{GM} & \multicolumn{3}{c}{MSBN} \\
    & CLCEP & DMCEP & PTCEP & CLCEP & DMCEP & PTCEP\\
  \hline 
  \hline
CLCEP  & 731 &  0 & 7 & 1130 &  1 & 10  \\
DMCEP  & 76  & 70 & 1 & 137  & 65 &  3  \\
PTCEP  & 177 &  0 & 3 & 16   &  1 &  0  \\
\end{tabular}
\end{table*}

\subsection{Eclipsing binaries}

Table \ref{ecl} shows a comparison between the OGLE sample of
eclipsing binaries and the samples obtained by our classifiers. We
have preferred not to include the subtype classification of eclipsing
binaries (EA/EB/EW) because, in our opinion, the boundaries between
them are not sufficiently well defined in terms of quantifiable
criteria and thus result in large error rates not justified in terms
of real classification errors.

\begin{figure*}
  \centering
  \includegraphics[angle=-90,scale=0.6]{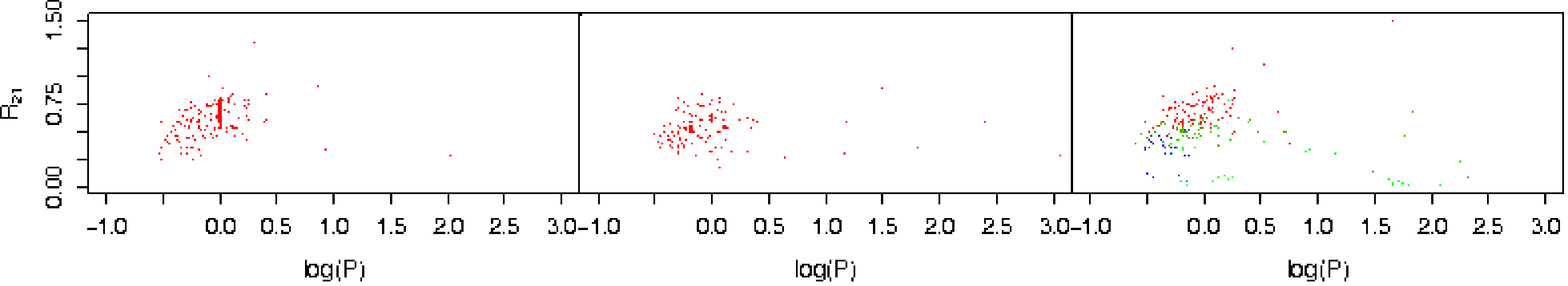}
  \caption{The $R_{21}-\log(P)$ plane of eclipsing binaries for the
    SMC. From left to right, the MSBN and GM samples and the OGLE catalogue.}
  \label{ecl-f1-r21}
\end{figure*}

\begin{figure*}
  \centering
  \includegraphics[angle=-90,scale=0.6]{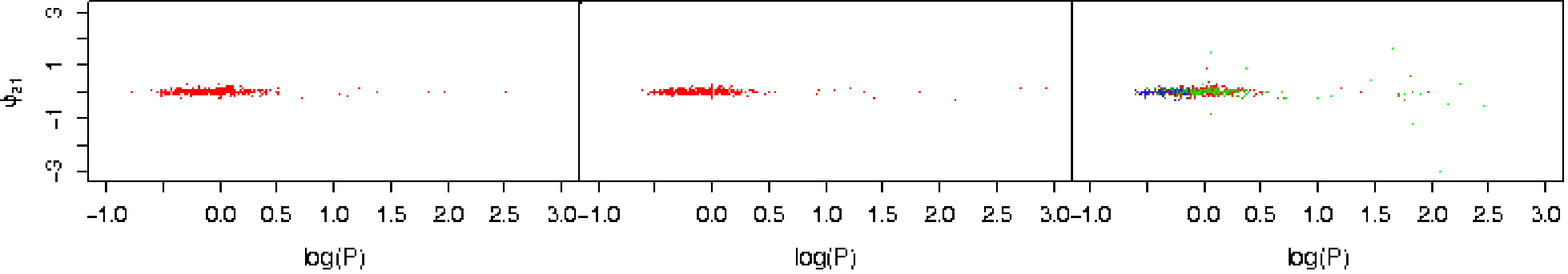}
  \caption{The $\phi_{21}-\log(P)$ plane of eclipsing binaries for the
    SMC. From left to right, the MSBN and GM samples and the OGLE
    catalogue.}
  \label{ecl-f1-phi21}
\end{figure*}

The good performance of the classifiers for this problematic class is
remarkable. Figures \ref{ecl-f1-r21} and
\ref{ecl-f1-phi21} corresponds to SMC objects classified as eclipsing
binaries with a probability above 90\% (for the MSBN classifier)
because the LMC eclipsing variables were used in the training set and
thus, performance estimates based on the same cases used for training
would have a strong optimistic bias. The MSBN classifiers recovers
75\% of the OGLE sample without incorporating colour information (73\%
using $B-V$ and $V-I$ and 43\% adding 2MASS colours) but, most
remarkably, it recovers 97\% of the bulge sample by \cite{Groenewegen:2005}
(95\% using $B-V$ and $V-I$ and 92\% adding 2MASS colours). These
percentages are even larger than those obtained for the LMC on a set
of systems used to train the classifier, as explained above.

As was the case with the double mode Cepheids, having used OGLE
observations of eclipsing binaries in the definition or training set
results in overfitting and a strong tendency to classify other variability types as eclipsing
binaries. This can be detected as a sizable
number of objects similar to RR~Lyrae stars and classical Cepheids
mistakenly classified as eclipsing binaries. They are easily detected
by the large phase differences between the various harmonics (these
objects do not appear in Figure \ref{ecl-f1-r21} because they have
class probabilities well below 90\%). 

The lack of systems with sinusoidal light curves and low $R_{21}$
ratio, specially around $\log(P)\approx 0$ is also evident from the
plots. This hypothesis is confirmed by two facts: the distribution of
the $R_{21}$ ratio amongst OGLE eclipsing binaries misclassified by
the MSBN classifier (though multimodal) has the strongest component
below $R_{21}=0.2$; second, the astonishing true positive detection
rate in the \cite{Groenewegen:2005} sample is due to its being
composed exclusively of detached systems (see Figure
\ref{bulecl-f1-r21}), because its main objective was to obtain
candidates for distance determination.

\begin{figure*}
  \centering
  \includegraphics[angle=-90,scale=0.45]{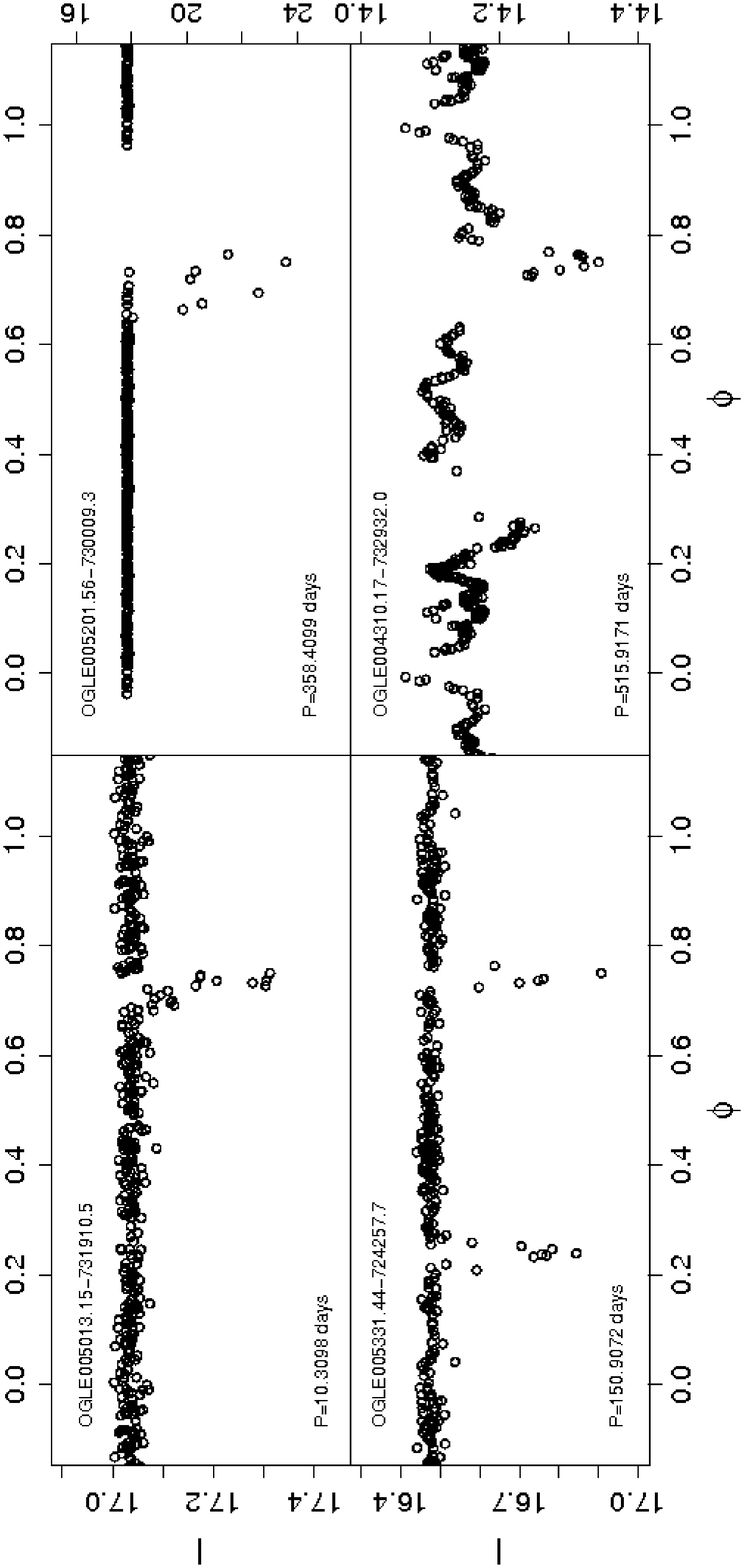}
  \caption{Example light curves of new systems classified as eclipsing
  binaries and not in the reference samples.}
  \label{new-ecl}
\end{figure*}

As with previous variability types, the classifiers provide candidate
lists that include objects not in the published reference samples. In
this case, the 90\%-confidence lists comprise 3122 candidates in the
LMC, 1216 in the SMC and 14610 in the Galactic bulge. Of these, 990
are new candidates in the LMC not in any of the published lists (330
and 11739 in the SMC and Galactic bulge respectively). As a check for
these new candidates, we have plotted some of the systems with the
longest periods and the largest $R_{21}$ ratios amongst the SMC
candidates (see Figure \ref{new-ecl}). On the left column plots we
show confirmed candidates of the category of eclipsing binaries while
the rightmost column shows one possible example of instrumental
effects (top; the dimming of the star always associated with the end of
a series of observations) and one example of a more complicated system
with various causes contributing to the light curve variability. As
with all previous categories, we do not claim that all these new
sources have to be treated as confirmed cases but rather as strong
candidates upon which further selection criteria can be applied in
order to obtain manageable candidate lists.

\begin{table*}
\caption{Number of eclipsing binary systems according to the OGLE and
Groenewegen catalogues and correctly identified by the Gaussian
Mixtures (GM) and multistage Bayesian networks (MSBN) classifiers
presented here. The table lists the number of systems in the two
catalogues with a clear counterpart in the OGLE variability database
and, subsequently, the fraction of these with available visible and
visible+2MASS colours.}
\label{ecl}
\centering
\begin{tabular}{ccccccccccc}
  \hline
  Catalogue  & Source & \multicolumn{3}{c}{Potential detections} & \multicolumn{2}{c}{GM} & \multicolumn{3}{c}{MSBN} \\
  \hline 
& & NC & +(B)VI & +JHK& NC & +(B)VI & NC & +(B)VI & +JHK\\ 
  \hline 
  \hline
OGLE eclipsing binaries        & LMC   &  2631 & 2467 & 210  & 1613 & 1528 & 2296 & 2072 &  150 \\
OGLE eclipsing binaries        & SMC   &  1387 & 1316 & 153  &  824 &  809 & 1045 &  967 &   65 \\
Groenewegen (2005) eclipsing binaries & LMC   &   173 &  162 &  27  &   80 &   77 &  132 &  108 &   10 \\
Groenewegen (2005) eclipsing binaries & SMC   &    16 &   15 &   8  &    8 &    8 &    9 &    8 &    1 \\
Groenewegen (2005) eclipsing binaries & bulge &  3034 & 2132 & 1260 & 2599 & 1295 & 2951 & 2016 & 1159 \\
\end{tabular}
\end{table*}


\subsection{Long period variables}

\begin{figure*}
   \centering
   \includegraphics[angle=-90,scale=0.45]{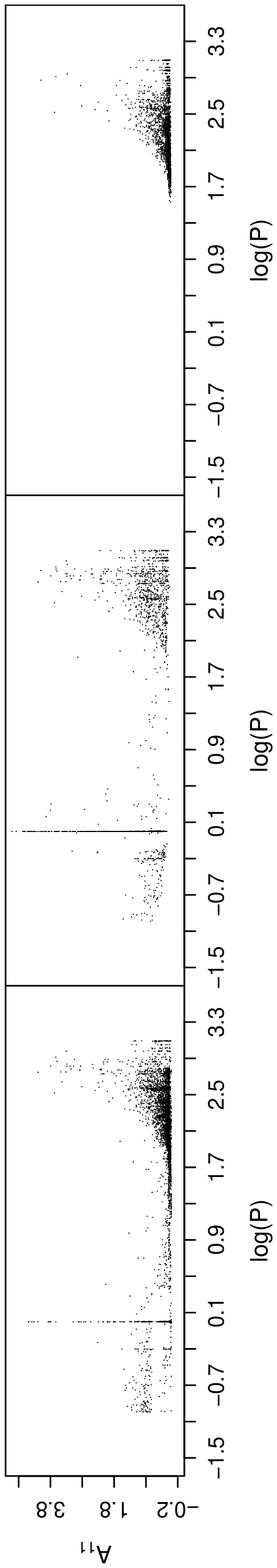}
   \caption{The $A_{11}-\log(P)$ plane for long period variables in
   the LMC according to (from left to right) the multistage and GM
   classifiers and the OGLE team sample.}
   \label{lpv-f1-a11}
\end{figure*}

Long period variables (LPVs) constitute the class where the most
significant discrepancies are found. As shown in Table \ref{lpv}, the
MSBN classifier barely recovers 50\% of the LMC OGLE sample of Mira
and Semiregular variables. The reason is two-fold: first, many of the
OGLE long period variables (17\% and 45\% in the OGLE LMC and Bulge
samples respectively) are missed in the frequency calculation step
where the sampling frequency ($\approx 1$ c/d) prevails over the
stellar pulsation, thus providing first and subsequent frequencies in
error. Second, there is a lack of low amplitude Miras and semiregular
stars with periods of less than 150 days in the training set, and those
are the main contribution to the missing LPVs. Figure \ref{TS-OGLE}
shows a comparison between the first frequency amplitude of Miras and
semiregulars in the training set and in the OGLE LMC sample. In this
regime, the number of examples is so low that it is indeed less than
that of the LBV or Periodically Variable B- and A-type supergiant
(PVSG) classes, the main contributors to the False Negative Rate
(misclassified Mira and Semiregular stars according to the OGLE
sample). Therefore, there is a clear need to extend the training set
representation of the Mira and Semiregular classes in this region of
the parameter space. The situation is different for the
\cite{2005MNRAS.364..117M} and \cite{2005AandA...443..143G} candidate
lists where the true positive rates increase to 87\%. We
interpret this increase in performance as a confirmation of the
hypothesis put forward above given the absence of low amplitude
variables with periods below $\log P\approx 2.2$ in these
lists. Unfortunately, the lack of low period-low amplitude Miras and
semiregulars is not visible in Figure \ref{lpv-f1-a11} due to the
crowd of stars in the plot.

As expected, the inclusion of Johnson photometry in the inference
process corrects the low performance of the classifiers in the OGLE
LMC case and increases the TPR up to 94\% (98\% when 2MASS
photometry is included). This effect can be easily understood given
the strong relevance (in the sense commonly accepted by the
Statistical Learning community) of these attributes. Surprisingly
though, it also results in a small performance degradation (2--5\%) in
the bulge samples by \cite{2005MNRAS.364..117M} and \cite{2005AandA...443..143G}.

Using a confidence threshold of 90\%, we find 67 new candidates in the
LMC and 990 in the Galactic Bulge. As in all previous cases, visual
inspection of the position of the new candidates in several 2D
projections confirms the adequacy of their parameters for the class
definitions in the training set and reference samples. Random
inspection of some candidates indicates that most of the new
candidates are semiregular pulsators often affected by long term
trends in the mean brightness and several frequency components.

\begin{table*}
\caption{Number of long period variables (LPV) according to the
\cite{2005MNRAS.364..117M} and \cite{2005AandA...443..143G}
catalogues, and correctly identified by the Gaussian Mixtures (GM) and
multistage Bayesian networks (MSBN) classifiers presented here. The
table lists the number of systems in the two catalogues with a clear
counterpart in the OGLE variability database and, subsequently, the
fraction of these with available visible and visible+2MASS colours.}
\label{lpv} 
\centering
\begin{tabular}{ccccccccccc}
  \hline
  Catalogue  & Source & \multicolumn{3}{c}{Potential detections} & \multicolumn{2}{c}{GM} & \multicolumn{3}{c}{MSBN} \\
  \hline 
& & NC & +(B)VI & +JHK& NC & +(B)VI & NC & +(B)VI & +JHK\\ 
  \hline 
  \hline
OGLE LPV            & LMC      & 3472 & 2735 & 2552 &  407 & 2060 & 1718 & 2576 & 2508 \\
OGLE LPV            & bulge    &  273 &  129 &   90 &   84 &   85 &   69 &   65 &   67 \\
Miras (Matsunaga et al. 2005)   & bulge    & 1882 & 1284 &  733 & 1498 & 1186 & 1642 & 1052 &  627 \\
Miras (Groenewegen $\&$ Blommaert, 2005) & bulge    & 1999 & 1276 &  737 & 1648 & 1178 & 1734 & 1038 &  632 \\
\end{tabular}
\end{table*}

\subsection{Multiperiodic variables}
It is clear that both classifiers perform well for the majority of the
classes considered above. However, most of these classes contain
monoperiodic (radial) pulsators, or eclipsing binaries. Our
classification scheme also included several multiperiodic classes.
Multiperiodic variables are amongst the most scientifically
interesting classes in relation to asteroseismic studies of the
stellar structure and evolution, see e.g. \cite{Kurtz:2006} for a
review. Nevertheless, they have not been thoroughly studied in the
OGLE variable databases.

\subsubsection{Pulsating B-stars in the Magellanic clouds}

We could not compare our results for those classes with existing
results in such an extensive way. These classes have been much less
studied up to now, mainly because their detection is less
obvious in the OGLE data. Since they are relevant for
asteroseismology, we present here the results obtained with both
classifiers for $3$ classes of massive intrinsically bright
multiperiodic pulsators: $\beta$-Cephei stars (BCEP), slowly pulsating
B-stars (SPB), and periodically variable super giants (PVSG). We limit
ourselves to these classes, since the other well-known multiperiodic
classes contain much fainter stars, making their detection even more
difficult in the OGLE data for the Magellanic clouds.  Because
single-band light curve information is usually not sufficient to
identify those objects in an unambiguous way, we only consider here
the classification results obtained with the additional colour
attributes B-V (and V-I) included for both classifiers. We also place
the new candidate variables in the HR diagram. This could be done only
for the LMC and SMC variables, since B-V colours, V magnitudes and
distances are only available for those objects. For the Bulge data,
only V-I and $2MASS$ colours are available, and the distance is
unknown. Moreover, the V-I colours for the Bulge have proven to be
less reliable, as mentioned earlier. However, the Bulge sample is
larger and contains brighter objects, so detection of those variables
(based on their light curve) is more likely in this sample (if they
are present). We present some of the best candidates in the Bulge in
the next section, by showing their phase plots (made with the dominant
frequency we detected) and listing some of their light curve
parameters. The samples are much too large to check all the candidates
(this is out of the scope of this work), but the full classification
results with both classifiers will be made available electronically.\\

The best candidate pulsators are shown in the HR-diagrams for both the
Small and the Large Magellanic cloud. The distances used to construct
the diagrams are as follows: $D(SMC)=60.6 \pm 2.97$ kpc
\citep{Hilditch:2005}, and $D(LMC)=48.1 \pm 3.70$ kpc
\citep{Macri:2006}. To convert the $V$ magnitudes of the objects into
absolute luminosities $\log (L/L_{\sun})$, we used the value of $4.75$
for the Sun's absolute bolometric magnitude. Bolometric corrections
and effective temperatures ($\log T_{eff}$) were obtained using the
corrected empirical transformations described in
\cite{Flower:1996}. Typical errors for $\log (L/L_{\sun})$ and $\log
T_{eff}$ have been derived, taking the uncertainties on the distance,
the V magnitudes and the B-V colours into account. Theoretical
instability strips for $\beta$-Cephei \citep{Stankov:2005}, SPB
\citep{DeCat:2002} and PVSG stars (Lefever et al. 2007) are shown. For
details on the derivation of the strips, we refer to
\cite{Miglio:2007}, \cite{Saio:2006}, and references therein. The PVSG
instability strip is for post-TAMS models
with non-radial mode degree values $l=1$ and $l=2$. The SPB and BCEP
instability strips are obtained with the OP opacity tables (giving the
widest strips), with metallicity values Z ranging from $0.005$ to
$0.02$, and non-radial mode degree values $l=0$ to $3$. Overshooting
is included ($\alpha = 0.2$ Hp), and stellar masses up to $18M_{\sun}$
were considered. Only main sequence models were included, and an
initial hydrogen mass fraction $X=0.7$ has been used. We plot
instability strips for different Z values, to show how the instability
domains are expected to shrink when Z decreases, and to show the
difference in metallicity between the LMC and the SMC. For the plots
of the results for the LMC, the SPB and BCEP instability strips are
shown for $Z=0.02$ (outer borders) and $Z=0.01$ (inner borders). For
the plots of the results for the SMC, the SPB and BCEP instability
strips are shown again for $Z=0.02$ (outer borders), and also the SPB
instability strip for $Z=0.005$ (inner borders). The BCEP instability
strip for $Z=0.005$ disappears \citep{Miglio:2007}. The PVSG
instability strip in both cases corresponds to $Z=0.02$. The position
in the HR diagram of the new candidates found with our classifiers,
relative to these instability strips, provides a reliability check of
the excitation models.\\

\begin{figure*}
   \centering
   \includegraphics[width=21cm,angle=-90,scale=0.7]{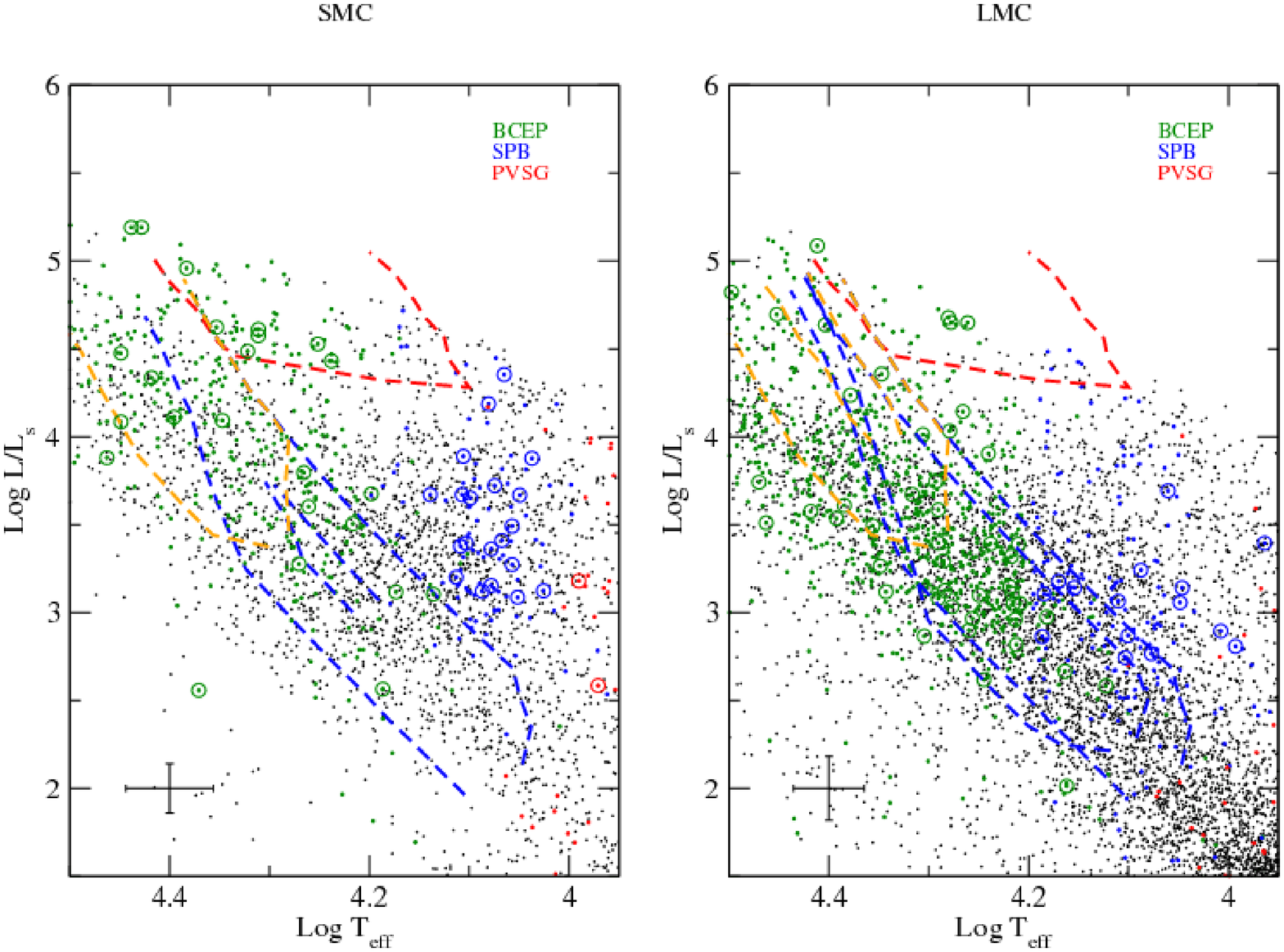}
   \caption{HR-diagram for both the SMC and LMC. The black dots
represent the total sample of variable stars for which colours were
available. The coloured dots represent those variables classified as
BCEP, SPB, and PVSG with the MSBN method. A lower limit of $0.5$ was
used for the class probabilities. The encircled dots (in the
respective class colours) represent objects classified as such with
both classifiers. The BCEP instability strips are plotted in orange
for visibility. For the right panel, the SPB and BCEP
instability strips are shown for $Z=0.02$ (outer borders) and $Z=0.01$
(inner borders). For the left panel, the SPB and BCEP instability
strips are shown again for $Z=0.02$ (outer borders), and also the SPB
instability strip for $Z=0.005$ (inner borders). The PVSG instability
strip corresponds to $Z=0.02$.}
   \label{HR-BN-0.5}
\end{figure*}

\begin{figure*}
   \centering
   \includegraphics[width=21cm,angle=-90,scale=0.7]{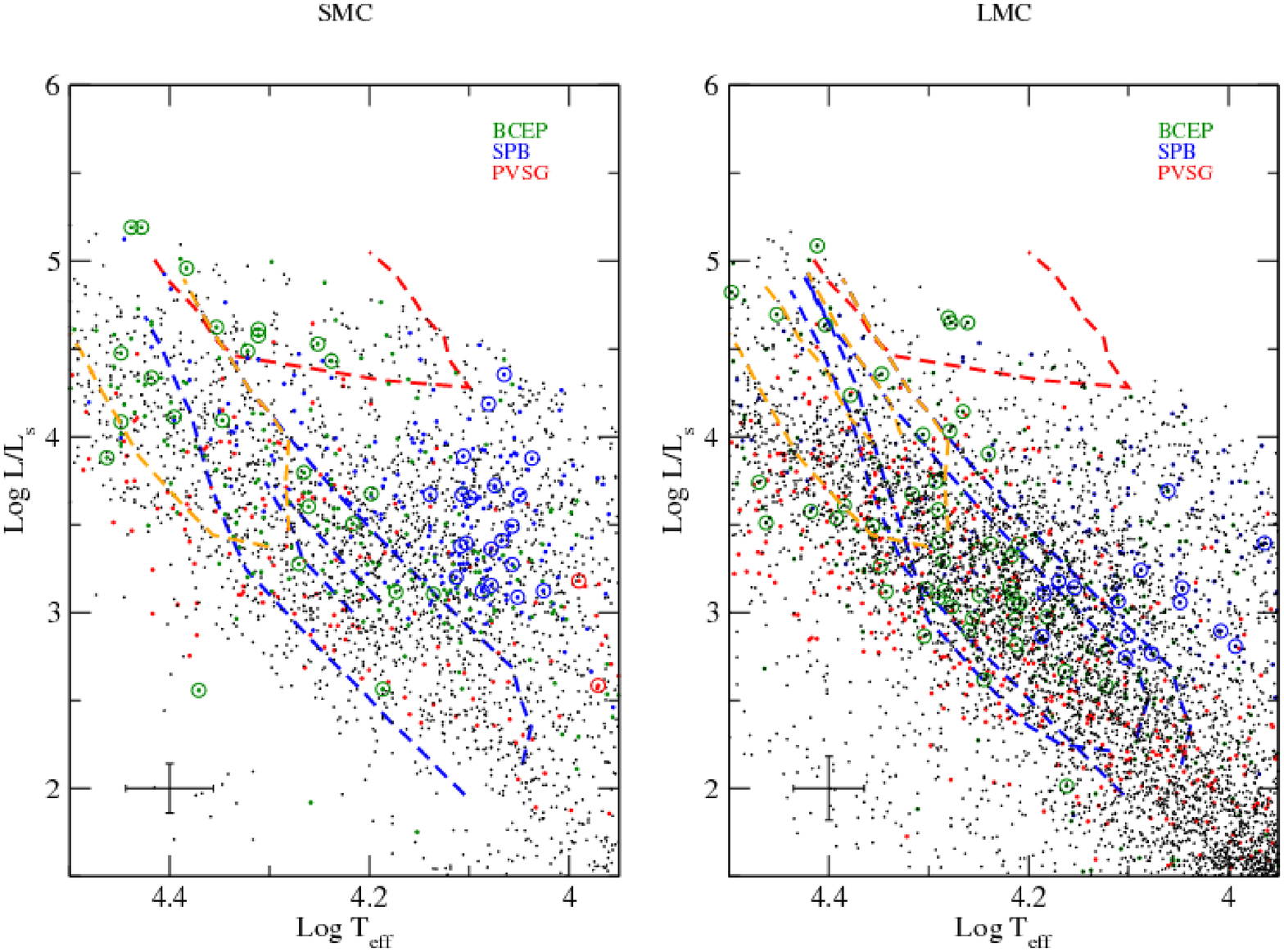}
   \caption{HR-diagram for both the SMC and LMC. The black dots
represent the total sample of variable stars for which colours were
available. The coloured dots represent those variables classified as
BCEP, SPB, and PVSG with the GM method. An upper limit of $3.5$ was
used for the Mahalanobis distance to the class centers. The
encircled dots (in the respective class colours) represent objects
classified as such with both classifiers. The BCEP instability strips
are plotted in orange for visibility. For the right panel, the SPB and
BCEP instability strips are shown for $Z=0.02$ (outer borders) and
$Z=0.01$ (inner borders). For the left panel, the SPB and BCEP
instability strips are shown again for $Z=0.02$ (outer borders), and
also the SPB instability strip for $Z=0.005$ (inner borders). The PVSG
instability strip corresponds to $Z=0.02$.}
   \label{HR-GM-3.5}
\end{figure*}

The whole sample of variable stars in the LMC and SMC with colours
available is shown in Figures \ref{HR-BN-0.5} and \ref{HR-GM-3.5}
(small black dots).

The new candidate pulsators for the $3$ B-type classes, and the
corresponding instability strips, are shown in colour. Note that the
BCEP instability strip is shown in orange (BCEP candidates are in
green), for visibility. Objects having the same class label with both
classifiers are encircled.\\ Since every object will be assigned to
one of the classes in our supervised classification scheme,
contamination in the classification results is to be expected, e.g.,
not all stars classified as belonging to one of the BCEP, SPB, or PVSG
classes will be real members of those classes. This is not a drawback,
however, since our class assignments are probabilistic, and allow us
to impose limits on the class probabilities. This way, we can select
the most probable candidates only.\\ The MSBN classifier provides
relative probabilities for an object to belong to any of the
classes. Figure \ref{HR-BN-0.5} shows all the objects having a
probability of belonging to the BCEP, SPB, or PVSG classes higher than
$0.5$, obtained with this classifier. Note that most SPB candidates
are situated above their instability domains (higher luminosity),
taking into account the errors bars. Their position on the temperature
scale is within the expected range, because the B-V colour was used as
a classification attribute. Objects far from this pre-defined range
are given a low class-probability and will not be present in our
selections.\\ The GM classifier provides relative probabilities, and,
in addition, the Mahalanobis distance to the center of the most
probable class. This distance can effectively be used to retain only
the objects that are not too far from the class center in a
statistical sense. It can be used together with the probabilities, in
order to select the best candidates. For the GM classifier, using only
the probability values is usually insufficient to select the best
candidates. Consider the case e.g., where the probability for one
class is $99\%$. This high probability value seems to indicate a very
certain class assignment. However, these are only relative
probabilities, and, even though the probability for the class is very
high, the object might still be very far away from the class
center. If this is the case, the Mahalanobis distance will have a
large value, and one has to conclude that the object is not a good
candidate to belong to the class after all. To guide us in choosing a
meaningful cutoff value for the Mahalanobis distance $D$, we can use
the fact that $D^2$ is chi-square distributed for multinormally
distributed classification parameters (the basis of the GM
classifier). The number of degrees of freedom $p$ is equal to the
number of classification attributes. Given the Mahalanobis distance
$D$ to the class, we can use this property to test the likelihood of
finding a distance larger than $D$, under the assumption that the
object belongs to the class. Note that for $p>2$, which is the case
for the GM classifier, the chi-square distribution will not be
monotonically decreasing with increasing value of $D^2$. This means
that very small values of $D$ are unlikely as well, and we should
perform a two-tailed hypothesis test.\\ Figure \ref{HR-GM-3.5} shows
the HR diagrams with the results of the GM classifier, for the SMC and
LMC, again with the variables classified as BCEP, SPB, PVSG, and their
respective instability strips shown in colours. All these candidate
variables have a Mahalanobis distance to the class center of less than
$3.5$ (dimensionless, similar to a distance in terms of sigma in the
one dimensional case). Objects having the same class label with both
classifiers are encircled. The same remarks as for the MSBN results
apply here: most SPB candidates are situated at higher luminosities
than expected for this type of variable.\\

The g-mode and p-mode pulsations in SPB and BCEP stars, respectively,
are caused by the $\kappa$-mechanism, acting in the partial ionization
zones of iron-group elements. This mechanism thus strongly depends on
the presence of those heavy elements, and hence on the metallicity of
the stellar environment. It was previously believed that the BCEP and
SPB instability strips nearly disappear for metallicities Z smaller
than $0.006$ and $0.01$ \citep{Pamyatnykh:1999}. However, the recent
results presented in \cite{Miglio:2007}, and used in this work, show
that an SPB instability strip can still exist for Z as low as
$0.005$. They do not predict BCEP pulsations at such a low metallicity
value, though. Since the metallicity of the SMC is estimated to be
between $Z=0.001$ and $Z=0.004$ \citep{Maeder:1999}, we would not
expect to find any BCEP or SPB pulsations here. However, several
independent investigations have shown that SPB and BCEP pulsators are
nevertheless present in low metallicity environments such as the LMC
and even the SMC. Examples are given in \cite {Koaczkowski:2004},
\cite{Pigulski:2002}, \cite{Karoff:2008} and \cite{Diago:2008}. Our
classification results for the OGLE LMC and SMC data support those
conclusions and suggest that even more candidates than found so far
exist. In total, we find 15 SPB and 48 BCEP candidates in the LMC, and
20 SPB and 24 BCEP candidates in the SMC. As is expected, more
pulsators are found in the metal-richer LMC. Note that a large number
of BCEP candidates are situated in the higher parts of the SPB
instability strips, both for the SMC and LMC. Overlap between the
instability strips is present in that area, and stars can show similar
pulsation characteristics there. The relatively large errors on the
position in the HR diagram (see the crosses in the plots) should be
kept in mind also. As mentioned above, we see that SPB candidates
appear at higher luminosities than expected, taking into account the
error bars. This is the case for both the LMC and SMC, and with both
the MSBN and GM classification results. Since the Magellanic clouds
contain evolved stars, we suggest that some of these SPB candidates
could in fact be B-type PVSG stars. In \cite{Waelkens:1998}, it was
suggested that the pulsations in those stars could be gravity modes
excited by the $\kappa$-mechanism, similar to the BCEP and SPB
stars. This is confirmed in \cite{Lefever:2007}, where a sample of
B-type PVSG stars is investigated in detail. Typical pulsation periods
for those variables are in the range $1-20$ days, so an overlap with
the typical period range for SPB stars is present. One may wonder why
those objects are then not classified as PVSG with our
classifiers. The PVSG class is a very heterogeneous class, containing
both B-type and A-type pulsators (note that the shown PVSG instability
strip is only for B-type stars). Moreover, they show pulsations over a
wide range of frequencies and amplitudes. This translates into a large
spread of this class in our classification parameter space. The PVSG
class overlaps with the SPB class in parameter space, but has a lower
probability density of objects at the locations overlapping with the
SPB class.  This implies that a potentially good PVSG candidate, but
with properties close to those of SPB stars, will most likely be
classified as SPB and not as PVSG. Candidate PVSG variables are shown
in Figure \ref{HR-BN-0.5} and Figure \ref{HR-GM-3.5}. There is a large
discrepancy between the numbers found by the MSBN and the GM
classifiers. This is a consequence of the poor definition of this
class. A visual check of the phase plots did not reveal convincing
candidates, in addition to the high-luminosity BCEP and SPB
candidates.

The SPB and BCEP candidates present in our selection lists and having
the same classification with both classifiers are most likely good
candidates. For those objects, we made phase plots with the dominant
frequencies ($f_1$) and list some of their light curve properties.
Note that the typical pulsation frequencies for SPB stars are situated
around $1$ c/d. The $1$ c/d frequency is unfortunately also a spurious
frequency often detected in the OGLE data, due to the daily gaps in
the observations (the OGLE window function). Since this frequency is
often significant, care must be taken not to interprete these as real
pulsation frequencies. We could exclude the most likely spurious
detections by checking the phase plots: if the plots show clear gaps,
we are probably dealing with a spurious frequency (though
in some cases, we might have a real pulsation frequency very
close to $1$ c/d).\\

Figure \ref{BCEP-SMC} and Figure \ref{SPB-SMC} show phase plots of
candidate BCEP and SPB stars in the SMC. The OGLE identifier and the
value of the dominant frequency are shown. Some of their properties
are listed in Table \ref{BCEP-SMC-param} and Table \ref{SPB-SMC-param}
respectively. Figures \ref{BCEP-LMC-1} to \ref{SPB-LMC} show the phase
plots of candidate BCEP and SPB stars in the LMC data. Their
properties are listed in Table \ref{BCEP-LMC-param} and Table
\ref{SPB-LMC-param}. The tables also list the value of the second
detected frequency $f_2$, one of the classification attributes used.

\subsubsection{The Galactic bulge}

To the best of our knowledge, the OGLE team only produced a candidate
list for the class of $\delta$ Scuti pulsators, in the first field of
the bulge and, unfortunately, only of the high amplitude candidates,
usually monoperiodic (see for example McNamara, 2000). Ten out of 11
systems listed in the catalogue by Mizerski and available to us are
correctly identified as $\delta$ Scuti stars by the MSBN classifier
and the eleventh (bul\_sc1\_1323) has a period of 6.7 hours, which is
slightly above the range of periods found for this class. The system
is classified as RRD. With the GM classifier, 7 out of 11 systems are
classified as $\delta$ Scuti stars.

Pigulski (private communication) has kindly provided us with candidate
lists of several types of multiperiodic pulsators prior to
publication, as well as an extended list of high amplitude $\delta$
Scuti (HADS) stars across all OGLE bulge fields
\citep{2006MmSAI..77..223P}. We have applied the same procedure
described above to these lists in order to assess the performance of
the classifiers in detecting multiperiodic pulsators. In the
following, we describe the results obtained with the time series
attributes alone since the inclusion of $V-I$ in the bulge has proved
detrimental to the classifiers, probably due to insufficient dereddening.

Table \ref{pigulski} shows the main contributors to the confusion
matrix constructed by assuming Pigulski's class assignments. His
results are grouped in three catalogues: the high amplitude $\delta$
Scuti stars (HADS) group, the mixed slowly pulsating B/ $\gamma$
Doradus group, and the $\beta$ Cephei/$\delta$ Scuti group. Again, the
classifiers are capable of retrieving a significant fraction of the
HADS candidates (63-78\% with the GM and MSBN classifier
respectively). These numbers decrease for the mixed groups (11-61\%
for the BCEP/DSCUT list and 70-37\% for the SPB/GDOR one with the GM
and MSBN classifier respectively). Note the low correspondence with
the BCEP/DSCUT list for the GM results and with the GDOR/SPB list for
the MSBN results. This confusion is inherently connected to the
physical properties for the stars in these classes, which imply
overlap in the characteristics of their pulsations. An example is the
occurence of both short-period p-modes and long-period g-modes in BCEP
stars (e.g. Handler et al. 2004, 2006) and the only vague separation
of the p-mode frequencies of evolved BCEP and DSCUT stars, from the
g-mode frequencies of young SPB and GDOR stars, respectively,
particularly when frequency shifts due to rotation are taken into
account.

\begin{table*}
\caption{Summary of the class assignments for objects in Pigulski's
lists (private communication) for both the GM and the MSBN classifier. }
\label{pigulski} 
\centering
\renewcommand{\tabcolsep}{1.1mm}
\begin{tabular}{cc|ccccc|ccccc}
  \hline
  &&\multicolumn{5}{c}{GM} & \multicolumn{5}{c}{MSBN} \\
  \hline
    Pigulski class & Potential detections & PVSG & BCEP & DSCUT & GDOR & SPB & PVSG & BCEP & DSCUT & GDOR & SPB\\
  \hline
  \hline

HADS           & 190& 10  & 6  &119   &2   &0   &   10 &    2 & 147 &   0  &  1 \\
BCEP-DSCUT     & 225& 3  &12   &12    &73     &16    &   12 &   28 & 109 &  16  & 11 \\
SPB-GDOR       & 623&22   & 27  & 4   &  194   & 239   &  149 &    7 &41 &  37  &191 \\

\end{tabular}
\end{table*}

\subsubsection{New candidates in the Bulge}

Here, we present a selection of Bulge objects classified as DSCUT,
BCEP, SPB or GDOR, with both classifiers, and not present in the
respective combination lists made by Pigulski. Figures \ref{SPB-BULGE}
to \ref{DSCUT-BULGE-4} show their phase plots made with $f_1$. The
OGLE Bulge identifiers are shown, and the values of $f_1$ in cycles
per day. Light curve parameters and V-I colour indices are listed in
Tables \ref{SPB-BULGE-param} to \ref{DSCUT-BULGE-param-2}. The most
obvious spurious detections (having a value of $f_1$ very close to $1$
c/d) were removed from our selections. We stress that these are
candidate lists obtained with probabilistic class assignments. Further
investigation is needed to reach more certainty about the true nature
of those objects. Significant overlap is present between the pulsation
properties of the GDOR/SPB and BCEP/DSCUT classes, which is the reason
why Pigulski did not make the distinction in his lists. We expect this
to be reflected in our candidate lists as well, e.g. some SPB
candidates might be GDORs and vice versa, and the same for the
BCEP/DSCUT classes. Apart from some inherent overlap between these
classes, this is mainly a limitation of the current classification
attributes that we can use (e.g. the absence of a good colour), and the
quality of the light curves.\\ As opposed to selections made with
extractor methods, we can have objects in our list having rather
atypical light curve parameters for that particular class. These can
be borderline cases, and in some cases, misidentifications. As was
mentioned earlier, however, stronger limits can be imposed on the class
probabilities and/or the Mahalanobis distance, to retain only the most
typical candidates. In doing so, the samples will be purer, but, on
the other hand, interesting border cases can be missed.

\section{Conclusions.}

In the past few years, the world of astronomy has seen a revolution
taking place with the advent of massive sky surveys and large scale
detectors. This revolution cannot be fully exploited unless automatic
methods are devised in order to preprocess the otherwise unmanageably
large databases. Otherwise, the efforts of the astronomical community
will have to focus on repetitive uninteresting data processing rather
than in the solution of the scientific questions that motivate the
efforts. In this work we have presented a scenario with many
interesting open questions for research (distance estimator
calibration, stellar interiors, galactic evolution...), i.e. that of
stellar variability, where automatic procedures for data processing
can help astronomers concentrate on the solution to these problems. We
have developed automatic classifiers that, in a matter of seconds or
minutes, can automatically assign class probabilities to hundreds of
thousands of variable objects, and we have proved that these
probabilities are highly reliable for the set of classical variables
best studied in the literature. These experiments are repeatable and
thus free from human subjectivity. The classifiers show minor
discrepancies with the classifications used as a reference in this
work (as explained in previous sections) and these discrepancies, when
due to the classifiers themselves, need to be corrected for. Until
then, users of the publicly available classifiers have to be aware of
these minor pitfalls when interpreting their results.

The results presented here suggest that further steps can be taken in
the analysis of the resulting samples. Two obvious steps are the
search for correlations between subsets of attributes not necessarily
of dimension 2, and the study of density plots and clustering results
in order to explore the substructure within each variability
class. This is the subject of ongoing research in the framework of the
CoRoT, Kepler and Gaia missions.

The training set and the classifiers are only the first operational
versions developed for the optimization of on-going and future
databases such as CoRoT, Kepler or Gaia. Obviously, both the training
set and the classifiers will greatly benefit from the analysis of
these future databases, especially for those classes underrepresented
in terms of the real prevalences. This is where the improvement and
correction of the discrepancies mentioned in the previous paragraph
will take place. They must be oriented towards obtaining a class
definition (training) set that better reproduces the real probability
densities in parameter space (the probability of a variable object of
class $C_k$ having a certain set of attributes such as frequencies,
amplitudes, phase differences, colours, etc). Furthermore, it must be
made more robust against overfitting by combining data from various
surveys/instruments in such a way that the sampling properties
(including measurement errors) have as little an impact on the
inference process as possible. We believe that this paper is a
crucial starting point in the sense that we have proved the validity
of the classifier predictions, and, at the same time, we have
identified and pointed out the source of its limitations, thus showing
the path to more complete and accurate classifiers. Obviously, it is
in the non-periodic and rarer classes that there is more room for
improvement.

Finally, there is ongoing development of new versions of the
classifiers adapted to handle spectral information making use of VSOP
data \citep{2007A&A...470.1201D} and including one of the features of
Bayesian Networks that make them especially suitable for their
integration in the framework of Virtual Observatories, i.e. their
capacity to draw inferences based on incomplete (missing) data. We
strongly believe that the probabilistic foundations of these models
(at the basis of these capabilities) provide astronomers with
explanations of the inference process very much in line with the
reasoning usually used in astronomy.\\

In this work We have concentrated on the validation of the developed
classifiers, using the OGLE database. This database contains a large
number of light curves of different variability types. Existing
extractor-type results for the classical pulsators and eclipsing
binaries allowed us to judge the quality of our classification
results. Our classifiers also identified candidate new members for
some of those classes. Little had been done up to now on the
multiperiodic pulsators, the most interesting targets from an
asteroseismological point of view. The OGLE data are not optimally
suited to study those variables, but some types could be
studied and discovered. Our classifiers have identified $107$
candidate B-type pulsators (SPB, BCEP and PVSG) in the Magellanic
clouds. Those candidates were placed on the HR diagram, to see how they
are situated with respect to the instability strips of B-type
pulsators. This allowed us to conclude that the present instability
computations are incomplete and that their improvement probably needs
new input physics. In practice, we provide here a list of new
candidate variables of multiperiodic classes (DSCUT, BCEP, SPB and
GDOR), including several in the Bulge. A more in-depth analysis of
these candidates is needed, but this is outside the scope of this
classification work.

\begin{acknowledgements}

This work was made possible thanks to support from the Belgian PRODEX
programme under grant PEA C90199 (CoRoT Mission Data Exploitation II),
from the research Council of Leuven University under grant GOA/2003/04
and from the Spanish Ministerio de Educaci\'on y Ciencia through grant
AYA2005-04286. This research has made use of the Spanish Virtual
Observatory supported from the Spanish MEC through grants
AyA2005-04286, AyA2005-24102-E. This publication makes use of data
products from the Two Micron All Sky Survey, which is a joint project
of the University of Massachusetts and the Infrared Processing and
Analysis Center/California Institute of Technology, funded by the
National Aeronautics and Space Administration and the National Science
Foundation. We are very grateful to A. Pigulski for showing us his list
of candidate pulsators prior to publication. Finally, we are very
grateful to the referee, Dr. Soszynski, for his constructive comments
and suggestions on how to improve the original manuscript.

\end{acknowledgements}

\bibliographystyle{aa.bst}
\bibliography{references-code}

\Online

\begin{figure*}
   \centering
   \includegraphics[angle=-90,scale=0.6]{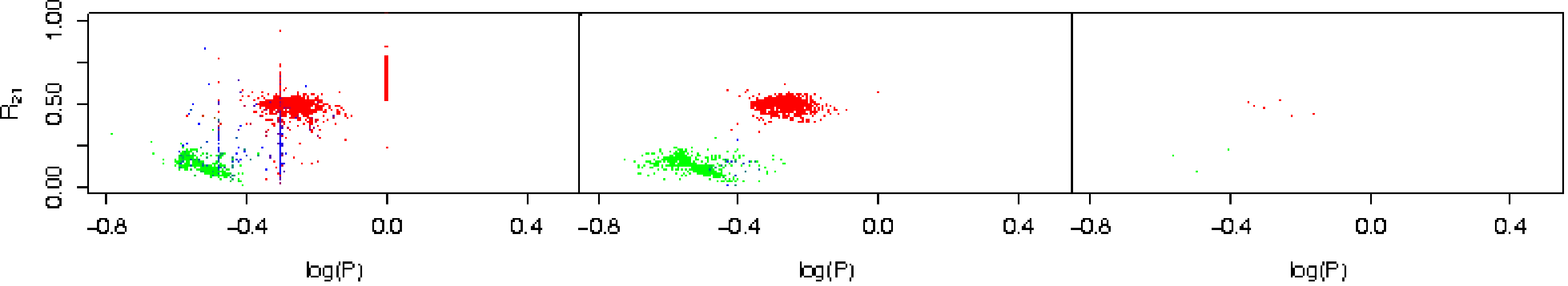}
   \caption{The $R_{21}-\log(P)$ plane of RRAB (red), RRC (green) and
   RRD stars (blue) in the Galactic Bulge, according to the multistage Bayesian
   networks (left) and Gaussian Mixtures classifiers (middle) and the
   OGLE catalogue (right).}
   \label{bulge-rrlyr-f1-r21}
\end{figure*}

\begin{figure*}
   \centering
   \includegraphics[angle=-90,scale=0.6]{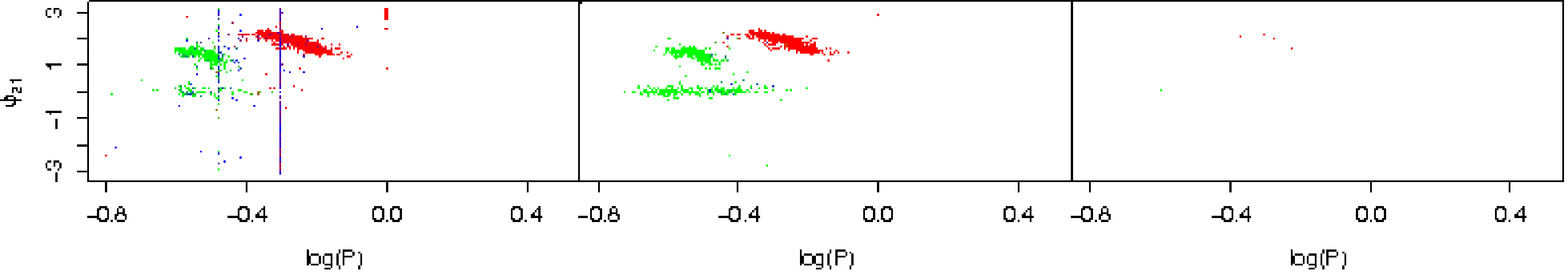}
   \caption{The $\phi_{21}-\log(P)$ plane of RRAB (red), RRC (green)
   and RRD stars (blue) in the Galactic Bulge, according to the
   multistage Bayesian networks (left) and Gaussian Mixtures
   classifiers (middle) and the OGLE catalogue (right).}
   \label{bulge-rrlyr-f1-phi21}
\end{figure*}

\begin{figure*}
   \centering
   \includegraphics[angle=-90,scale=0.6]{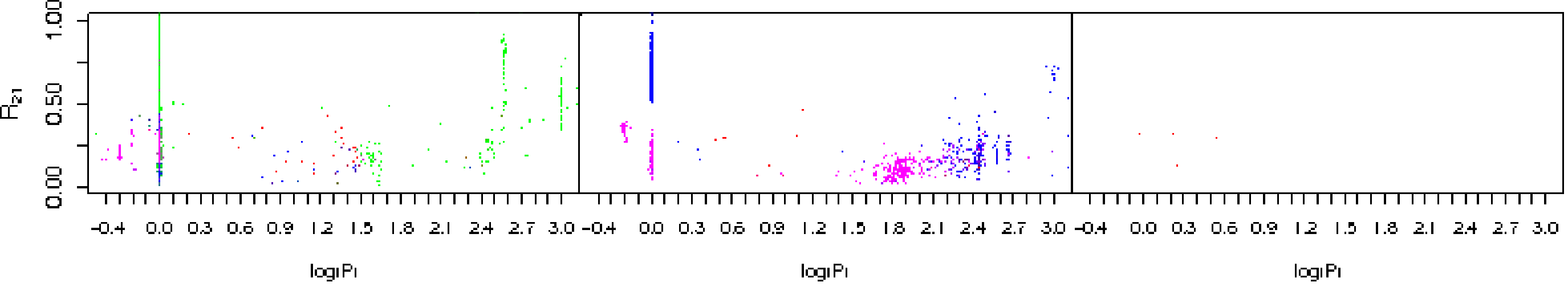}
   \caption{The $R_{21}-\log(P)$ plane of classical Cepheids (red),
   RVTAU (green), PTCEP (blue) and DMCEP (magenta) in the Galactic
   Bulge according to the multistage and GM classifiers and the OGLE
   team sample (only fundamental and first overtone classical Cepheids
   in red, and DMCEP in magenta).}
   \label{bulcep-f1-r21}
\end{figure*}

\begin{figure*}
   \centering
   \includegraphics[angle=-90,scale=0.6]{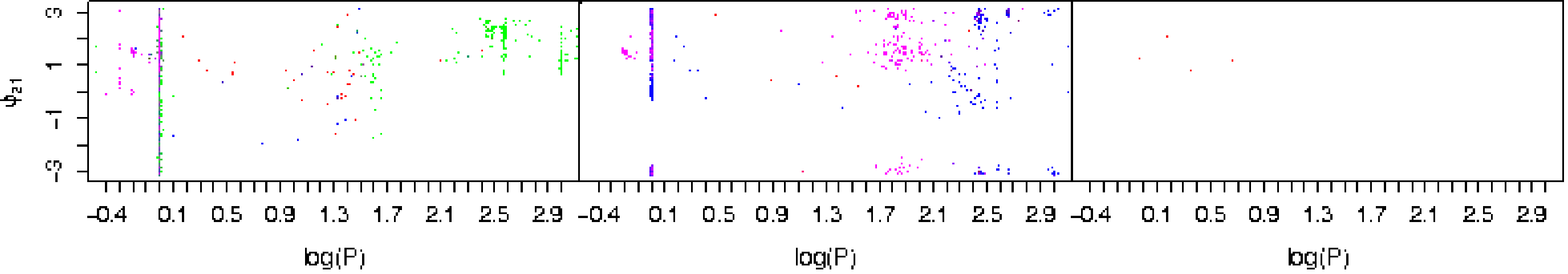}
   \caption{The $\phi_{21}-\log(P)$ plane of classical Cepheids (red),
   RVTAU (green), PTCEP (blue) and DMCEP (magenta) in the Galactic
   Bulge according to the multistage and GM classifiers and the OGLE
   team sample (only fundamental and first overtone classical Cepheids
   in red, and DMCEP in magenta).}
   \label{bulcep-f1-phi21}
\end{figure*}

\begin{figure*}
   \centering
   \includegraphics[angle=-90,scale=0.6]{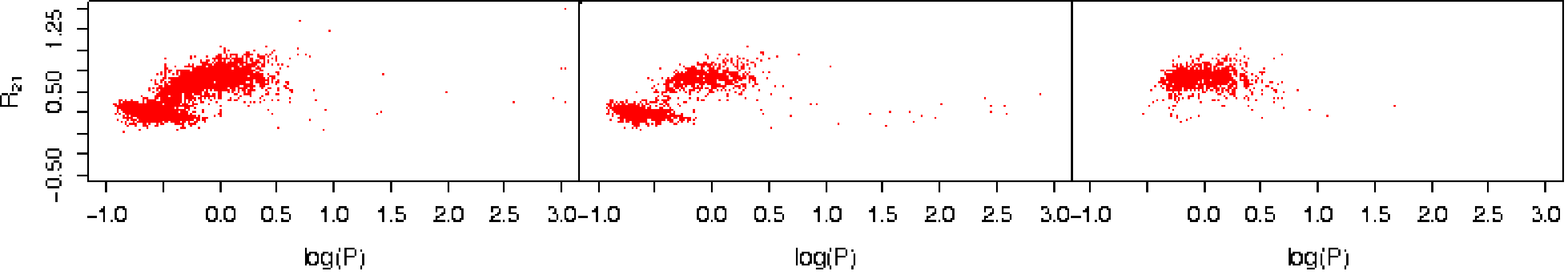}
   \caption{The $R_{21}-\log(P)$ plane of eclipsing binaries for the
    Galactic Bulge. From left to right, the MSBN and GM samples and
    the OGLE catalogue.}
   \label{bulecl-f1-r21}
\end{figure*}

\begin{figure*}
   \centering
   \includegraphics[angle=-90,scale=0.6]{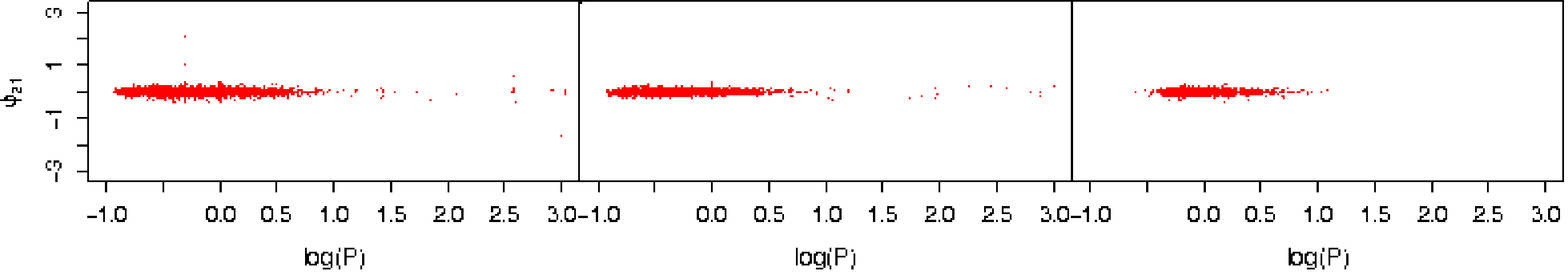}
   \caption{The $\phi_{21}-\log(P)$ plane of eclipsing binaries for
    the Galactic Bulge. From left to right, the MSBN and GM samples and the
    OGLE catalogue.}
   \label{bulecl-f1-phi21}
\end{figure*}

\begin{figure*}
   \centering
   \includegraphics[angle=-90,scale=0.3]{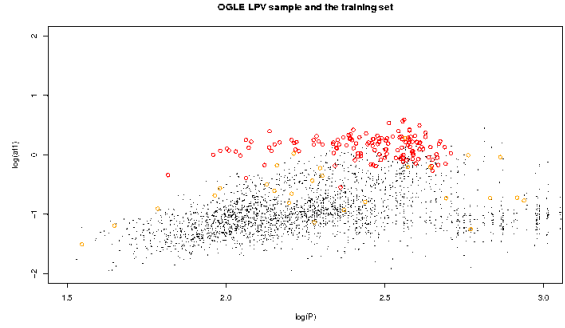}
   \caption{First frequency amplitude vs. $\log(P)$ of Miras (red) and
   SRs (orange) in the training set and in the OGLE sample (black).}
   \label{TS-OGLE}
\end{figure*}

\begin{figure*}
   \centering
   \includegraphics[angle=-90,scale=0.4]{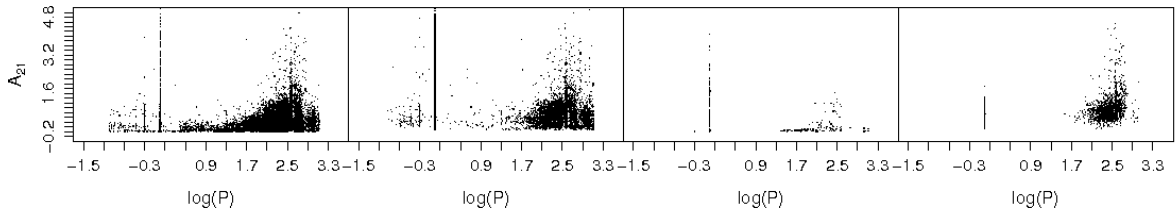}
   \caption{The $A_{11}-\log(P)$ plane of long period variables for
    the Galactic Bulge. From left to right, the MSBN and GM samples and the
    Mizerski and Groenewegen catalogues.}
   \label{bullpv-f1-a11}
\end{figure*}

\begin{figure*}
   \centering
   \includegraphics[width=16.5cm,scale=0.7]{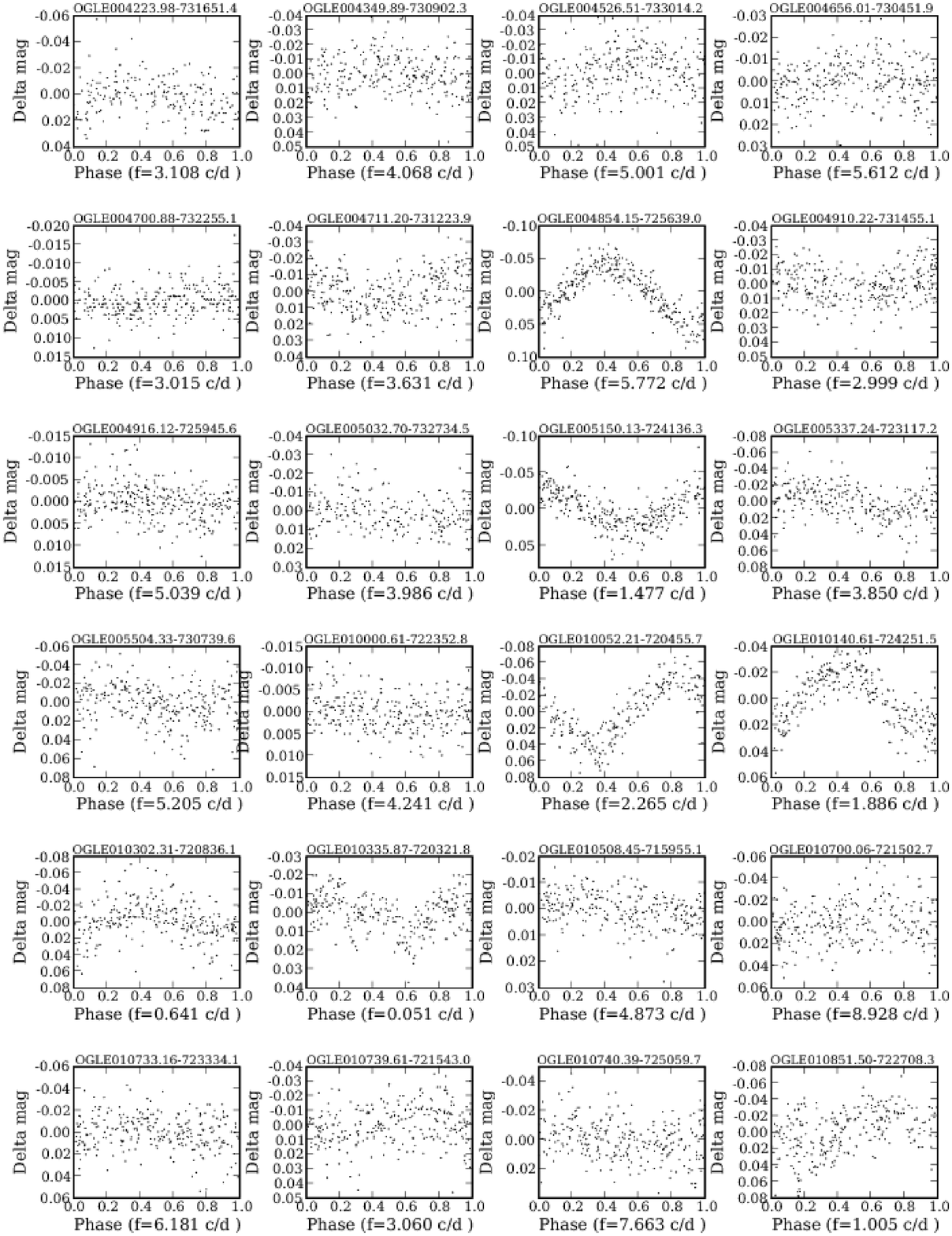}
   \caption{Phase plots of variables in the SMC classified as BCEP with both the MSBN and the GM method. The OGLE identifier is shown, and the dominant frequency, used to fold the light curves, in units of cycles per day (c/d).}
   \label{BCEP-SMC}
\end{figure*}
\begin{figure*}
   \centering
   \includegraphics[width=16.5cm,scale=0.7]{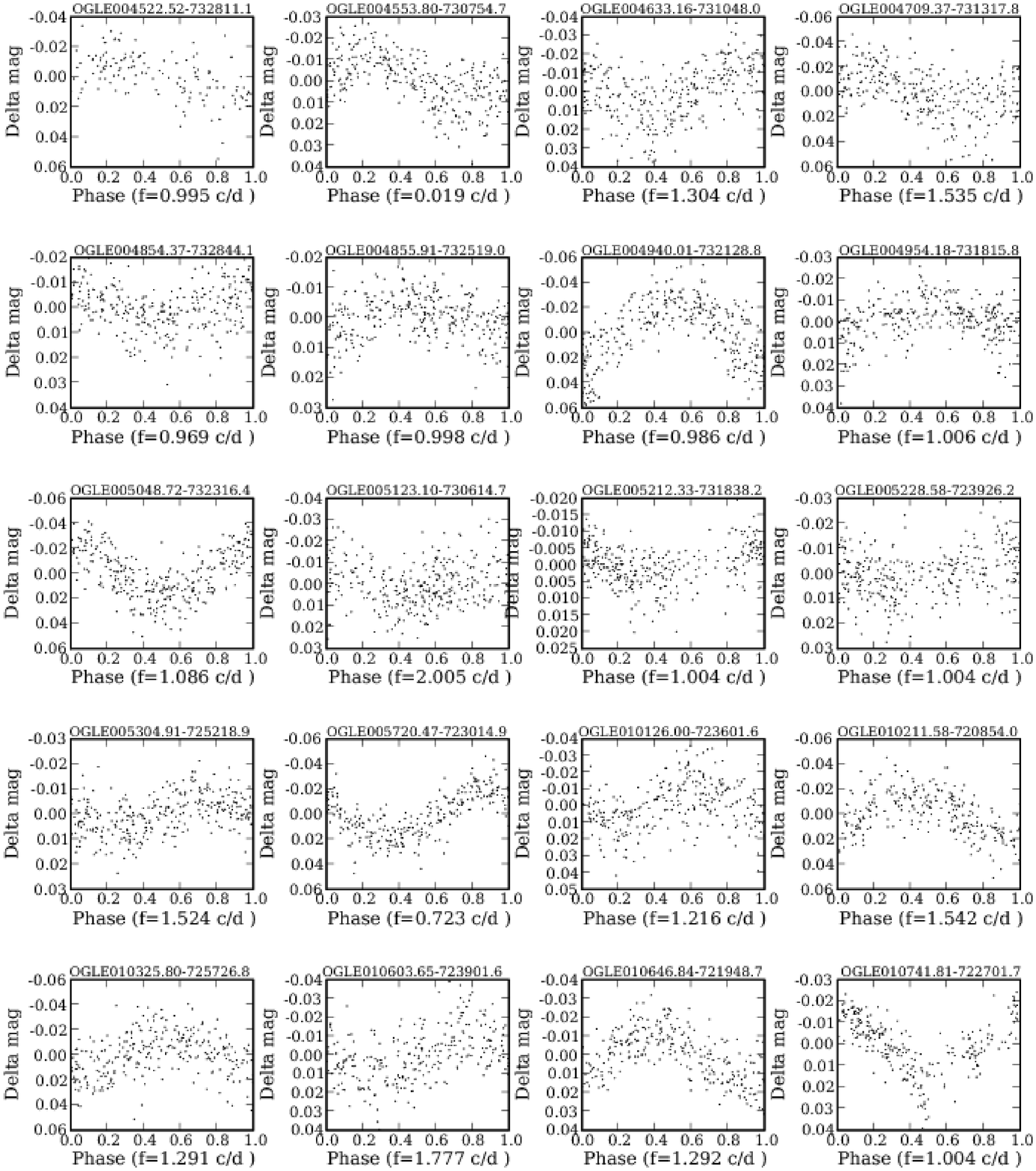}
   \caption{Phase plots of variables in the SMC classified as SPB with both the MSBN and the GM method. The OGLE identifier is shown, and the dominant frequency, used to fold the light curves, in units of cycles per day (c/d).}
   \label{SPB-SMC}
\end{figure*}
\begin{figure*}
   \centering
   \includegraphics[width=16.5cm,scale=0.7]{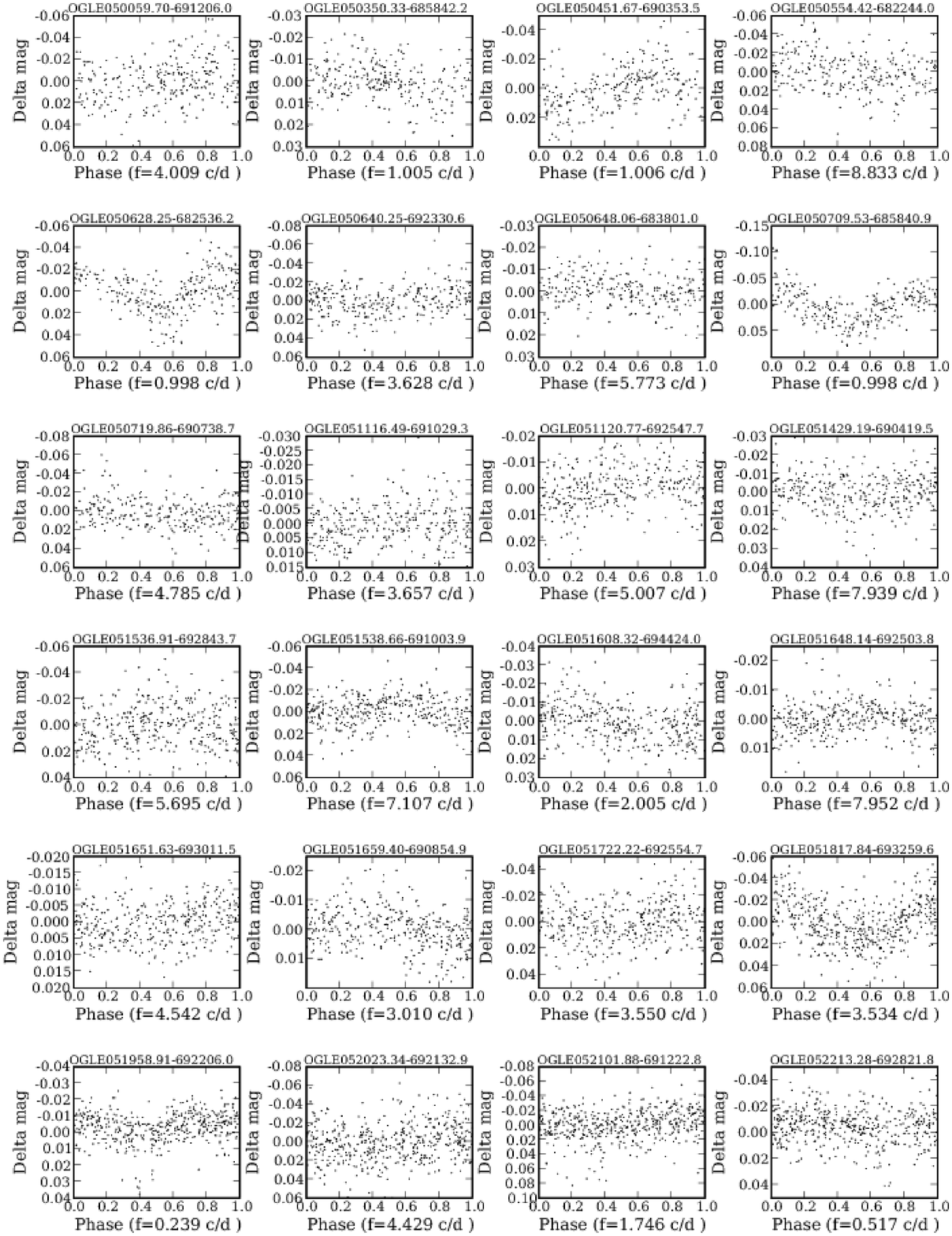}
   \caption{Phase plots of variables in the LMC classified as BCEP with both the MSBN and the GM method. The OGLE identifier is shown, and the dominant frequency, used to fold the light curves, in units of cycles per day (c/d).}
   \label{BCEP-LMC-1}
\end{figure*}
\begin{figure*}
   \centering
   \includegraphics[width=16.5cm,scale=0.7]{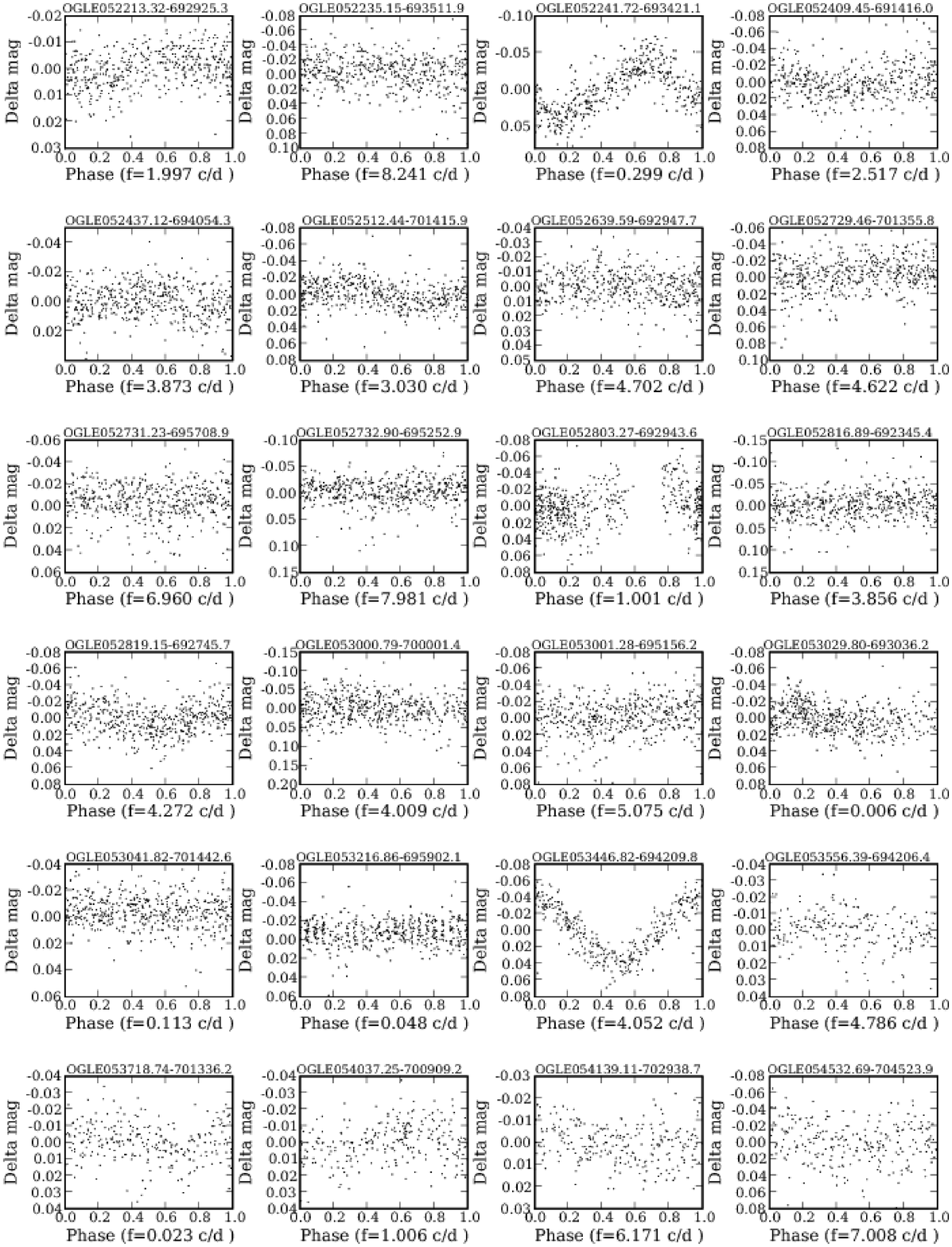}
   \caption{Phase plots of variables in the LMC classified as BCEP with both the MSBN and the GM method. The OGLE identifier is shown, and the dominant frequency, used to fold the light curves, in units of cycles per day (c/d).}
   \label{BCEP-LMC-2}
\end{figure*}
\begin{figure*}
   \centering
   \includegraphics[width=16.5cm,scale=0.7]{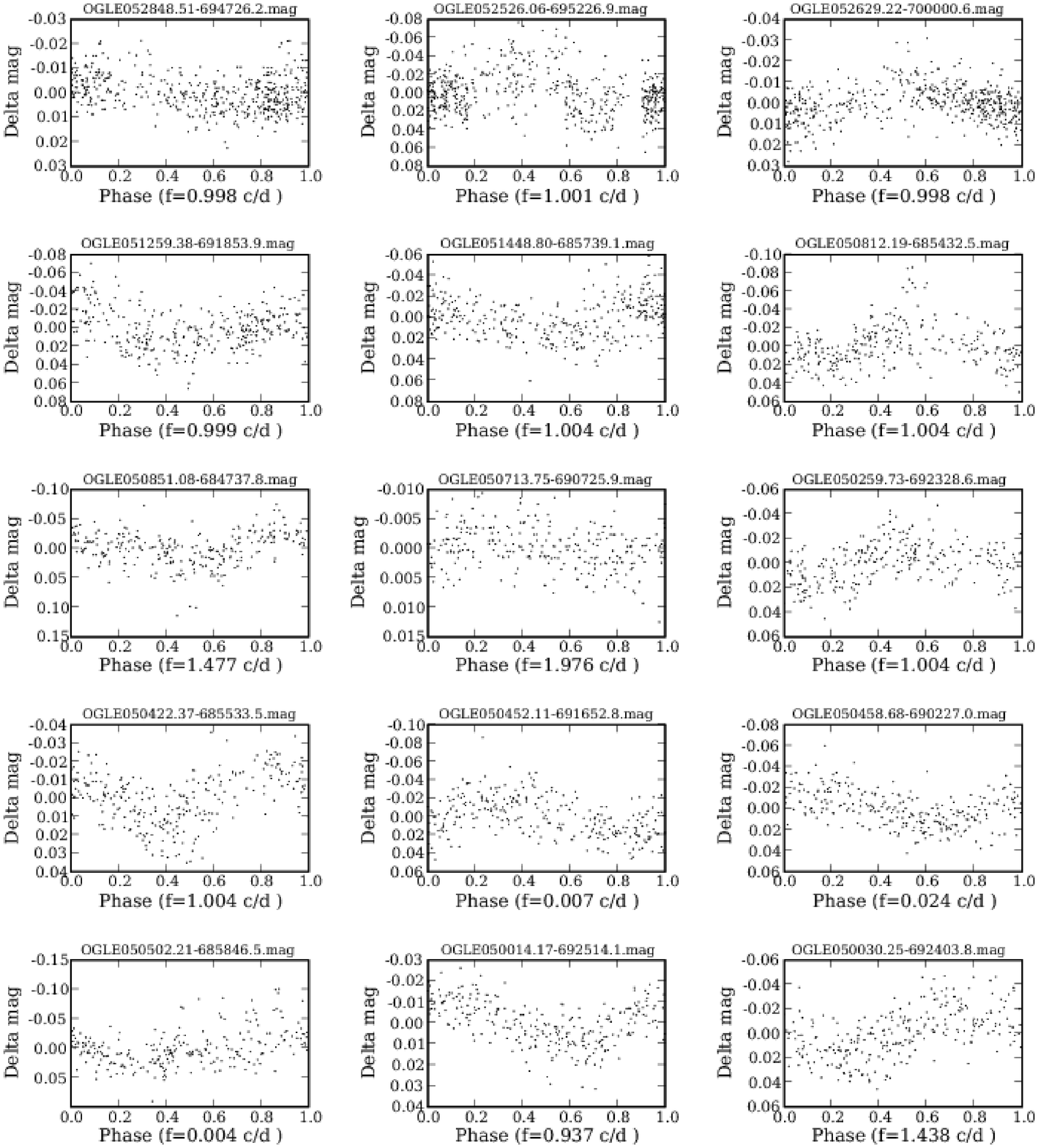}
   \caption{Phase plots of variables in the LMC classified as SPB with both the MSBN and the GM method. The OGLE identifier is shown, and the dominant frequency, used to fold the light curves, in units of cycles per day (c/d).}
   \label{SPB-LMC}
\end{figure*}

\clearpage
\begin{table*}
\caption{Basic light curve and physical properties of SMC stars
classified as BCEP with both the MSBN and the GM method. The dominant
frequency $f_1$, the second frequency $f_2$, the effective temperature
$\log T_{eff}$ and the luminosity $\log (L/L_{\sun})$ are listed. The
estimated precision on the frequencies is about $0.001$ c/d or
smaller. Note that $\log T_{eff}$ and $\log (L/L_{\sun})$ are listed
with more digits than the estimated precision, with the only purpose
to allow readers to locate the objects in the HR diagrams. The same
remark applies to the following tables also. Several of these stars
might be evolved pulsators, termed PVSG here, rather than BCEP.}
\renewcommand{\tabcolsep}{1.1mm}
\begin{tabular}{ccccc}
\hline
Object identifier&$f_1$ ($c/d$)&$f_2$ ($c/d$)&$\log T_{eff}$&$\log (L/L_{\sun})$\\
\hline
OGLE004223.98-731651.4&$3.108 $&$3.124 $&$4.31 $&$4.58 $\\
OGLE004349.89-730902.3&$4.068 $&$7.151 $&$4.38 $&$4.96 $\\
OGLE004526.51-733014.2&$5.001 $&$7.521 $&$4.19 $&$2.56 $\\
OGLE004656.01-730451.9&$5.612 $&$8.576 $&$4.42 $&$4.34 $\\
OGLE004700.88-732255.1&$3.015 $&$9.123 $&$4.25 $&$4.53 $\\
OGLE004711.20-731223.9&$3.631 $&$3.735 $&$4.22 $&$3.51 $\\
OGLE004854.15-725639.0&$5.772 $&$1.006 $&$4.27 $&$3.28 $\\
OGLE004910.22-731455.1&$2.999 $&$0.135 $&$4.31 $&$4.61 $\\
OGLE004916.12-725945.6&$5.039 $&$4.986 $&$4.37 $&$2.56 $\\
OGLE005032.70-732734.5&$3.986 $&$5.609 $&$4.35 $&$4.09 $\\
OGLE005150.13-724136.3&$1.477 $&$0.739 $&$4.46 $&$3.88 $\\
OGLE005337.24-723117.2&$3.850 $&$0.001 $&$4.26 $&$3.60 $\\
OGLE005504.33-730739.6&$5.205 $&$5.410 $&$4.14 $&$3.10 $\\
OGLE010000.61-722352.8&$4.241 $&$3.008 $&$4.35 $&$4.62 $\\
OGLE010052.21-720455.7&$2.265 $&$2.006 $&$4.40 $&$4.11 $\\
OGLE010140.61-724251.5&$1.886 $&$0.943 $&$4.45 $&$4.08 $\\
OGLE010302.31-720836.1&$0.641 $&$0.298 $&$4.32 $&$4.49 $\\
OGLE010335.87-720321.8&$0.051 $&$0.355 $&$4.43 $&$5.19 $\\
OGLE010508.45-715955.1&$4.873 $&$2.251 $&$4.20 $&$3.67 $\\
OGLE010700.06-721502.7&$8.928 $&$7.743 $&$4.27 $&$3.80 $\\
OGLE010733.16-723334.1&$6.181 $&$1.409 $&$4.17 $&$3.12 $\\
OGLE010739.61-721543.0&$3.060 $&$1.002 $&$4.24 $&$4.43 $\\
OGLE010740.39-725059.7&$7.663 $&$1.057 $&$4.44 $&$5.19 $\\
OGLE010851.50-722708.3&$1.005 $&$0.010 $&$4.45 $&$4.48 $\\

\hline
\end{tabular}

\label{BCEP-SMC-param}
\end{table*}

\begin{table*}
\caption{Basic light curve and physical properties of SMC stars classified as SPB with both the MSBN and the GM method. The dominant frequency $f_1$, the second frequency $f_2$, the effective temperature $\log T_{eff}$ and the luminosity $\log (L/L_{\sun})$ are listed. Several of these stars might be evolved pulsators, termed PVSG here, rather than SPB.}
\renewcommand{\tabcolsep}{1.1mm}
\begin{tabular}{ccccc}
\hline
Object identifier&$f_1$ ($c/d$)&$f_2$ ($c/d$)&$\log T_{eff}$&$\log (L/L_{\sun})$\\
\hline
OGLE004522.52-732811.1&$0.995 $&$6.800 $&$4.06 $&$3.49 $\\
OGLE004553.80-730754.7&$0.019 $&$0.286 $&$4.07 $&$3.72 $\\
OGLE004633.16-731048.0&$1.304 $&$2.792 $&$4.07 $&$3.41 $\\
OGLE004709.37-731317.8&$1.535 $&$0.488 $&$4.10 $&$3.65 $\\
OGLE004854.37-732844.1&$0.969 $&$1.001 $&$4.04 $&$3.88 $\\
OGLE004855.91-732519.0&$0.998 $&$1.005 $&$4.11 $&$3.89 $\\
OGLE004940.01-732128.8&$0.986 $&$0.493 $&$4.08 $&$3.36 $\\
OGLE004954.18-731815.8&$1.006 $&$0.002 $&$4.05 $&$3.67 $\\
OGLE005048.72-732316.4&$1.086 $&$1.163 $&$4.10 $&$3.39 $\\
OGLE005123.10-730614.7&$2.005 $&$3.127 $&$4.05 $&$3.09 $\\
OGLE005212.33-731838.2&$1.004 $&$6.006 $&$4.08 $&$4.19 $\\
OGLE005228.58-723926.2&$1.004 $&$0.273 $&$4.11 $&$3.38 $\\
OGLE005304.91-725218.9&$1.524 $&$2.254 $&$4.07 $&$4.35 $\\
OGLE005720.47-723014.9&$0.723 $&$1.620 $&$4.06 $&$3.27 $\\
OGLE010126.00-723601.6&$1.216 $&$0.998 $&$4.11 $&$3.20 $\\
OGLE010211.58-720854.0&$1.542 $&$0.001 $&$4.09 $&$3.12 $\\
OGLE010325.80-725726.8&$1.291 $&$1.368 $&$4.03 $&$3.12 $\\
OGLE010603.65-723901.6&$1.777 $&$0.003 $&$4.08 $&$3.15 $\\
OGLE010646.84-721948.7&$1.292 $&$0.999 $&$4.14 $&$3.67 $\\
OGLE010741.81-722701.7&$1.004 $&$3.009 $&$4.11 $&$3.67 $\\
\hline
\end{tabular}

\label{SPB-SMC-param}
\end{table*}

\begin{table*}
\caption{Basic light curve and physical properties of LMC stars classified as BCEP with both the MSBN and the GM method. The dominant frequency $f_1$, the second frequency $f_2$, the effective temperature $\log T_{eff}$ and the luminosity $\log (L/L_{\sun})$ are listed. Several of these stars might be evolved pulsators, termed PVSG here, rather than BCEP.}
\renewcommand{\tabcolsep}{1.1mm}
\begin{tabular}{ccccc}
\hline
Object identifier&$f_1$ ($c/d$)&$f_2$ ($c/d$)&$\log T_{eff}$&$\log (L/L_{\sun})$\\
\hline
OGLE053446.82-694209.8&$4.052 $&$8.120 $&$4.29 $&$3.29 $\\
OGLE053000.79-700001.4&$4.009 $&$2.991 $&$4.24 $&$2.62 $\\
OGLE053001.28-695156.2&$5.075 $&$6.642 $&$4.21 $&$2.82 $\\
OGLE053029.80-693036.2&$0.006 $&$0.001 $&$4.46 $&$3.51 $\\
OGLE053041.82-701442.6&$0.113 $&$0.045 $&$4.22 $&$3.11 $\\
OGLE053216.86-695902.1&$0.048 $&$0.357 $&$4.42 $&$3.58 $\\
OGLE052729.46-701355.8&$4.622 $&$2.018 $&$4.12 $&$2.58 $\\
OGLE052731.23-695708.9&$6.960 $&$4.350 $&$4.28 $&$3.12 $\\
OGLE052732.90-695252.9&$7.981 $&$4.201 $&$4.26 $&$2.90 $\\
OGLE052803.27-692943.6&$1.001 $&$5.273 $&$4.34 $&$3.12 $\\
OGLE052816.89-692345.4&$3.856 $&$3.831 $&$4.30 $&$2.86 $\\
OGLE052819.15-692745.7&$4.272 $&$0.725 $&$4.16 $&$2.01 $\\
OGLE052512.44-701415.9&$3.030 $&$1.010 $&$4.47 $&$3.74 $\\
OGLE052639.59-692947.7&$4.702 $&$4.124 $&$4.36 $&$3.49 $\\
OGLE052235.15-693511.9&$8.241 $&$0.011 $&$4.28 $&$3.03 $\\
OGLE052241.72-693421.1&$0.299 $&$9.976 $&$4.21 $&$2.96 $\\
OGLE052409.45-691416.0&$2.517 $&$2.537 $&$4.22 $&$3.05 $\\
OGLE052437.12-694054.3&$3.873 $&$2.005 $&$4.38 $&$3.61 $\\
OGLE052023.34-692132.9&$4.429 $&$4.440 $&$4.29 $&$3.08 $\\
OGLE052101.88-691222.8&$1.746 $&$0.040 $&$4.25 $&$3.10 $\\
OGLE052213.28-692821.8&$0.517 $&$0.960 $&$4.22 $&$3.15 $\\
OGLE052213.32-692925.3&$1.997 $&$2.040 $&$4.28 $&$4.68 $\\
OGLE051817.84-693259.6&$3.534 $&$0.001 $&$4.39 $&$3.53 $\\
OGLE051958.91-692206.0&$0.239 $&$0.359 $&$4.32 $&$3.67 $\\
OGLE051536.91-692843.7&$5.695 $&$9.192 $&$4.18 $&$2.97 $\\
OGLE051538.66-691003.9&$7.107 $&$0.002 $&$4.22 $&$3.32 $\\
OGLE051608.32-694424.0&$2.005 $&$0.205 $&$4.38 $&$4.24 $\\
OGLE051648.14-692503.8&$7.952 $&$8.938 $&$4.50 $&$4.82 $\\
OGLE051651.63-693011.5&$4.542 $&$7.476 $&$4.40 $&$4.64 $\\
OGLE051659.40-690854.9&$3.010 $&$6.453 $&$4.28 $&$4.03 $\\
OGLE051722.22-692554.7&$3.550 $&$1.003 $&$4.35 $&$3.26 $\\
OGLE051429.19-690419.5&$7.939 $&$0.100 $&$4.29 $&$3.58 $\\
OGLE051116.49-691029.3&$3.657 $&$5.256 $&$4.45 $&$4.70 $\\
OGLE051120.77-692547.7&$5.007 $&$1.005 $&$4.21 $&$3.39 $\\
OGLE050640.25-692330.6&$3.628 $&$3.860 $&$4.21 $&$3.05 $\\
OGLE050554.42-682244.0&$8.833 $&$9.227 $&$4.29 $&$3.40 $\\
OGLE050628.25-682536.2&$0.998 $&$2.013 $&$4.31 $&$4.01 $\\
OGLE050648.06-683801.0&$5.773 $&$3.042 $&$4.28 $&$4.65 $\\
OGLE050709.53-685840.9&$0.998 $&$0.112 $&$4.24 $&$3.90 $\\
OGLE050719.86-690738.7&$4.785 $&$6.116 $&$4.30 $&$3.13 $\\
OGLE050350.33-685842.2&$1.005 $&$1.186 $&$4.35 $&$4.36 $\\
OGLE050451.67-690353.5&$1.006 $&$1.970 $&$4.24 $&$3.39 $\\
OGLE050059.70-691206.0&$4.009 $&$6.219 $&$4.26 $&$2.97 $\\
OGLE053556.39-694206.4&$4.786 $&$3.779 $&$4.41 $&$5.09 $\\
OGLE053718.74-701336.2&$0.023 $&$1.011 $&$4.26 $&$4.65 $\\
OGLE054037.25-700909.2&$1.006 $&$1.852 $&$4.29 $&$3.74 $\\
OGLE054139.11-702938.7&$6.171 $&$2.096 $&$4.27 $&$4.14 $\\
OGLE054532.69-704523.9&$7.008 $&$4.788 $&$4.16 $&$2.67 $\\
\hline
\end{tabular}

\label{BCEP-LMC-param}
\end{table*}

\begin{table*}
\caption{Basic light curve and physical properties of LMC stars classified as SPB with both the MSBN and the GM method. The dominant frequency $f_1$, the second frequency $f_2$, the effective temperature $\log T_{eff}$ and the luminosity $\log (L/L_{\sun})$ are listed. Several of these stars might be evolved pulsators, termed PVSG here, rather than SPB.}
\renewcommand{\tabcolsep}{1.1mm}
\begin{tabular}{ccccc}
\hline
Object identifier&$f_1$ ($c/d$)&$f_2$ ($c/d$)&$\log T_{eff}$&$\log (L/L_{\sun})$\\
\hline
OGLE052848.51-694726.2&$0.998 $&$2.006 $&$4.09 $&$3.24 $\\
OGLE052526.06-695226.9&$1.001 $&$8.362 $&$4.19 $&$2.86 $\\
OGLE052629.22-700000.6&$0.998 $&$1.003 $&$3.96 $&$3.40 $\\
OGLE051259.38-691853.9&$0.999 $&$1.014 $&$4.08 $&$2.77 $\\
OGLE051448.80-685739.1&$1.004 $&$0.992 $&$4.18 $&$3.11 $\\
OGLE050812.19-685432.5&$1.004 $&$1.005 $&$4.17 $&$3.18 $\\
OGLE050851.08-684737.8&$1.477 $&$8.872 $&$4.10 $&$2.74 $\\
OGLE050713.75-690725.9&$1.976 $&$6.631 $&$4.06 $&$3.69 $\\
OGLE050259.73-692328.6&$1.004 $&$0.999 $&$4.05 $&$3.06 $\\
OGLE050422.37-685533.5&$1.004 $&$1.006 $&$3.99 $&$2.81 $\\
OGLE050452.11-691652.8&$0.007 $&$0.990 $&$4.10 $&$2.87 $\\
OGLE050458.68-690227.0&$0.024 $&$0.049 $&$4.01 $&$2.89 $\\
OGLE050502.21-685846.5&$0.999 $&$0.031 $&$4.11 $&$3.06 $\\
OGLE050014.17-692514.1&$0.937 $&$2.082 $&$4.05 $&$3.14 $\\
OGLE050030.25-692403.8&$1.438 $&$3.278 $&$4.15 $&$3.14 $\\
\hline
\end{tabular}

\label{SPB-LMC-param}
\end{table*}

\clearpage
\begin{figure*}
   \centering
   \includegraphics[width=16.5cm,scale=0.7]{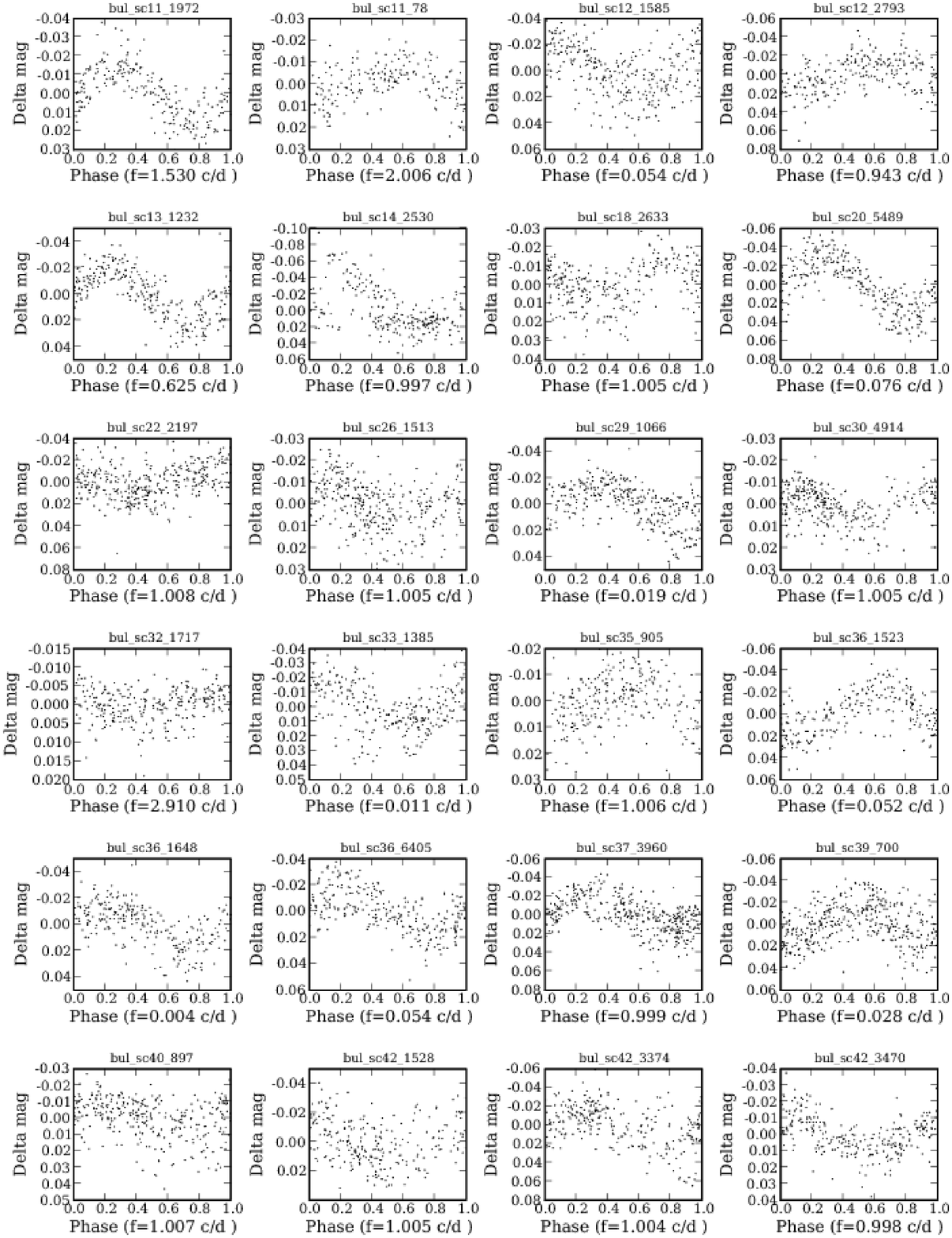}
   \caption{Phase plots of variables in the Galactic Bulge classified as SPB with both the MSBN and the GM method, and not present in the list of Pigulski. The OGLE identifier is shown, and the dominant frequency, used to fold the light curves, in units of cycles per day (c/d).}
   \label{SPB-BULGE}
\end{figure*}

\begin{figure*}
   \centering
   \includegraphics[width=16.5cm,scale=0.7]{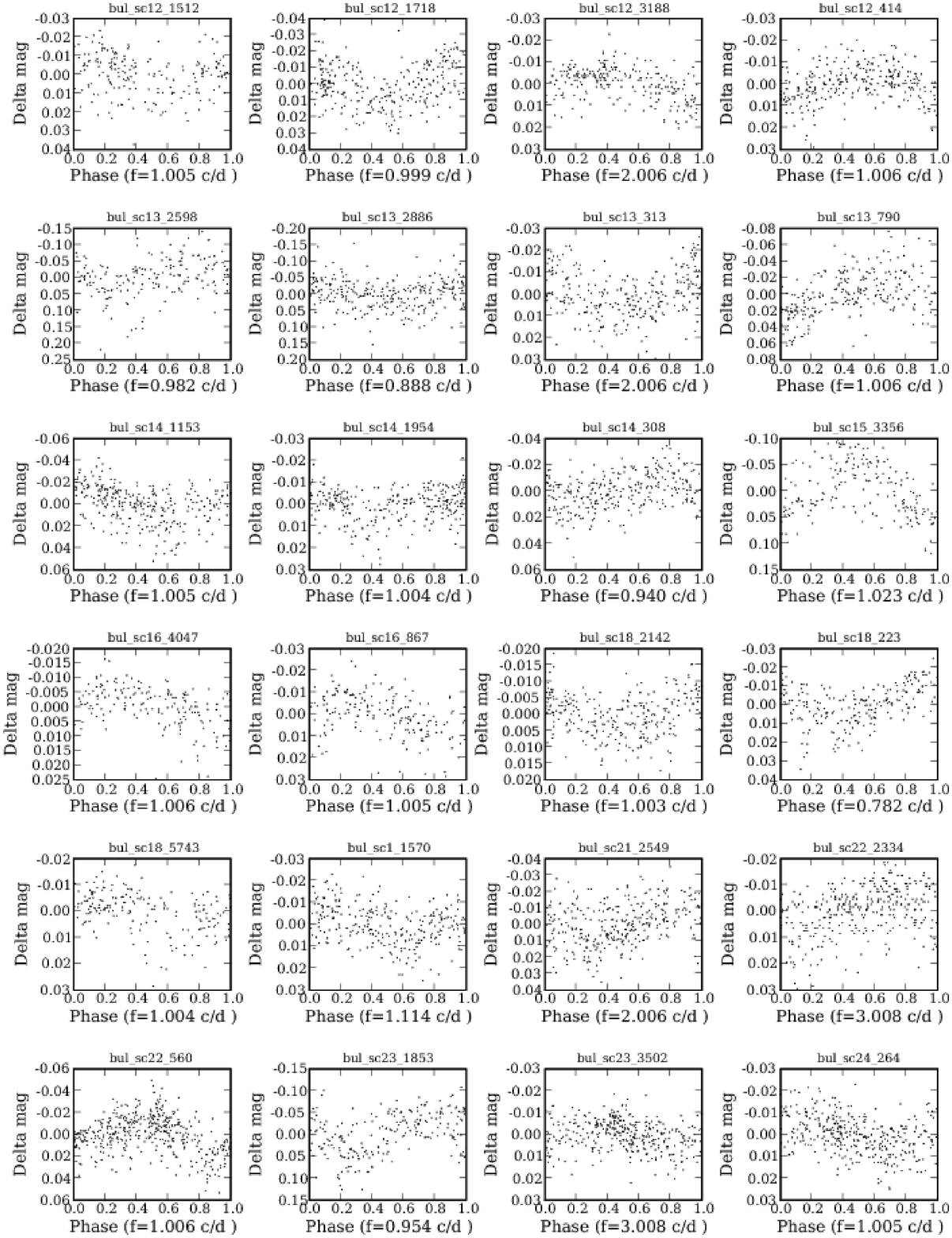}
   \caption{Phase plots of variables in the Galactic Bulge classified as GDOR with both the MSBN and the GM method, and not present in the list of Pigulski. The OGLE identifier is shown, and the dominant frequency, used to fold the light curves, in units of cycles per day (c/d).}
   \label{GDOR-BULGE-1}
\end{figure*}

\begin{figure*}
   \centering
   \includegraphics[width=16.5cm,scale=0.7]{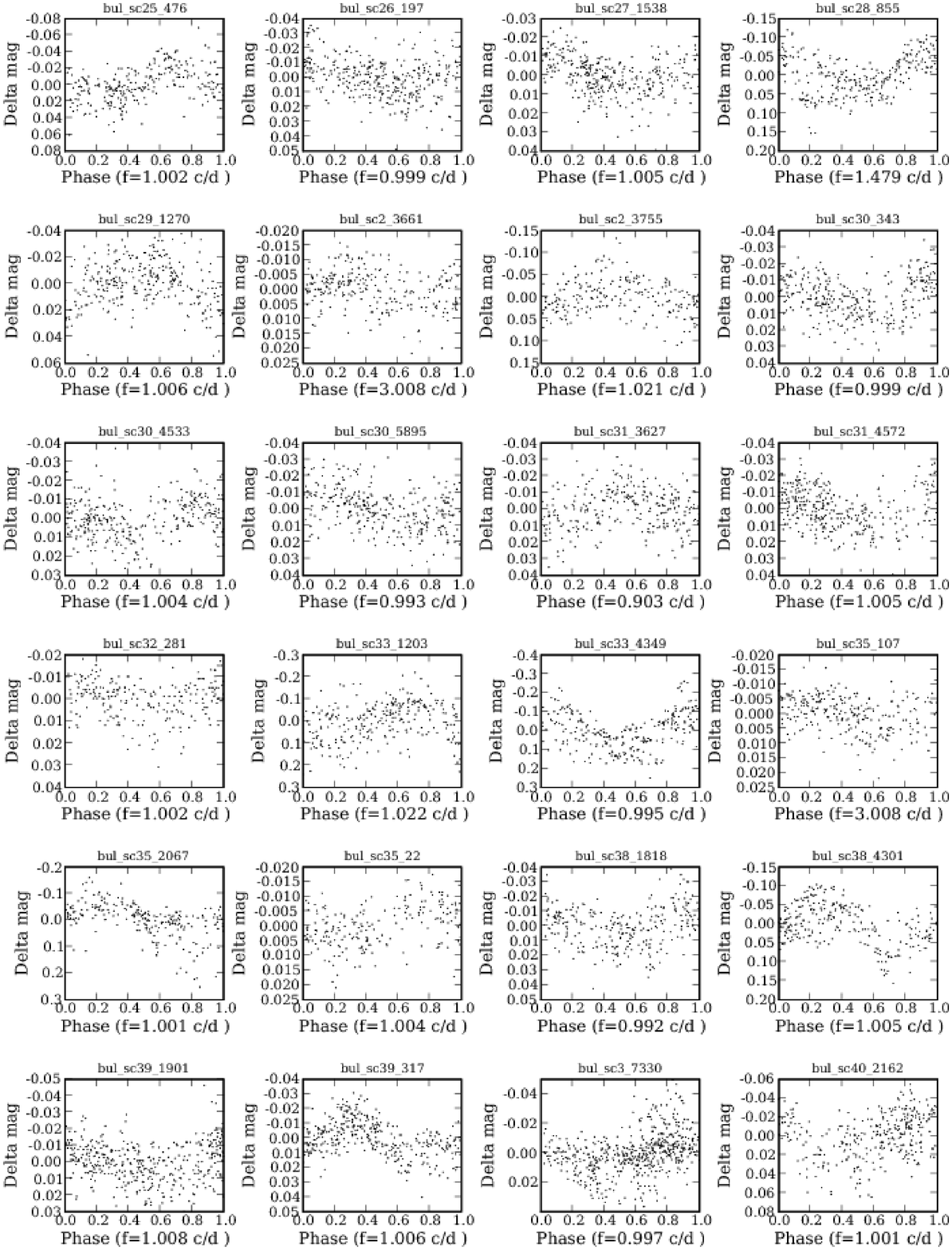}
   \caption{Phase plots of variables in the Galactic Bulge classified as GDOR with both the MSBN and the GM method, and not present in the list of Pigulski. The OGLE identifier is shown, and the dominant frequency, used to fold the light curves, in units of cycles per day (c/d).}
   \label{GDOR-BULGE-2}
\end{figure*}

\begin{figure*}
   \centering
   \includegraphics[width=16.5cm,scale=0.7]{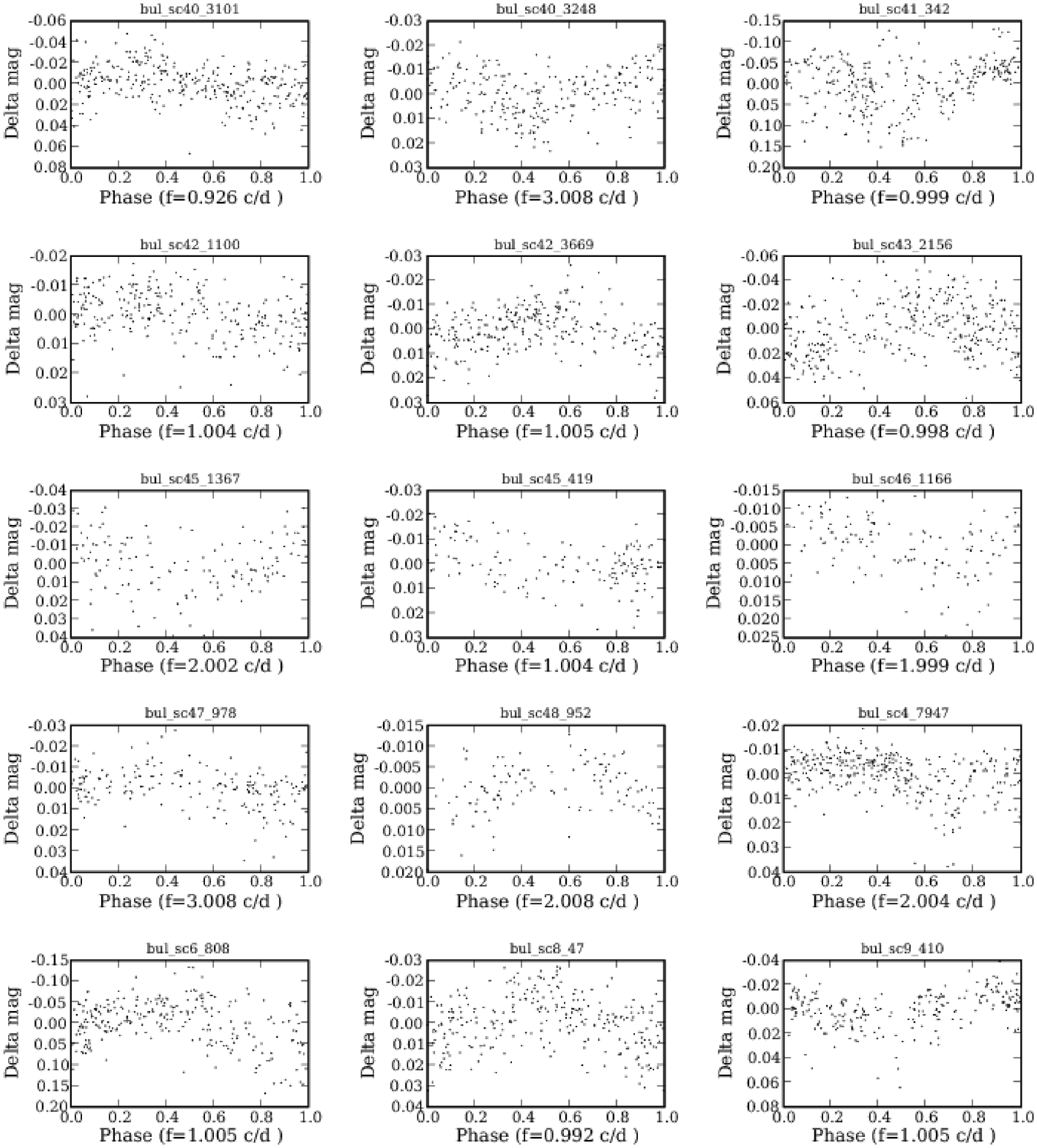}
   \caption{Phase plots of variables in the Galactic Bulge classified as GDOR with both the MSBN and the GM method, and not present in the list of Pigulski. The OGLE identifier is shown, and the dominant frequency, used to fold the light curves, in units of cycles per day (c/d).}
   \label{GDOR-BULGE-3}
\end{figure*}

\begin{figure*}
   \centering
   \includegraphics[width=16.5cm,scale=0.7]{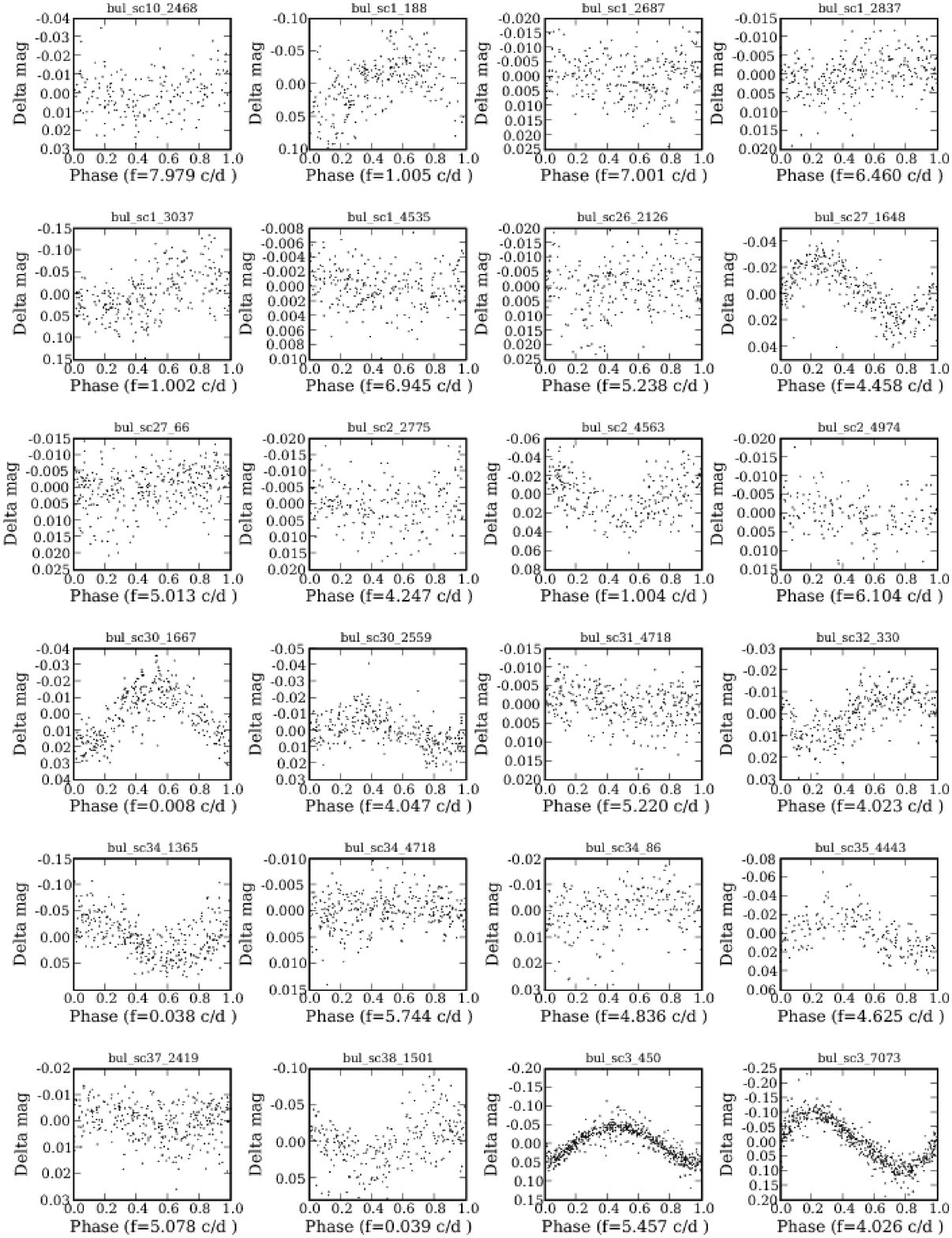}
   \caption{Phase plots of variables in the Galactic Bulge classified as BCEP with both the MSBN and the GM method, and not present in the list of Pigulski. The OGLE identifier is shown, and the dominant frequency, used to fold the light curves, in units of cycles per day (c/d).}
   \label{BCEP-BULGE-1}
\end{figure*}

\begin{figure*}
   \centering
   \includegraphics[width=16.5cm,scale=0.7]{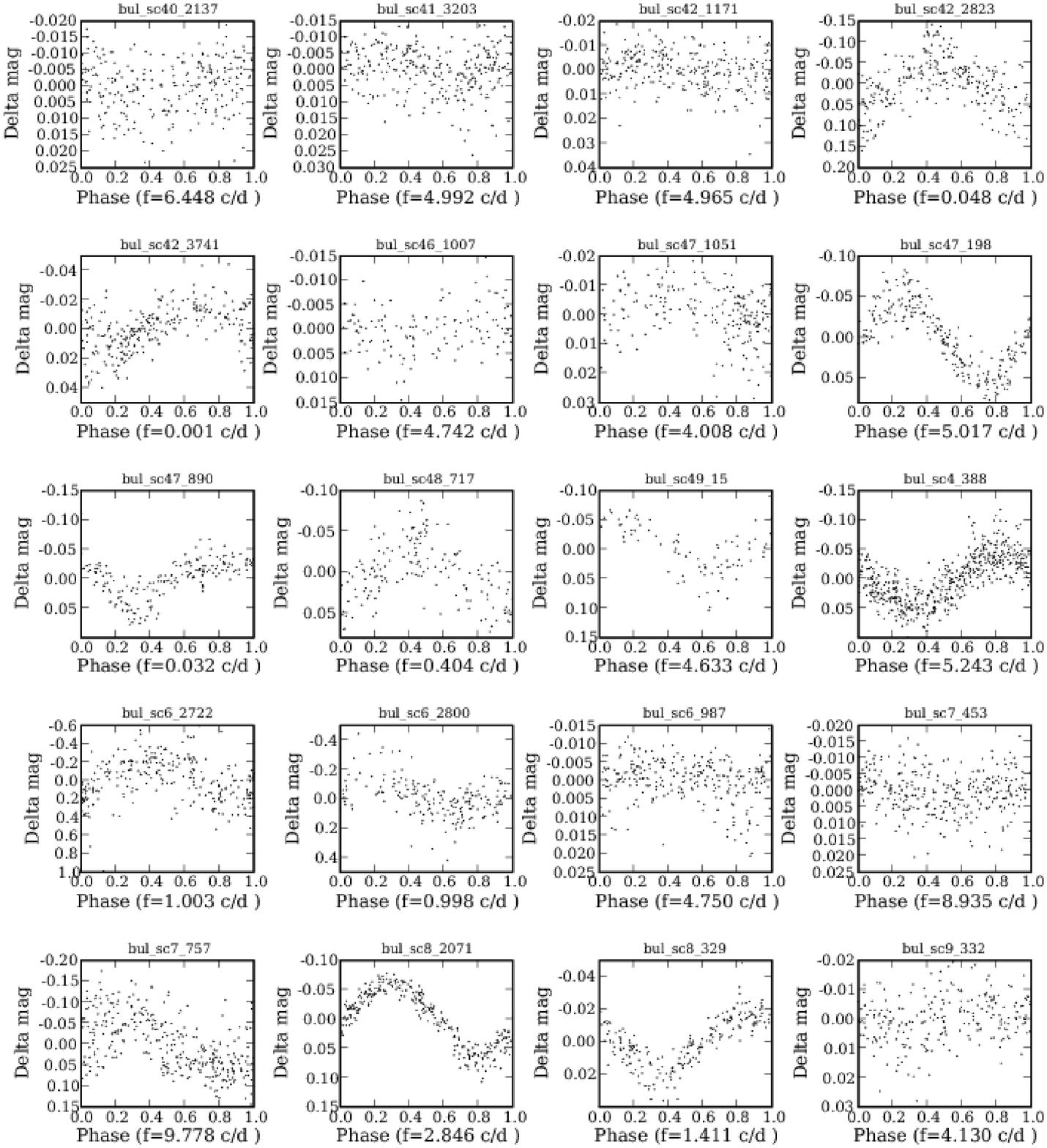}
   \caption{Phase plots of variables in the Galactic Bulge classified as BCEP with both the MSBN and the GM method, and not present in the list of Pigulski. The OGLE identifier is shown, and the dominant frequency, used to fold the light curves, in units of cycles per day (c/d).}
   \label{BCEP-BULGE-2}
\end{figure*}

\begin{figure*}
   \centering
   \includegraphics[width=16.5cm,scale=0.7]{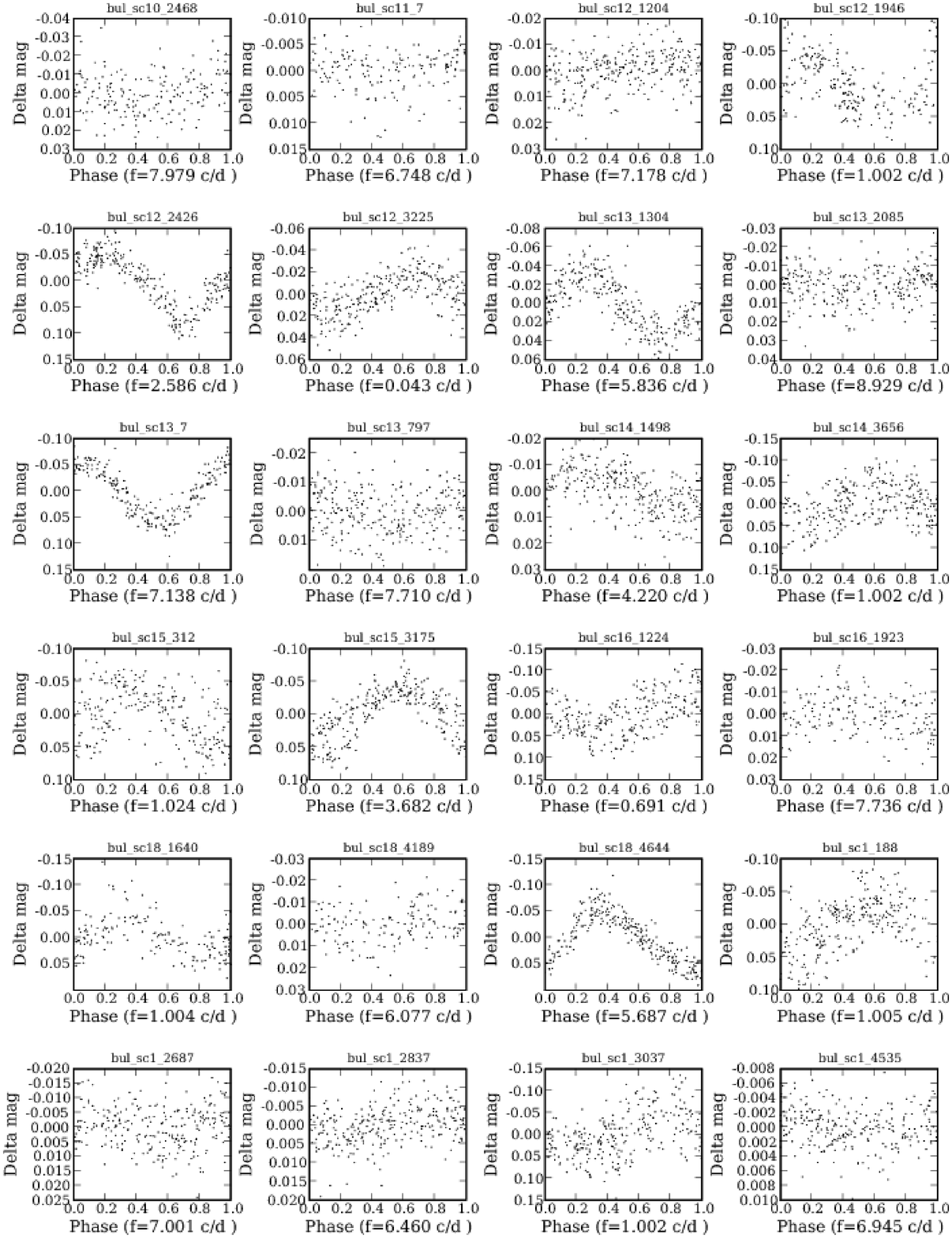}
   \caption{Phase plots of variables in the Galactic Bulge classified as DSCUT with both the MSBN and the GM method, and not present in the list of Pigulski. The OGLE identifier is shown, and the dominant frequency, used to fold the light curves, in units of cycles per day (c/d).}
   \label{DSCUT-BULGE-1}
\end{figure*}

\begin{figure*}
   \centering
   \includegraphics[width=16.5cm,scale=0.7]{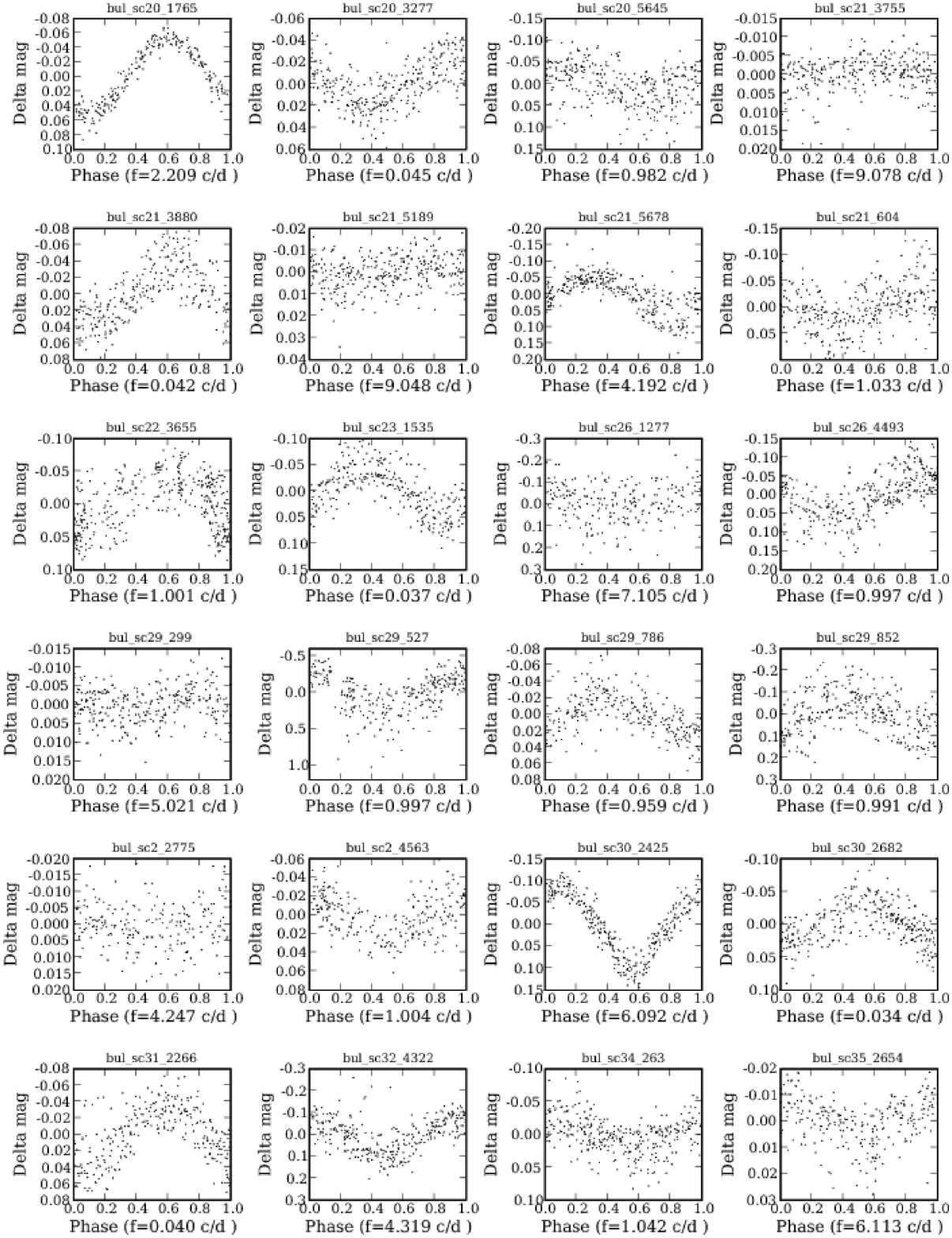}
   \caption{Phase plots of variables in the Galactic Bulge classified as DSCUT with both the MSBN and the GM method, and not present in the list of Pigulski. The OGLE identifier is shown, and the dominant frequency, used to fold the light curves, in units of cycles per day (c/d).}
   \label{DSCUT-BULGE-2}
\end{figure*}

\begin{figure*}
   \centering
   \includegraphics[width=16.5cm,scale=0.7]{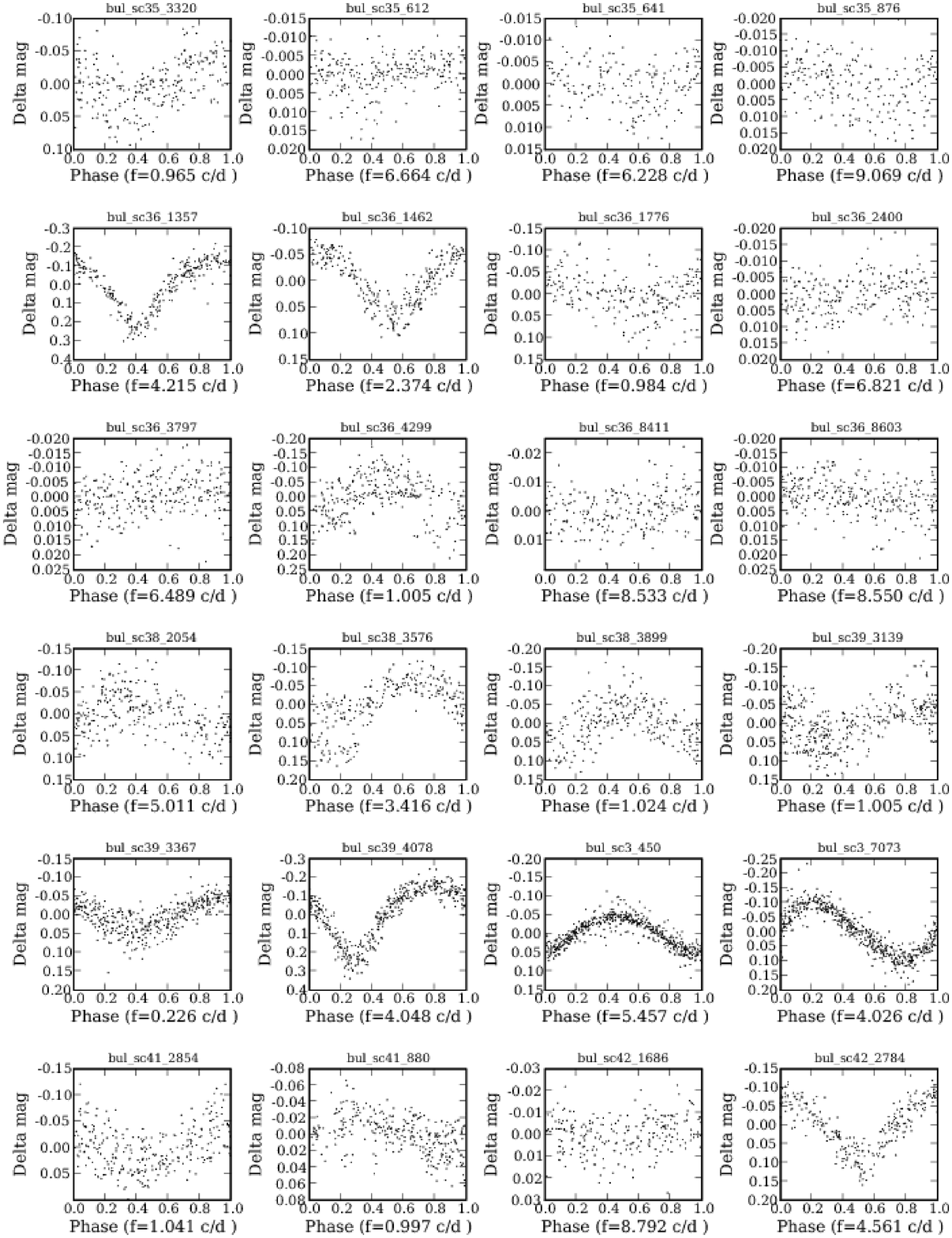}
   \caption{Phase plots of variables in the Galactic Bulge classified as DSCUT with both the MSBN and the GM method, and not present in the list of Pigulski. The OGLE identifier is shown, and the dominant frequency, used to fold the light curves, in units of cycles per day (c/d).}
   \label{DSCUT-BULGE-3}
\end{figure*}

\begin{figure*}
   \centering
   \includegraphics[width=16.5cm,scale=0.7]{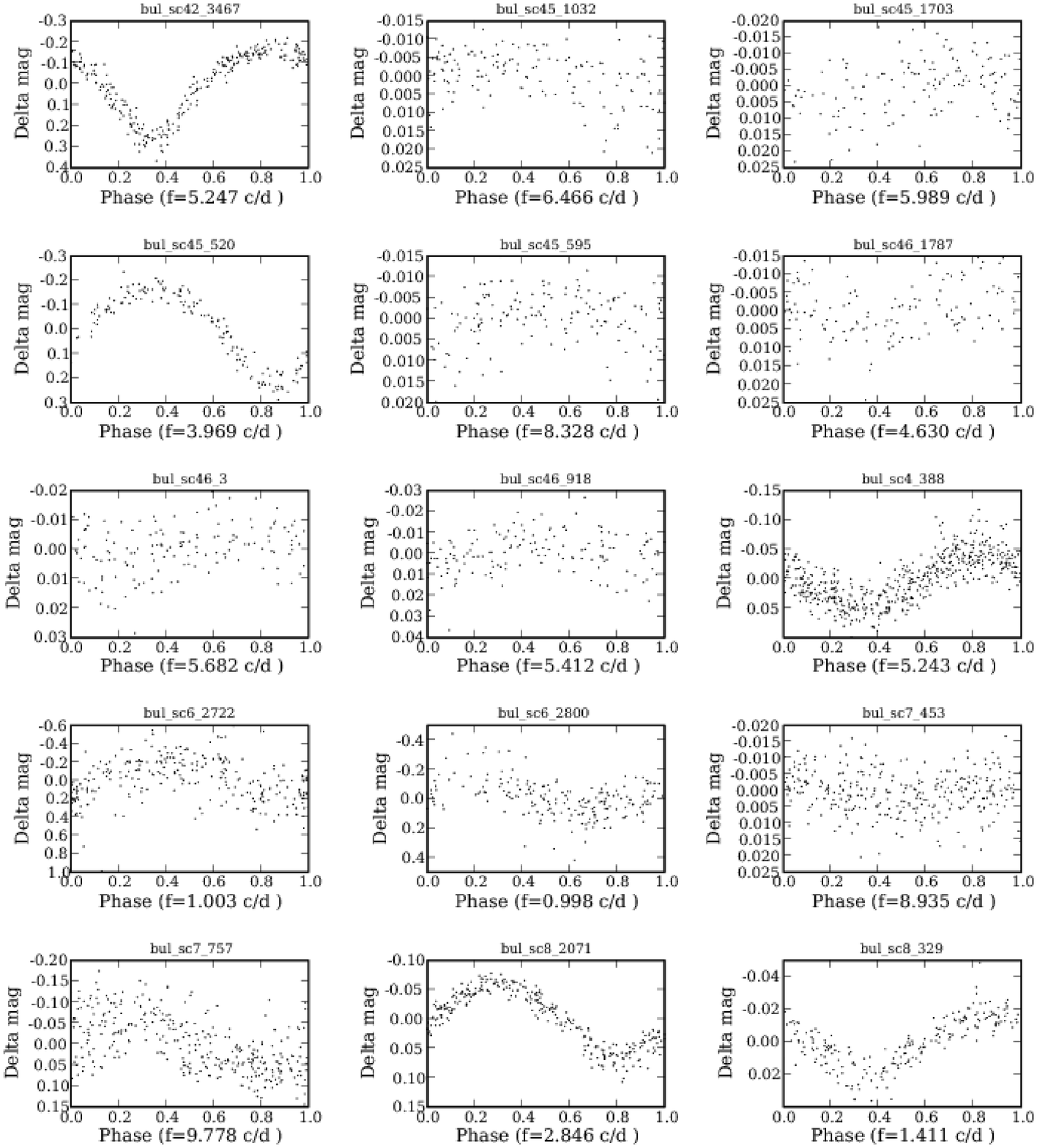}
   \caption{Phase plots of variables in the Galactic Bulge classified as DSCUT with both the MSBN and the GM method, and not present in the list of Pigulski. The OGLE identifier is shown, and the dominant frequency, used to fold the light curves, in units of cycles per day (c/d).}
   \label{DSCUT-BULGE-4}
\end{figure*}

\clearpage
\begin{table}
\caption{Basic light curve and physical properties of Galactic Bulge stars classified as SPB with both the MSBN and the GM method. The dominant frequency $f_1$, the second frequency $f_2$ and the V-I colour index are listed. Several of these stars might be evolved pulsators, termed PVSG here, rather than SPB.}
\renewcommand{\tabcolsep}{1.1mm}
\begin{tabular}{cccc}
\hline
Object identifier&$f_1$ ($c/d$)&$f_2$ ($c/d$)&V-I (mag)\\
\hline
bul\_sc11\_1972&$1.530 $&$1.768 $&$-0.13$\\
bul\_sc11\_78&$2.006 $&$3.852 $&$-0.16$\\
bul\_sc12\_1585&$0.054 $&$0.040 $&$0.01 $\\
bul\_sc12\_2793&$0.943 $&$0.057 $&$0.07 $\\
bul\_sc13\_1232&$0.625 $&$0.313 $&$-0.03$\\
bul\_sc14\_2530&$0.997 $&$2.005 $&$-0.17$\\
bul\_sc18\_2633&$1.005 $&$0.002 $&$0.05 $\\
bul\_sc20\_5489&$0.076 $&$0.038 $&$-0.01$\\
bul\_sc22\_2197&$1.008 $&$1.028 $&$-0.04$\\
bul\_sc26\_1513&$1.005 $&$0.001 $&$-0.15$\\
bul\_sc29\_1066&$0.019 $&$0.001 $&$-0.01$\\
bul\_sc30\_4914&$1.005 $&$0.001 $&$-0.06$\\
bul\_sc32\_1717&$2.910 $&$7.858 $&$-0.12$\\
bul\_sc33\_1385&$0.011 $&$0.020 $&$0.05 $\\
bul\_sc35\_905&$1.006 $&$0.980 $&$-0.13$\\
bul\_sc36\_1523&$0.052 $&$0.052 $&$-0.13$\\
bul\_sc36\_1648&$0.004 $&$0.001 $&$0.07 $\\
bul\_sc36\_6405&$0.054 $&$0.056 $&$0.03 $\\
bul\_sc37\_3960&$0.999 $&$0.002 $&$-0.01$\\
bul\_sc39\_700&$0.028 $&$0.003 $&$-0.02$\\
bul\_sc40\_897&$1.007 $&$1.005 $&$0.04 $\\
bul\_sc42\_1528&$1.005 $&$1.044 $&$0.08 $\\
bul\_sc42\_3374&$1.004 $&$0.997 $&$-0.03$\\
bul\_sc42\_3470&$0.998 $&$1.999 $&$-0.13$\\
\hline
\end{tabular}

\label{SPB-BULGE-param}
\end{table}

\begin{table}
\caption{Basic light curve and physical properties of Galactic Bulge stars classified as GDOR with both the MSBN and the GM method. The dominant frequency $f_1$, the second frequency $f_2$ and the V-I colour index are listed.}
\renewcommand{\tabcolsep}{1.1mm}
\begin{tabular}{cccc}
\hline
Object identifier&$f_1$ ($c/d$)&$f_2$ ($c/d$)&V-I (mag)\\
\hline
bul\_sc1\_1570&$1.114 $&$0.103 $&$0.41 $\\
bul\_sc2\_3661&$3.008 $&$7.808 $&$0.39 $\\
bul\_sc2\_3755&$1.021 $&$0.003 $&$0.47 $\\
bul\_sc3\_7330&$0.997 $&$0.470 $&$0.35 $\\
bul\_sc4\_7947&$2.004 $&$0.011 $&$0.53 $\\
bul\_sc6\_808&$1.005 $&$0.020 $&$0.50 $\\
bul\_sc8\_47&$0.992 $&$0.005 $&$0.49 $\\
bul\_sc9\_410&$1.005 $&$0.065 $&$0.40 $\\
bul\_sc12\_1512&$1.005 $&$2.089 $&$0.49 $\\
bul\_sc12\_1718&$0.999 $&$1.006 $&$0.42 $\\
bul\_sc12\_3188&$2.006 $&$0.542 $&$0.42 $\\
bul\_sc12\_414&$1.006 $&$2.005 $&$0.43 $\\
bul\_sc13\_2598&$0.982 $&$2.004 $&$0.46 $\\
bul\_sc13\_2886&$0.888 $&$6.009 $&$0.54 $\\
bul\_sc13\_313&$2.006 $&$1.002 $&$0.47 $\\
bul\_sc13\_790&$1.006 $&$0.009 $&$0.53 $\\
bul\_sc14\_1153&$1.005 $&$0.998 $&$0.32 $\\
bul\_sc14\_308&$0.940 $&$1.005 $&$0.53 $\\
bul\_sc14\_1954&$1.004 $&$3.673 $&$0.50 $\\
bul\_sc15\_3356&$1.023 $&$1.006 $&$0.36 $\\
bul\_sc16\_867&$1.005 $&$5.711 $&$0.54 $\\
bul\_sc16\_4047&$1.006 $&$7.194 $&$0.46 $\\
bul\_sc18\_2142&$1.003 $&$3.008 $&$0.51 $\\
bul\_sc18\_223&$0.782 $&$0.784 $&$0.54 $\\
bul\_sc18\_5743&$1.004 $&$6.058 $&$0.48 $\\
bul\_sc21\_2549&$2.006 $&$0.043 $&$0.52 $\\
bul\_sc22\_2334&$3.008 $&$0.999 $&$0.52 $\\
bul\_sc22\_560&$1.006 $&$0.050 $&$0.46 $\\
bul\_sc23\_3502&$3.008 $&$0.001 $&$0.46 $\\
bul\_sc23\_1853&$0.954 $&$0.048 $&$0.49 $\\
bul\_sc24\_264&$1.005 $&$8.774 $&$0.47 $\\
bul\_sc25\_476&$1.002 $&$0.988 $&$0.32 $\\
bul\_sc26\_197&$0.999 $&$0.948 $&$0.47 $\\
bul\_sc27\_1538&$1.005 $&$0.096 $&$0.44 $\\
bul\_sc28\_855&$1.479 $&$1.480 $&$0.33 $\\
bul\_sc29\_1270&$1.006 $&$0.007 $&$0.46 $\\
bul\_sc30\_343&$0.999 $&$0.193 $&$0.38 $\\
bul\_sc30\_4533&$1.004 $&$3.971 $&$0.35 $\\
bul\_sc30\_5895&$0.993 $&$0.066 $&$0.54 $\\
bul\_sc31\_3627&$0.903 $&$0.075 $&$0.34 $\\
bul\_sc31\_4572&$1.005 $&$0.055 $&$0.50 $\\
bul\_sc32\_281&$1.002 $&$2.394 $&$0.53 $\\
bul\_sc33\_1203&$1.022 $&$0.032 $&$0.45 $\\
bul\_sc33\_4349&$0.995 $&$1.011 $&$0.34 $\\
bul\_sc35\_107&$3.008 $&$7.215 $&$0.50 $\\
bul\_sc35\_2067&$1.001 $&$1.005 $&$0.45 $\\
bul\_sc35\_22&$1.004 $&$4.042 $&$0.41 $\\
bul\_sc38\_1818&$0.992 $&$4.913 $&$0.51 $\\
bul\_sc38\_4301&$1.005 $&$0.990 $&$0.41 $\\
bul\_sc39\_1901&$1.008 $&$1.003 $&$0.42 $\\
bul\_sc39\_317&$1.006 $&$1.007 $&$0.45 $\\
bul\_sc40\_2162&$1.001 $&$0.019 $&$0.42 $\\
bul\_sc40\_3101&$0.926 $&$0.061 $&$0.47 $\\
bul\_sc40\_3248&$3.008 $&$9.208 $&$0.43 $\\
bul\_sc41\_342&$0.999 $&$0.011 $&$0.53 $\\
bul\_sc42\_1100&$1.004 $&$1.006 $&$0.51 $\\
bul\_sc42\_3669&$1.005 $&$8.812 $&$0.43 $\\
bul\_sc43\_2156&$0.998 $&$0.032 $&$0.36 $\\
bul\_sc45\_1367&$2.002 $&$6.468 $&$0.43 $\\
bul\_sc45\_419&$1.004 $&$5.367 $&$0.41 $\\
bul\_sc46\_1166&$1.999 $&$6.525 $&$0.44 $\\
bul\_sc47\_978&$3.008 $&$0.840 $&$0.46 $\\
bul\_sc48\_952&$2.008 $&$7.209 $&$0.46 $\\
\hline
\end{tabular}

\label{GDOR-BULGE-param}
\end{table}
\clearpage
\begin{table}
\caption{Basic light curve and physical properties of Galactic Bulge stars classified as BCEP with both the MSBN and the GM method. The dominant frequency $f_1$, the second frequency $f_2$ and the V-I colour index are listed. Several of these stars might be evolved pulsators, termed PVSG here, rather than BCEP.}
\renewcommand{\tabcolsep}{1.1mm}
\begin{tabular}{cccc}
\hline
Object identifier&$f_1$ ($c/d$)&$f_2$ ($c/d$)&V-I (mag)\\
\hline
bul\_sc1\_4535&$6.945 $&$6.122 $&$0.39 $\\
bul\_sc1\_2687&$7.001 $&$5.029 $&$0.40 $\\
bul\_sc1\_2837&$6.460 $&$1.006 $&$0.44 $\\
bul\_sc1\_3037&$1.002 $&$0.022 $&$0.49 $\\
bul\_sc1\_188&$1.005 $&$0.976 $&$0.27 $\\
bul\_sc2\_2775&$4.247 $&$8.952 $&$0.41 $\\
bul\_sc2\_4563&$1.004 $&$0.491 $&$0.35 $\\
bul\_sc2\_4974&$6.104 $&$7.784 $&$0.46 $\\
bul\_sc3\_450&$5.457 $&$0.440 $&$0.45 $\\
bul\_sc3\_7073&$4.026 $&$0.438 $&$0.45 $\\
bul\_sc4\_388&$5.243 $&$2.621 $&$0.51 $\\
bul\_sc6\_2722&$1.003 $&$0.005 $&$0.29 $\\
bul\_sc6\_2800&$0.998 $&$1.004 $&$0.33 $\\
bul\_sc6\_987&$4.750 $&$8.153 $&$0.47 $\\
bul\_sc7\_757&$9.778 $&$7.772 $&$0.31 $\\
bul\_sc7\_453&$8.935 $&$6.939 $&$0.52 $\\
bul\_sc8\_2071&$2.846 $&$3.972 $&$0.41 $\\
bul\_sc8\_329&$1.411 $&$1.003 $&$0.41 $\\
bul\_sc9\_332&$4.130 $&$2.854 $&$0.54 $\\
bul\_sc10\_2468&$7.979 $&$8.872 $&$0.42 $\\
bul\_sc26\_2126&$5.238 $&$8.780 $&$-0.28$\\
bul\_sc27\_1648&$4.458 $&$0.553 $&$-0.16$\\
bul\_sc27\_66&$5.013 $&$6.710 $&$-0.16$\\
bul\_sc30\_1667&$0.008 $&$1.002 $&$-0.19$\\
bul\_sc30\_2559&$4.047 $&$1.003 $&$-0.03$\\
bul\_sc31\_4718&$5.220 $&$9.492 $&$-0.18$\\
bul\_sc32\_330&$4.023 $&$0.004 $&$0.17 $\\
bul\_sc34\_1365&$0.038 $&$0.003 $&$-0.17$\\
bul\_sc34\_4718&$5.744 $&$1.404 $&$-0.19$\\
bul\_sc34\_86&$4.836 $&$0.005 $&$-0.14$\\
bul\_sc35\_4443&$4.625 $&$0.003 $&$-0.11$\\
bul\_sc37\_2419&$5.078 $&$1.005 $&$-0.24$\\
bul\_sc38\_1501&$0.039 $&$1.041 $&$-0.30$\\
bul\_sc40\_2137&$6.448 $&$8.439 $&$-0.14$\\
bul\_sc41\_3203&$4.992 $&$1.001 $&$-0.13$\\
bul\_sc42\_1171&$4.965 $&$1.224 $&$-0.26$\\
bul\_sc42\_2823&$0.048 $&$0.049 $&$-0.29$\\
bul\_sc42\_3741&$0.001 $&$0.003 $&$-0.36$\\
bul\_sc46\_1007&$4.742 $&$5.299 $&$-0.18$\\
bul\_sc47\_1051&$4.008 $&$0.998 $&$-0.34$\\
bul\_sc47\_198&$5.017 $&$4.016 $&$-0.29$\\
bul\_sc47\_890&$0.032 $&$0.031 $&$-0.21$\\
bul\_sc48\_717&$0.404 $&$0.001 $&$-0.26$\\
bul\_sc49\_15&$4.633 $&$2.316 $&$0.07 $\\
\hline
\end{tabular}

\label{BCEP-BULGE-param}
\end{table}

\begin{table}
\caption{Basic light curve and physical properties of Galactic Bulge stars classified as DSCUT with both the MSBN and the GM method. The dominant frequency $f_1$, the second frequency $f_2$ and the V-I colour index are listed.}
\renewcommand{\tabcolsep}{1.1mm}
\begin{tabular}{cccc}
\hline
Object identifier&$f_1$ ($c/d$)&$f_2$ ($c/d$)&V-I (mag)\\
\hline
bul\_sc1\_4535&$6.945 $&$6.122 $&$0.39 $\\
bul\_sc1\_2687&$7.001 $&$5.029 $&$0.40 $\\
bul\_sc1\_2837&$6.460 $&$1.006 $&$0.44 $\\
bul\_sc1\_3037&$1.002 $&$0.022 $&$0.49 $\\
bul\_sc1\_188&$1.005 $&$0.976 $&$0.27 $\\
bul\_sc2\_2775&$4.247 $&$8.952 $&$0.41 $\\
bul\_sc2\_4563&$1.004 $&$0.491 $&$0.35 $\\
bul\_sc3\_450&$5.457 $&$0.440 $&$0.45 $\\
bul\_sc3\_7073&$4.026 $&$0.438 $&$0.45 $\\
bul\_sc4\_388&$5.243 $&$2.621 $&$0.51 $\\
bul\_sc6\_2722&$1.003 $&$0.005 $&$0.29 $\\
bul\_sc6\_2800&$0.998 $&$1.004 $&$0.33 $\\
bul\_sc7\_757&$9.778 $&$7.772 $&$0.31 $\\
bul\_sc7\_453&$8.935 $&$6.939 $&$0.52 $\\
bul\_sc8\_2071&$2.846 $&$3.972 $&$0.41 $\\
bul\_sc8\_329&$1.411 $&$1.003 $&$0.41 $\\
bul\_sc9\_332&$4.130 $&$2.854 $&$0.54 $\\
bul\_sc10\_2468&$7.979 $&$8.872 $&$0.42 $\\
bul\_sc11\_7&$6.748 $&$2.538 $&$0.44 $\\
bul\_sc12\_1946&$1.002 $&$0.085 $&$0.23 $\\
bul\_sc12\_3225&$0.043 $&$0.048 $&$0.52 $\\
bul\_sc12\_1204&$7.178 $&$1.004 $&$0.47 $\\
bul\_sc12\_2426&$2.586 $&$1.353 $&$0.50 $\\
bul\_sc13\_2085&$8.929 $&$9.719 $&$0.45 $\\
bul\_sc13\_797&$7.710 $&$9.461 $&$0.37 $\\
bul\_sc13\_7&$7.138 $&$0.001 $&$0.53 $\\
bul\_sc13\_1304&$5.836 $&$0.001 $&$0.70 $\\
bul\_sc14\_1498&$4.220 $&$2.932 $&$0.34 $\\
bul\_sc14\_3656&$1.002 $&$0.023 $&$0.36 $\\
bul\_sc15\_312&$1.024 $&$0.004 $&$0.45 $\\
bul\_sc15\_3175&$3.682 $&$1.841 $&$0.23 $\\
bul\_sc16\_1224&$0.691 $&$0.345 $&$0.16 $\\
bul\_sc16\_1923&$7.736 $&$0.255 $&$0.53 $\\
bul\_sc18\_1640&$1.004 $&$0.992 $&$0.35 $\\
bul\_sc18\_4189&$6.077 $&$4.922 $&$0.51 $\\
bul\_sc18\_4644&$5.687 $&$9.645 $&$0.49 $\\
bul\_sc20\_1765&$2.209 $&$1.006 $&$0.45 $\\
bul\_sc20\_3277&$0.045 $&$1.028 $&$0.38 $\\
bul\_sc20\_5645&$0.982 $&$0.029 $&$0.49 $\\
bul\_sc21\_3755&$9.078 $&$8.027 $&$0.51 $\\
bul\_sc21\_3880&$0.042 $&$0.971 $&$0.40 $\\
bul\_sc21\_5678&$4.192 $&$2.096 $&$0.38 $\\
bul\_sc21\_5189&$9.048 $&$7.945 $&$0.54 $\\
bul\_sc21\_604&$1.033 $&$0.004 $&$0.35 $\\
bul\_sc22\_3655&$1.001 $&$0.038 $&$0.37 $\\
bul\_sc23\_1535&$0.037 $&$0.028 $&$0.40 $\\
bul\_sc26\_1277&$7.105 $&$8.102 $&$0.46 $\\
bul\_sc26\_4493&$0.997 $&$0.019 $&$0.24 $\\
bul\_sc29\_299&$5.021 $&$9.005 $&$0.47 $\\
bul\_sc29\_527&$0.997 $&$0.001 $&$0.14 $\\
bul\_sc29\_786&$0.959 $&$1.958 $&$0.40 $\\
bul\_sc29\_852&$0.991 $&$1.001 $&$0.26 $\\
bul\_sc30\_2425&$6.092 $&$3.046 $&$0.42 $\\
bul\_sc30\_2682&$0.034 $&$0.017 $&$0.37 $\\
bul\_sc31\_2266&$0.040 $&$0.020 $&$0.48 $\\
bul\_sc32\_4322&$4.319 $&$0.008 $&$0.31 $\\
bul\_sc34\_263&$1.042 $&$0.029 $&$0.20 $\\
bul\_sc35\_2654&$6.113 $&$0.001 $&$0.52 $\\
bul\_sc35\_3320&$0.965 $&$1.001 $&$0.27 $\\
bul\_sc35\_612&$6.664 $&$4.674 $&$0.46 $\\
bul\_sc35\_641&$6.228 $&$8.510 $&$0.42 $\\
bul\_sc35\_876&$9.069 $&$8.432 $&$0.43 $\\
\hline
\end{tabular}

\label{DSCUT-BULGE-param-1}
\end{table}

\clearpage

\begin{table}
\caption{Basic light curve and physical properties of Galactic Bulge stars classified as DSCUT with both the MSBN and the GM method (continued). The dominant frequency $f_1$, the second frequency $f_2$ and the V-I colour index are listed.}
\renewcommand{\tabcolsep}{1.1mm}
\begin{tabular}{cccc}
\hline
Object identifier&$f_1$ ($c/d$)&$f_2$ ($c/d$)&V-I (mag)\\
\hline
bul\_sc36\_1462&$2.374 $&$1.187 $&$0.42 $\\
bul\_sc36\_1357&$4.215 $&$7.326 $&$0.51 $\\
bul\_sc36\_1776&$0.984 $&$0.004 $&$0.48 $\\
bul\_sc36\_2400&$6.821 $&$6.853 $&$0.50 $\\
bul\_sc36\_3797&$6.489 $&$5.983 $&$0.35 $\\
bul\_sc36\_4299&$1.005 $&$0.999 $&$0.53 $\\
bul\_sc36\_8411&$8.533 $&$1.673 $&$0.50 $\\
bul\_sc36\_8603&$8.550 $&$5.868 $&$0.49 $\\
bul\_sc38\_2054&$5.011 $&$5.162 $&$0.47 $\\
bul\_sc38\_3576&$3.416 $&$1.708 $&$0.43 $\\
bul\_sc38\_3899&$1.024 $&$0.001 $&$0.18 $\\
bul\_sc39\_3139&$1.005 $&$0.016 $&$0.22 $\\
bul\_sc39\_3367&$0.226 $&$0.113 $&$0.40 $\\
bul\_sc39\_4078&$4.048 $&$1.003 $&$0.49 $\\
bul\_sc41\_2854&$1.041 $&$0.040 $&$0.46 $\\
bul\_sc41\_880&$0.997 $&$1.026 $&$0.51 $\\
bul\_sc42\_1686&$8.792 $&$5.969 $&$0.51 $\\
bul\_sc42\_2784&$4.561 $&$5.920 $&$0.46 $\\
bul\_sc42\_3467&$5.247 $&$0.001 $&$0.17 $\\
bul\_sc45\_1032&$6.466 $&$7.232 $&$0.44 $\\
bul\_sc45\_1703&$5.989 $&$4.237 $&$0.53 $\\
bul\_sc45\_520&$3.969 $&$3.687 $&$0.23 $\\
bul\_sc45\_595&$8.328 $&$1.674 $&$0.38 $\\
bul\_sc46\_1787&$4.630 $&$9.715 $&$0.48 $\\
bul\_sc46\_3&$5.682 $&$0.476 $&$0.48 $\\
bul\_sc46\_918&$5.412 $&$6.028 $&$0.41 $\\
\hline
\end{tabular}

\label{DSCUT-BULGE-param-2}
\end{table}

\end{document}